\documentclass[useAMS,usenatbib,usegraphicx]{mn2e}

\title[IRTF observations of white dwarfs with possible near-infrared excess]{IRTF observations of white dwarfs with possible\\ near-infrared excess}

\author[J. Farihi]{J. Farihi$^{1,2}$\\
$^{1}$Department of Physics \& Astronomy,
			University of Leicester,
			Leicester LE1 7RH, 
			UK; jf123@star.le.ac.uk\\
$^{2}$Visiting Astronomer at the NASA 
			Infrared Telescope Facility}

\begin{document}

\date{}

%\pagerange{\pageref{firstpage}--\pageref{lastpage}} \pubyear{2009}

\maketitle

\label{firstpage}

\begin{abstract}
Near-infrared photometry and spectroscopy is obtained for a heterogeneous sample of nearby 
white dwarfs with possible excess flux as identified primarily in the Two Micron All Sky Survey.  
Among the sample of 43 stars are a number of white dwarfs that are either metal-rich, magnetic, 
or binary suspects.  With a few notable exceptions in four (or possibly five) distinct categories, 
the newly obtained $JHK$ photometric data fail to corroborate the putative excesses, with $<K
_{\rm IRTF} - K_{\rm 2MASS}>=+0.31$ mag.  Where available, {\em GALEX} photometric data 
are used to better constrain the overall spectral energy distribution of the white dwarfs, enabling 
any excess near-infrared flux to stand out more readily against the expected stellar photosphere.  

With superior data, a near-infrared photometric excess is confirmed at three metal-rich white 
dwarfs and ruled out at nine others.  Several new binaries are confirmed or suggested; five 
white dwarf - red dwarf pairs and five double degenerates.  Four apparently single magnetic
white dwarfs -- two DA and two DQp -- display modest to strong near-infrared excess (relative 
to non-magnetic models), which may be better described as two effective temperatures owing 
to a redistribution of energy in highly magnetic or peculiar atmospheres.
\end{abstract}

\begin{keywords}
	binaries: general ---
	circumstellar matter---
	infrared: stars--
	stars: evolution---
	stars: low-mass, brown dwarfs ---
	white dwarfs
\end{keywords}

\section{INTRODUCTION}

The physical parameters of white dwarfs make them excellent targets for binary and multiple
system studies.  These faint stellar embers permit a relatively unobstructed view, across multiple 
wavelengths, of their most common companions, low mass main-sequence stars \citep{far05}.
Their Earth-sized radii offer the ultimate, natural, low contrast background for the near-infrared
detection of all types of intrinsically faint companions: cool white dwarfs \citep{zuc97}, ultracool 
main-sequence dwarfs \citep{pro83}, brown dwarfs \citep{bec88}, and planets \citep{bur08}.  
Furthermore, their typical warm to hot stellar effective temperatures give white dwarfs a distinct 
ultraviolet signature, often detectable against early-type main-sequence companions 
\citep{bur98}.

White dwarfs come with a hard lower limit on their total age, an attribute lacking in any given 
non-degenerate field star.  Together with an accurate mass estimate, the most likely total age 
of a white dwarf can be inferred from its most likely progenitor mass and lifetime \citep{wil09,
kal08,dob06}.  In this way, white dwarfs with substellar companions offer the most accessible,
and best empirical tests of brown dwarf and planetary cooling models at intermediate to older 
ages \citep{ste09,bur09,far08a}.

In recent years, the advantageous, compact nature of white dwarfs has been extended via
the discovery of dust within the Roche radius of more than one dozen stars \citep{far09,von07,
jur07}, the likely result of tidally-disrupted minor bodies perturbed into close approach.  This 
orbital phase space is essentially covered by the diameter of main-sequence stars, where a 
similar encounter would instead produce an impact.  While near-infrared excess emission is 
measured for about half the white dwarfs with circumstellar dust \citep{far09,kil06}, such warm 
dust is virtually unheard of at their main-sequence progenitors \citep{zuc08}.

This paper presents the results of an extensive photometric search for near-infrared excess
due to low mass stellar and substellar companions, cool white dwarf companions, and warm 
circumstellar dust.  Excess emission is confirmed at a few to several stars in each of these
categories, as well as in another group that may be associated with magnetism, peculiar 
atmospheric composition, or both.  The observations and data are presented in \S2 and 
\S3, with detailed results on individual objects given in \S4, and a brief discussion in \S5.

\begin{table*} 
\centering
\begin{minipage}{100mm}
\caption{SpeX Target Stars\label{tbl1}} 
\begin{tabular}{@{}lllrlr@{}}
\hline

WD		&Name	&Spectral		& $T_{\rm eff}^a$	&Data	&Reference\\
 		&		&Type		&(K)				&Type	&\\
 
\hline
 
0023$-$109	&G158-78 	&DA+DC+dM	&10400	&Phot		&1,2\\
0106$-$328	&HE 			&DAZ		&15700	&Phot		&3\\	
0108$+$277	&NLTT 3915 	&DA			&5300	&Phot		&4,5\\
0146$+$187	&GD 16 		&DAZB		&11500	&Phot		&6\\	
0155$+$069	&GD 20 		&DA			&20600	&Phot		&7\\
0156$+$155	&PG 			&DC			&9800	&Phot		&8\\
0235$+$064	&PG 			&DAZ+dM		&13500	&Phot		&9,10\\	
0253$+$508	&KPD 		&DAP		&20000	&Phot		&11\\
0257$-$005	&KUV 		&DAO+dM	&80900	&Phot		&1\\
0408$-$041	&GD 56 		&DAZ		&14400	&Phot		&3\\	
0518$+$333	&EG 43 		&DA+dM		&9500	&Phot		&1,12\\
0939$+$071	&PG			&UV+dF		&...		&Spec		&1\\
0956$+$045	&PG 			&DA			&18200	&Phot		&13\\
1000$+$220	&TON 1145 	&UV+dF		&...		&Phot/Spec	&14\\
1013$-$010	&G53-38 		&DA+DC		&8800	&Phot/Spec	&15,16\\	
1036$-$204	&LHS 2293 	&DQp		&7500	&Phot/Spec	&17\\
1108$+$325	&TON 60		&DA+dM		&63000	&Phot		&13\\
1133$+$489	&PG 			&DO+dM		&47500	&Phot		&18\\
1140$+$004	&SDSS 		&DA+dM		&14400	&Phot		&19,20\\
1156$+$132	&LP 494-12 	&DQp		&...		&Phot		&1\\
1225$-$079	&PG 			&DZA		&10500	&Phot/Spec	&21\\	
1254$+$345	&HS	 		&DAH		&15000	&Phot/Spec	&22\\
1330$+$015	&G62-46 		&DAH+DC	&6000	&Phot		&23\\
1339$+$346	&PG 			&DA			&16000	&Phot		&13\\
1350$-$090	&LP 907-37 	&DAP		&9500	&Phot/Spec	&13\\
1428$+$373	&PG 			&DA+DC		&14000	&Phot/Spec	&13\\	
1434$+$289	&TON 210	&DA			&33000	&Phot/Spec	&13\\
1455$+$298	&G166-58 	&DAZ		&7400	&Phot/Spec	&12,13\\
1457$-$086	&PG 			&DAZ		&20400	&Phot		&3\\
1619$+$525	&PG 			&DA+dM+dM	&18000	&Phot		&13\\
1626$+$368	&G180-57 	&DZA		&8400	&Spec		&24\\
1653$+$385	&NLTT 43806 	&DAZ		&5700	&Phot		&4\\
1845$+$683	&KUV 		&DA			&36000	&Phot		&25\\
2032$+$188	&GD 231		&DA+DC		&18500	&Phot		&26,27\\	
2144$-$079	&G26-31		&DBZ		&16500	&Phot		&28\\
2201$-$228	&HE 			&DB			&18000	&Phot		&29\\
2211$+$372	&LHS 3779 	&DC			&6000	&Phot		&30\\
2215$+$388	&GD 401 		&DZ			&8800	&Phot		&31\\
2216$+$484	&GD 402 		&DA+DC		&7000	&Phot		&32\\
2316$+$123	&KUV 		&DAP		&10400	&Phot		&11\\
2333$-$049	&G157-82 	&DA			&10500	&Phot		&33\\
2336$-$187	&G273-97 	&DA			&8100	&Phot		&33\\	
2354$+$159	&PG 			&DBZ		&24400	&Phot		&28\\	

\hline
\end{tabular}

$^a$ The effective temperatures in the fourth column are previously published values unless 
otherwise stated; these values do not always agree with the fits shown in the figures.\\

References:
1) This work;
2) \citealt{cat08};
3) \citealt{koe05a};
4) \citealt{kaw06};
5) \citealt{far09};
6) \citealt{koe05b};
7) \citealt{hom98};
8) \citealt{put97};
9) \citealt{zuc03};
10) \citealt{far08b};
11) \citealt{lie85};
12) \citealt{ber01};
13) \citealt{lie05a};
14) \citealt{gre86};
15) \citealt{pau06};
16) \citealt{nel05};
17) \citealt{lie03};
18) \citealt{dre96};
19) \citealt{kle04};
20) \citealt{ven02};
21) \citealt{wol02};
22) \citealt{hag87};
23) \citealt{ber93};
24) \citealt{duf07};
25) \citealt{nap99};
26) \citealt{ber92};
27) \citealt{mor05};
28) \citealt{vos07};
29) \citealt{jor01};
30) \citealt{gre90};
31) \citealt{dup93};
32) \citealt{ber90};
33) \citealt{koe01}

\end{minipage}
\end{table*}

\section{OBSERVATIONS \& DATA}

The majority of the 43 targets listed in Table \ref{tbl1} come from the white dwarf catalog of
\citet{mcc99}, and were selected because their Two Micron All Sky Survey (2MASS) $JHK_s$ 
photometry suggested near-infrared excess emission.  These are primarily white dwarf plus low 
mass stellar or substellar companion candidates, a significant fraction of which are taken from 
\citet{far05}, and \citet{wac03}.  A small number of these stars are known or suspected magnetic 
white dwarfs, but were not intentionally selected as such, and hence these targets overlap with
the sample analyzed by \citet{wel05}.  Another subset of targets were chosen from various 
literature sources as binary suspects based on optical spectroscopy and photometry; many 
of these are double degenerate suspects (e.g. \citealt{ber90}).  A final subset of targets are 
metal-rich white dwarfs with low or questionable quality 2MASS photometry, most of which 
are part of various {\em Spitzer} programs aimed at searching for circumstellar dust 
\citep{far09,far08b,jur07}.

Data were obtained at the NASA Infrared Telescope Facility (IRTF) using the medium-resolution
spectrograph and imager SpeX \citep{rayn03} on 2006 October $7-9$ and 2007 April $7-9$.  
Conditions were photometric or near-photometric for all of the imaging observations, while thin 
cirrus was present during some of the spectroscopy.  The instrument was used primarily for 
imaging, but also in its intended spectrographic configuration.  Science target images were 
taken at $JHK$ using individual exposure times that were typically 30 seconds.  A seven point 
dither pattern was repeated once or twice, resulting in total integration times between 3.5 and 
7.0 minutes.  Photometric standard stars were observed in a similar manner a few to several 
times during each night to measure the zero point in each passband.

Spectroscopy of select targets was performed with SpeX using the low-resolution prism mode
that covers the entire $0.8-2.5$ $\mu$m region in a single exposure setting.  The $0\farcs8$ slit 
was used to maximize gathered light at the expense of resolution, resulting in $R\approx100$
at $H$ band.  Individual exposures of 120 seconds were used at two positions along the slit,
repeated ten times for a total integration time of 20 minutes.  An A0V telluric standard star was
observed immediately following each science observation, as were spectral flat fields and arc 
lamp images.

Near-infrared images for 41 targets were reduced in the standard fashion, with long exposure 
sky frames normalized to serve as flat fields.  Images of each target at each bandpass were 
sky-subtracted by removing the median of the raw image stack, flat-fielded, then registered and 
recombined by averaging.  Photometry of science targets and standard stars was performed 
with aperture radii between $r=10$ and 20 pixels, corresponding to $1\farcs5-3\farcs0$, and 
generally measured at the widest possible radius while avoiding contamination from nearby 
objects or companions.  Where possible, 2MASS sources within the SpeX field of view were 
used to corroborate the photometric calibration.  Median extinction values for Mauna Kea were 
used to correct all photometry to airmass 1.00, and the zero point of each filter for each night 
was established by averaging the measurements for three to four standard stars.  The zero 
points for all nights during both observing runs were found to agree within a few percent.

Spectroscopic data for 11 targets were reduced with Spextool \citep{cus04}, including 
sky-subtraction, flat-fielding, spectral extraction and averaging.  Spextool also was used 
to perform wavelength calibration, telluric feature removal, sensitivity correction, and flux 
calibration.  However, due to slit losses and other sources of error, the final flux calibrations
of the science target spectra were established using photometry.

\section{ANALYSIS}

The resulting near-infrared photometry and spectroscopy for all 43 targets are listed in Table 
\ref{tbl2} and plotted together with the ultraviolet and optical spectral energy distributions in 
Figures \ref{fig1} -- \ref{fig52}, ordered by right ascension.  Optical photometric data plotted in 
the figures were taken from various literature sources, including but not limited to the Sloan 
Digital Sky Survey \citep{ade08}, DENIS \citep{den05}, the white dwarf catalog of \citet{mcc99} 
and references therein.  There are a few cases with photographic photometry \citep{sts06,cmc06,
zac05,mon03} where superior data were unavailable.  The SDSS fluxes were assumed to have 
5\% errors for the following reasons: 1) the quoted photometric errors in SDSS DR7 are often 
less than 0.01 mag, which is unlikely to be realistic; 2) where two flux measurements of a single
source are available, they sometimes differ by 5\% or greater in a given bandpass; 3) many of
the flux measurements carry middle-grade photometric quality flags; and 4) several objects
cataloged as point sources are in fact members of close double or multiple systems, implying 
further photometric error.  Additionally, available {\em GALEX} far- and near-ultraviolet fluxes are 
included in the plots; a weighted average was taken where multiple measurements are given.  
These fluxes are uncorrected for extinction, and minimum 10\% errors have been assigned.  
Where available, 2MASS $JHK_s$ photometry \citep{skr06} is plotted alongside the IRTF data 
for direct comparison.   

The optical and ultraviolet photometry for all stars were fitted with model spectra of the 
appropriate effective temperatures, which sometimes deviated from previously published
literature values.  Hydrogen atmosphere stars were fitted with log $g=8.0$ white dwarf spectral 
models of \citet{koe09}, while blackbody spectra were used for helium-rich stars.  In most cases, 
the model $T_{\rm eff}$ could be adjusted to match the near-infrared photometry, consistent with 
emission from the stellar photosphere.  In cases where a potential near-infrared excess was 
indicated in a single or multiple bandpasses, those data were not employed to adjust the model 
fits.  The white dwarf mass (via log $g$) was not varied in the model fits, as the focus of the study 
was to establish the presence or absence of a near-infrared excess, which is best achieved by 
constraining the stellar $T_{\rm eff}$.  While white dwarf colors are, to second order, affected by 
log $g$, errors resulting from fixing log $g$ in the model fits should be of the same order or 
smaller than the uncertainty in $T_{\rm eff}$ itself \citep{ber95b}.  Where utilized for distance 
estimates to binary and multiple system components (Table \ref{tbl3}), white dwarf model 
parameters (masses, radii, colors, and absolute magnitudes) were taken from \citet{ber95a,
ber95b}, while empirical low mass stellar parameters were evaluated using the colors and 
absolute magnitudes given in \citet{kir94}.

\begin{table*} 
\centering
\begin{minipage}{115mm}
\caption{Near-Infrared Photometry\label{tbl2}} 
\begin{tabular}{@{}lccccccr@{}}
\hline

WD	&$J_{\rm 2MASS}$	&$H_{\rm 2MASS}$	&$K_{\rm 2MASS}$	&$J_{\rm IRTF}$	&$H_{\rm IRTF}$	&$K_{\rm IRTF}$	&Excess\\
	&(mag)			&(mag)			&(mag)			&(mag)			&(mag)			&(mag)			&Type$^a$\\
 
\hline
 
0023$-$109		&16.05(08)	&15.84(17)	&15.68(25)	&15.97	&15.85	&15.93		&1\\
0106$-$328		&15.75(08)	&15.78(17)	&...     		&15.75	&15.78	&15.92		&...\\
0108$+$277		&15.22(06)	&15.04(08)	&14.87(13)	&15.03	&14.79	&14.72		&...\\
0146$+$187		&15.80(08)	&15.53(10)	&15.36(16)	&15.55	&15.51	&15.29		&2\\
0155$+$069		&15.81(08)	&15.47(13)	&...     		&15.75	&15.78	&15.96		&...\\
0156$+$155		&15.74(08)	&15.44(11)	&...     		&15.60	&15.56	&15.57		&...\\
0235$+$064		&15.69(07)	&15.91(20)	&...     		&15.73	&15.77	&15.87		&...\\
0253$+$508		&15.58(06)	&15.31(11)	&15.07(12)	&15.44	&15.36	&15.38		&3\\
0257$-$005		&16.77(14)	&16.34(21)	&15.53(20)	&16.52	&16.18	&15.96		&4\\
0408$-$041		&15.87(06)	&15.99(13)	&15.44(18)	&15.85	&15.75	&15.13		&2\\
0518$+$333		&15.37(09)	&15.12(21)	&14.81(16)	&15.91	&15.87	&15.89		&1\\
0939$+$071		&13.99(03)	&13.68(04)	&13.60(03)	&...     	&...     	&...     		&...\\
0956$+$045A		&...     		&...     		&...     		&16.41	&16.38	&16.45		&...\\
0956$+$045B$^b$	&14.66(04)	&14.13(05)	&13.84(06)	&14.67	&14.21	&13.85		&...\\
1000$+$220		&16.28(10)	&15.97(21)	&15.67(23)	&16.21	&16.21	&16.33		&...\\
1013$-$010		&14.79(03)	&14.61(06)	&14.71(12)	&14.80	&14.65	&14.64		&1\\
1036$-$204		&14.63(03)	&14.35(04)	&14.04(07)	&14.62	&14.31	&14.09		&5\\
1108$+$325		&15.80(07)	&15.19(08)	&15.23(18)	&15.77	&15.29	&15.13		&4\\
1133$+$489		&16.19(09)	&15.68(14)	&15.41(16)	&16.16	&15.75	&15.55		&4\\
1140$+$004		&15.55(07)	&15.06(08)	&15.12(16)	&15.60	&15.12	&14.83		&4\\
1156$+$132		&16.95(15)	&...     		&...     		&17.32	&17.38	&17.08		&5\\
1225$-$079		&14.88(04)	&14.92(08)	&14.85(12)	&14.78	&14.80	&14.80		&...\\
1254$+$345		&16.70(13)	&16.14(19)	&...     		&16.58	&16.50	&16.56		&...\\
1330$+$015		&16.40(12)	&16.30(21)	&...     		&16.31	&16.27	&16.24		&1\\
1339$+$346		&14.09(03)	&13.70(03)	&13.59(03)	&16.06	&16.07	&16.15		&...\\
1350$-$090		&...     		&14.15(05)	&...     		&14.23	&14.12	&14.12		&...\\
1428$+$373		&15.80(06)	&15.57(09)	&15.58(23)	&15.54	&15.61	&15.74		&...\\
1434$+$289		&16.51(12)	&16.33(20)	&15.92(29)	&16.43	&16.52	&16.70		&...\\
1455$+$298		&14.97(05)	&14.61(08)	&14.74(13)	&14.85	&14.74	&14.72		&...\\
1457$-$086		&16.04(10)	&16.21(23)	&15.62(23)	&16.08	&16.10	&16.03		&2\\
1619$+$525AB	&...     		&...     		&...     		&15.34	&15.10	&14.95		&4\\
1619$+$525C$^b$	&14.17(03)	&13.55(04)	&13.43(04)	&14.18	&13.60	&13.39		&...\\
1626$+$368		&13.64(02)	&13.65(03)	&13.58(04)	&...     	&...     	&...     		&...\\
1653$+$385		&15.53(06)	&15.35(11)	&15.27(15)	&15.62	&15.36	&15.25		&...\\
1845$+$683		&16.07(09)	&16.28(22)	&...     		&15.98	&16.24	&16.30		&...\\
2032$+$188		&15.80(10)	&15.53(17)	&15.26(23)	&15.68	&15.71	&15.73		&...\\
2144$-$079		&15.03(05)	&15.21(10)	&14.90(14)	&15.14	&15.21	&15.31		&...\\
2201$-$228		&16.28(10)	&15.97(21)	&15.67(23)	&16.21	&16.21	&16.33		&...\\
2211$+$372		&16.25(10)	&16.06(19)	&15.62(24)	&16.12	&16.05	&15.89		&...\\
2215$+$388		&16.14(10)	&...     		&...     		&16.02	&16.01	&16.03		&...\\
2216$+$484		&15.48(06)	&15.30(10)	&15.32(16)	&15.36	&15.25	&15.18		&1\\
2316$+$123		&15.47(06)	&15.72(15)	&15.14(19)	&15.49	&15.51	&15.56		&3\\
2333$-$049		&15.73(05)	&15.72(11)	&15.48(19)	&15.71	&15.74	&15.66		&...\\
2336$-$187		&15.06(04)	&14.94(06)	&14.68(09)	&15.04	&14.92	&14.89		&...\\
2354$+$159		&16.22(12)	&16.02(17)	&15.52(24)	&16.19	&16.21	&16.27		&...\\

\hline
\end{tabular}

Photometric errors for the IRTF data are 5\%.\\

$^a$ The near-infrared excesses listed in the final column are associated with: 
1) a cool white dwarf companion; 2) warm circumstellar dust; 3) a magnetic DA, relative to 
non-magnetic models; 4) a red dwarf companion;  5) a magnetic DQ, relative to non-magnetic 
models.\\

$^b$ Not a white dwarf, but a spatially resolved red dwarf companion

\end{minipage}
\end{table*}

\section{RESULTS FOR INDIVIDUAL OBJECTS}

\paragraph*{{\em 0023$-$109}}.  G158-78 is a likely triple system.  \citet{egg65} first reported 
a co-moving, cool companion, G158-77, henceforth designated G158-78B.  The wide pair are 
separated by $59\farcs4$ at position angle $330\degr$ based on analysis of archival plates 
dating from 1954.7, 1983.6 and 1998.8.  Modern astrometry yields a proper motion of $(72,-205)$ 
mas yr$^{-1}$ for the white dwarf \citep{sal03}, and $(74,-206)$ mas yr$^{-1}$ for the secondary 
\citep{zac04}.  G158-78B has $V-K=3.5$ \citep{rei96}, indicating a spectral type of M0V at a 
photometric distance of 92 pc \citep{kir94}.

\citet{rei96} reported radial velocity variations in the white dwarf G158-78A, with at least one 
value clearly incompatible with the radial velocity of G158-78B.  While the broad-band optical 
colors of the primary suggest a 7000 K star, high quality spectroscopic data reveal an effective
temperature near 10000 K \citep{cat08,kle04}.  This higher temperature, necessary to explain 
the {\em GALEX} data for this white dwarf, implies the red optical and near-infrared fluxes are 
greater than expected for a single star.  Figure \ref{fig1} shows the complete spectral energy 
distribution is well reproduced by the addition of a low-luminosity companion with temperature 
around 6500 K.  The implied tertiary star, G158-78C, can only be another white dwarf.

\paragraph*{{\em 0106$-$328}}.  The coordinates of this warm DAZ star are inaccurate in 
SIMBAD, but correct to within several arcseconds in the current online version of \citet{mcc08}.  
Its 2MASS position is $01^{\rm h} 08^{\rm m} 36.03^{\rm s}$ $-32\degr 37\arcmin 43\farcs6$ 
(J2000).  There is no photometric nor spectroscopic $K$-band excess at this polluted white 
dwarf \citep{kil08}.

\paragraph*{{\em 0108$+$277}}.  NLTT 3915 is located a few arcseconds away from a 
background star \citep{far09} that is perhaps responsible (by contaminating sky apertures) 
for the lower fluxes reported in 2MASS versus the IRTF values.

\paragraph*{{\em 0146$+$187}}.  GD 16 has a mid-infrared excess from warm circumstellar 
dust \citep{far09}.  The IRTF photometry disagrees somewhat with 2MASS at $J$ band, but is 
consistent with the white dwarf effective temperature \citep{koe05b}.  The {\em GALEX} detection 
of GD 16 appears on the edge of the array, and is unlikely to be reliable.

\paragraph*{{\em 0155$+$069, 0156$+$155, 1434$+$289, 2144$-$079, \& 2211$+$372}}.  
The apparent 2MASS excesses at these stars are spurious.

\paragraph*{{\em 0235$+$064}}.  This PG star is located several arcseconds from a 
common-proper motion companion M dwarf.  The companion has likely caused some prior 
studies of the white dwarf to underestimate its effective temperature via contamination of optical 
data \citep{far08b}.  All data presented here is unaffected by the light of the dM companion, and 
a model fit to the full spectral energy distribution yields $T_{\rm eff}\approx13500$ K.  Assuming 
log $g=8.0$ places the system at 62 pc, consistent with a companion spectral type of M3.5 
($V-K=4.9$) at that distance.

\paragraph*{{\em 0253$+$508}}.  The ultraviolet and optical data for this star are well-modeled 
with a $T_{\rm eff}=20000$ K hydrogen atmosphere, non-magnetic white dwarf \citep{lie85,
dow83}, as can be seen in Figure \ref{fig8}.  However, the $V-K=-0.16$ color is suggestive of a 
much cooler star, near 10000 K.  Whether the $JHK$ excess (relative to non-magnetic models) 
emission is photospheric in nature remains to be seen, but the fluxes are not consistent with 
a low mass star or brown dwarf \citep{far05}, nor circumstellar dust \citep{kil06}.  It is possible 
that the highly magnetic stellar atmosphere is causing an emergent two-temperature appearance 
and further investigation is warranted.

\paragraph*{{\em 0257$-$005}}.  This system is resolved into two components separated by 
$1\farcs0$ in {\em HST} / ACS observations (Farihi et al. 2009, in preparation).  The binary is 
also partially resolved in the IRTF observations, but only photometry for the composite double 
was possible.  Near-infrared data alone does not strongly constrain the companion parameters, 
but the ACS data yield $I-K=2.6$ and an estimated spectral type of M4.5 at 490 pc.  

\citet{eis06} list the primary star as type sdO with $T_{\rm eff}=69000$ K and log $g=5.9$, yet 
only white dwarfs should attain such high temperatures.  These parameters may be due to a 
problematic automated fit to the SDSS spectrum and photometry; a mild red continuum and a 
few late-type stellar features are present in the data beyond 6000 \AA.  Visual examination of 
the spectrum indicates a likely DAO white dwarf, a hypothesis corroborated by the implied 
distance to the M dwarf companion; at 490 pc, the primary would be intrinsically too faint for 
a horizontal branch star \citep{lis05}.

Following the method of \citet{reb07}, a fit to the SDSS spectrum yields  $T_{\rm eff}=80900
\pm7600$ K and log $g=7.13\pm0.34$, verifying KUV 0257$-$005 is a DAO white dwarf.  
Allowing for some mild extinction in the optical yields a photometric distance of $1040\pm
330$ pc, which can only be matched by the cool companion if it is an equal luminosity double
at 700 pc.  Further observations are required to test this scenario.

\paragraph*{{\em 0408$-$041}}.  GD 56 has a rare, strong infrared excess due to circumstellar 
dust that manifests beginning at $H$ band \citep{kil06}.  The IRTF photometry corroborates this 
finding, and improves significantly upon 2MASS data.

\paragraph*{{\em 0518$+$333}}.  EG 43 (G86-B1B, NLTT 14920) is part of a common-proper 
motion binary with an M dwarf (G86-B1A, NLTT 14919; \citealt{sal03}) $7\farcs4$ distant.  The 
white dwarf itself is a double degenerate suspect, but that assessment was based on $T_{\rm eff}
\approx7800$ K \citep{ber01}, and almost certainly a result of $VRIJHK$ photometry contaminated 
by the M dwarf companion.  This is a good possibility, as the M dwarf outshines the white dwarf at 
$B$ and $V$ bands by around 0.4 and 1.8 magnitudes, respectively \citep{rei96,mas95}.  
Furthermore, the IRTF data differ quite significantly from the near-infrared data of \citet{ber01}, 
as can be seen in Figure \ref{fig11}.

Using the $UBV$ photometry from \citet{mas95}, a 9500 K DA white dwarf model appears to 
reproduce the $UBVJHK$ data rather well.  At the parallax distance of 65.4 pc ($\pi=15.3\pm
2.2$ mas; \citealt{van95}), the white dwarf should have $M_V=12.0\pm0.3$ mag, a range that 
includes log $g=8.0$ at this higher effective temperature.  Therefore, the white dwarf should no 
longer be considered a double degenerate suspect.

The M dwarf companion has $V-K=4.5$, consistent with a spectral type of M3.  However,
the implied $M_K=5.7$ mag at 65.4 pc is brighter than expected for this spectral type.  Indeed, 
this star is a visual binary separated by $0\farcs2$ with {\em HST} / ACS, making the system a 
triple (Farihi et al. 2009, in preparation).  Assuming equal luminosity gives each star in 
G86-B1AC $M_K=6.5$ mag, consistent with M3 spectral types implied by their color.

\paragraph*{{\em 0939$+$071}}.  This star has persisted among white dwarf lists over the years 
lacking any high quality observations in the literature.  Its status as a degenerate star rests firstly 
with \citet{gre84}, who lists it as a weak-lined DA of temperature near 7200 K, and secondly upon
\citet{gre86}, who list the star as DC in the same temperature class, with a blue photoelectric
color index of $U-B=-0.55$.  However, the $V=14.9$ mag star has a very small proper motion
on the order of 10 mas yr$^{-1}$ \citep{ade08,far05}, and its optical spectrum reveals calcium
H and K lines (A. Gianninas 2008, private communication; see Figure \ref{fig12}).  Together,
these data indicate a main-sequence F-type star, with the former more likely based on the 
strength of the Balmer lines compared to Kurucz models\footnote{http://www.stsci.edu/hst/observatory/cdbs/k93models.html}

Interestingly, the star has a clear blue-ultraviolet excess at {\em GALEX} and optical wavelengths
that cannot be reproduced by a single temperature component, and is most readily explained 
by a hot subdwarf or very hot white dwarf companion.  However, for this to be the case, the
far-ultraviolet flux would have to suffer from a significant degree of extinction.  Figures \ref{fig12} 
-- \ref{fig14} display various data for this star, which strongly support both an ultraviolet and blue 
optical excess, relative to a main-sequence F or G-type star.  High-resolution observations of 
this object taken by the SPY survey with VLT / UVES also reveal a clear blue-ultraviolet excess 
(R. Napiwotzki 2009, private communication).

\paragraph*{{\em 0956$+$045}}.  This PG white dwarf is part of a visual double separated by 
$2\farcs0$ \citep{far05}, and the M dwarf companion dominates the binary light redward of 0.8 
$\mu$m.  Spatially-resolved $JHK$ photometry from the IRTF matches the predictions of an 
18000 K DA model \citep{lie05a}.  SDSS $griz$ photometry of PG 0956$+$045B provides new 
color information, and when combined with its near-infrared photometry, indicate an updated 
spectral type of M5 \citep{boc07}.

\paragraph*{{\em 1000$+$220}}.  TON 1145 appears similar to PG 0939$+$071 in a few ways; 
its data are shown in Figures \ref{fig16}--\ref{fig18}.  First, its blue optical spectrum appears to be 
F-type with similar line strengths.  Second, while a main-sequence stellar model fits nearly all the 
photometric data, there may be an ultraviolet excess, but unfortunately, there are no {\em GALEX} 
data for this star.  Third, it has a proper motion of order 10 mas yr$^{-1}$, meaning it is quite
distant.  A late A-type or early F-type star fits most of the photometric data excepting the SDSS 
$ug$ fluxes, perhaps indicating a hot subdwarf or very hot white dwarf companion.

\paragraph*{{\em 1013$-$010}}.  G53-38 is a confirmed double degenerate, a single-lined 
spectroscopic binary \citep{nel05}.  From Figure \ref{fig19}, there appears to be photometric 
evidence of the hidden white dwarf companion.  \citet{pau06} give 8800 K for the primary 
based on Balmer line spectroscopy, that underpredicts the red optical and near-infrared fluxes 
substantially.  A combination of a 9000 K DA white dwarf and a $T{\rm eff}\sim6000$ K DC white 
dwarf is able to reproduce the entire spectral energy distribution quite well.

\paragraph*{{\em 1036$-$204}}.  LHS 2293 is a DQ peculiar white dwarf that displays two clearly 
disparate effective temperatures, even while excluding the heavily absorbed $4000-6000$ \AA 
\ region \citep{jor02,sch99}.  The {\em GALEX} near-ultraviolet and optical $URI$ photometry can 
be fitted with a 7500 K star, consistent with the analysis of \cite{lie78}, while the $IJHK$ photometry 
can be reproduced by a 4300 K blackbody.  This could be a spectacular example of energy 
redistribution in a highly unusual atmosphere; clearly more data are warranted for this highly
magnetic white dwarf.

\paragraph*{{\em 1108$+$325}}.  This is a visual binary separated by $0\farcs2$ as imaged with 
{\em HST} / ACS (Farihi et al. 2009, in preparation).  The improved $JHK$ data here indicate 
$I-K=2.1$ and a dM3 companion, consistent with the estimated white dwarf distance of 560 pc 
\citep{lie05a}.

\paragraph*{{\em 1133$+$489}}.  \citet{far06} resolve this binary at $0\farcs1$ with {\em HST}
/ ACS.  The improved $JHK$ data here indicate $I-K=2.7$ for PG 1133$+$489B, and a dM5 
companion, consistent with a 370 pc distance to a log $g=8.0$ white dwarf with $T_{\rm eff}=
47500$ K.

\paragraph*{{\em 1140$+$004}}.  The optical spectrum of this star reveals two strong 
components, a white dwarf in the blue and an M dwarf in the red.  The companion spectral type 
has been estimated previously to be M4 \citep{ray03} and M6 \citep{sil06}.  The IRTF near-infrared 
data permit a broad-band color analysis of the companion that yields $I-K=2.6$ and an estimated 
spectral type of M5.  This is consistent with the 240 pc distance to a $T_{\rm eff}=14500$ K, 
log $g=8.0$ white dwarf \citep{kle04}.

\paragraph*{{\em 1156$+$132}}.  LP 494-12 is a DQ peculiar star with a relatively warm effective 
temperature.  Ongoing coverage of this region by the UKIRT Infrared Deep Sky Survey (UKIDSS) 
confirms the IRTF $K$-band excess, reporting $(H,K)=(17.32,17.08)$ mag for this white dwarf.  The 
nature of the excess emission may be related to its unusual atmosphere or a strong magnetic field 
(or both), and could be a redistribution of emergent energy as potentially seen in LHS 2293.  If 
correct, the excess at LP 494-12 is still unusual as it only manifests beyond the $H$ band.

\paragraph*{{\em 1225$-$079}}.  A strong 8600 \AA \ calcium absorption feature is clearly 
detected in this DZA white dwarf.  The individual lines of the triplet are unresolved at the resolution 
of the SpeX prism.

\paragraph*{{\em 1254$+$345}}.  Using non-magnetic DA models, an effective temperature 
of 9500 K is estimated for this highly magnetic white dwarf (see Figure \ref{fig27}).  This is 
considerably lower than the previous estimate of 15000 K \citep{hag87}, and is consistent 
with the slope of its optical, SDSS spectrum.

\paragraph*{{\em 1330$+$015}}.  G62-46 is a likely DAH+DC binary based on spectra that 
reveal diluted, Zeeman-split lines of H$\alpha$ \citep{ber93}.  Figure \ref{fig28} essentially 
reproduces the double degenerate model fit of \citet{ber93} with 9500 K DC, and 6000 K DA 
components.  However, the $I$-band flux differs significantly from the SDSS $iz$ photometry, 
that appear to be corroborated by DENIS $I$-band data.  Using the SDSS and IRTF data, the full 
spectral energy distribution cannot be modeled within the photometric errors as either a single or 
double source; the resulting composite either overpredicts the $HK$, or underpredicts the $iz$ 
fluxes.  The SDSS catalog lists a 'good' quality flag for this photometry, and hence variability may 
be present.  A less likely possibility is near-infrared flux suppression in the cooler DA component.
The {\em GALEX} near-ultraviolet and $u$-band fluxes may favor a DC component closer to 8000 K.

\paragraph*{{\em 1339$+$346}}.  Part of an apparent stellar triple at galactic latitude $|b|=
77\degr$, the white dwarf is not physically associated with the remaining pair of stars, currently 
located $2\farcs6$ and $21\farcs2$ distant.  This pair of cool stars appear to be comoving at 
$(27,-37)$ mas yr$^{-1}$ \citep{zac05}, while their SDSS and IRTF colors suggest they are 
main-sequence K stars at roughly 720 pc.

Utlizing an epoch 1950.4 POSS plate scan compared with the epoch 2007.4 IRTF $J$-band 
image, the white dwarf has clearly moved relative to the K stars by a few arcseconds.  From 
the motion relative to the nearest K star, a total proper motion of $(80,-72)$ mas yr$^{-1}$ is 
calculated for the white dwarf.

\paragraph*{{\em 1350$-$090}}.  The DA model fitted to the broad energy distribution 
in Figure \ref{fig31} yields an effective temperature around 1000 K lower than the 
spectroscopically derived value of \citet{lie05a}.  LP 907-37 is a DAP white dwarf with a 
kG magnetic field \citep{sch94}; the Zeeman split hydrogen lines \citep{koe05a} cannot be 
seen in the low resolution SpeX prism data.  The 2MASS catalog lacks $JK_s$ photometry 
for this $m<14.5$ mag white dwarf.

\paragraph*{{\em 1428$+$373}}.   The DA model fitted to the broad energy distribution in 
Figure \ref{fig32} yields an effective temperature around 1000 K lower than the spectroscopically 
derived value of \citet{lie05a}. 

\paragraph*{{\em 1455$+$298}}.  This is another case where the SDSS $iz$ photometry 
deviates significantly from the $I$-band flux of \citet{ber01}.  The SpeX prism data appear to 
agree well with the SDSS fluxes.  The apparent 2MASS $H$-band photometric excess for this 
white dwarf is spurious.

\paragraph*{{\em 1457$-$086}}.  The apparent 2MASS $K_s$-band excess at this metal-rich 
white dwarf is real, but significantly milder in the IRTF photometry.  Optical Balmer line spectroscopy 
of this star by two independent studies yields $T_{\rm eff}=21500$ K \citep{lie05a} and 20400 K 
\citep{koe05a}, yet such models cannot reproduce the $UBVIJH$ fluxes in Figure \ref{fig35}.  
A 20000 K DA model fitted to the $UBVI$ photometry predicts the measured $J$-band flux is 
in excess of the expected white dwarf photosphere.  The optical photometry of \citet{kil97} may 
be somewhat uncertain, but with only these data available, an 18000 K model reproduces all
the photometric fluxes excepting $K$ band; consistent with an excess due to warm dust 
\citep{far09}.

\paragraph*{{\em 1619$+$525}}.   This is a possible triple system \citep{far06} consisting of 
a white dwarf and a cool main-sequence companion separated by $0\farcs5$ at position angle 
$24\degr$ , together with another cool companion at $2\farcs6$ and position angle $283\degr$.  
The IRTF images spatially resolve PG 1619$+$525AB from C, and when combined with their 
{\em HST} / ACS F814W photometry, imply earlier companion spectral types than previously 
suspected based on the nominal 100 pc photometric distance to the white dwarf \citep{far06,lie05a}.

However, the likely companions both yield consistent photometric distances near 300 pc: 
PG 1619$+$525B has $K=15.3$ mag, $I-K=2.8$, while PG 1619$+$525C has $K=13.4$ 
mag, $I-K=1.9$, compatible with spectral types M5 and M2 respectively.  Common proper 
motion has not been established for these stars, yet a physical association is probable based 
on their proximity and high galactic latitude \citep{far06}.  The probability of a chance alignment 
with an M dwarf at 300 pc within a cylinder of $r=0\farcs5$ and length 100 pc centered on the 
white dwarf is less than 1 in $5\times10^5$, while for two M dwarfs within $r=2\farcs6$ the 
probability is less than 1 in $2\times10^6$.  The odds would be somewhat greater for distant 
background stars with reddened $I$-band magnitudes, yet ultimately unlikely.

Examination of epoch 1954.5 POSS I plates reveals little appreciable motion relative to SDSS 
images taken in 2004.5; USNO-B1 gives a proper motion for the composite triple of $(34,-12)$ 
mas yr$^{-1}$, favoring a distant system ($v_{\rm tan}=17$ km s$^{-1}$ at 100 pc versus 51 km
s$^{-1}$ at 300 pc).  Yet the white dwarf parameters derived via Balmer line spectroscopy do
not permit such a distant interpretation, as it would imply a temperature inconsistent with the
$UBV$ photometry and slope of the optical spectrum, or an extremely low surface gravity.
For the time being, the physical association of these three stars remains tentative.

\paragraph*{{\em 1626$+$368}}.  G180-57 is a DZA with a strong 8600 \AA \ calcium 
absorption feature detected in its spectrum.  The individual lines of the triplet are unresolved 
at the resolution of the SpeX prism.  It is unclear why the SDSS photometry differs significantly 
with available optical fluxes from \citet{ber01} and several similar measurements \citep{mcc99}; 
variability appears possible in this star.

\paragraph*{{\em 1653$+$385}}.  NLTT 43806 is listed among nearby white dwarfs with a 
photometric distance of 15 pc \citep{hol08,kaw06}, based on an estimated $V=15.9$ mag 
\citep{sal03}.  The SDSS photometry implies a distance greater than 20 pc, with $g=17.0$ mag 
($V\approx16.8$ mag) and $d=24$ pc.

\paragraph*{{\em 1845$+$683}}.  This white dwarf is neither a binary nor binary suspect.  
\citet{gre00} incorrectly identified this extreme ultraviolet source with a nearby, unrelated 
near-infrared source \citep{far06}.

\paragraph*{{\em 2032$+$188}}.  GD 231 is a double degenerate \citep{mor05} and may 
display mild near-infrared excess due to its unseen companion, but the larger 2MASS excess 
is not confirmed.  Figure \ref{fig40} demonstrates that the entire spectral energy distribution can 
be decently matched by a 16500 K DA model, but this may simply reflect a good approximation 
of the composite light rather than an accurate effective temperature of either component.  
Assuming the 18500 K spectroscopic temperature derived by \citet{ber92} is correct, Figure 
\ref{fig41} predicts an excess at $JHK$ that is consistent with a 6000 K DC companion.

\paragraph*{{\em 2201$-$228}}.  This star was tentatively classified as a magnetic DA, but 
has since been correctly reclassified as DB \citep{jor01}.  The apparent 2MASS excess at this 
white dwarf is not confirmed; the 13000 K fit to its spectral energy distribution shown in Figure 
\ref{fig43} is the first time an effective temperature has been assessed for this DB star.

\paragraph*{{\em 2215$+$388}}.  The $U$-band photometry for this DZ star appears influenced 
by its strong Ca H and K absorption, and possibly other elements \citep{sio90}.

\paragraph*{{\em 2216$+$484}}.  GD 402 is a suspected DA+DC system based on the fact 
that its optical colors (and the shape of its optical spectrum) predict a significantly higher effective 
temperature than do its relatively weak Balmer lines \citep{ber90}.   Although the expanded spectral 
energy distribution -- including the IRTF near-infrared data -- can be nearly reproduced by a single 
7000 K component as shown in Figure \ref{fig46}, the photometry can also be well-modeled
with two white dwarf components of approximately 8000 K and 5000 K (Figure \ref{fig47}).

\paragraph*{{\em 2316$+$123}}.  This magnetic white dwarf has a spectral energy distribution 
that cannot be fitted by a single temperature (non-magnetic) white dwarf model.  Figure \ref{fig48} 
fits the combined ultraviolet and optical fluxes to yield a temperature near 13000 K, while Figure 
\ref{fig49} fits the combined optical and near-infrared colors with a temperature around 10500 K.  
It is unclear whether this difficulty is related to the magnetic nature of the star, or if there is a real 
$JHK$ photometric excess (relative to non-magnetic models), implying the possibility of binarity.

\paragraph*{{\em 2333$-$049}}.  This white dwarf may have a small $K$-band excess, but 
more data are needed to confirm or rule out this possibility; the $V$-band flux appears too bright 
for the model shown in Figure \ref{fig50}.

\paragraph*{{\em 2336$-$187}}.  G273-97 is by far the best near-infrared excess candidate 
based on its 2MASS data, which appear reliable at all wavelengths.  However, the IRTF 
observations demonstrate that the 2MASS catalog contains significant errors even at S/N $>10$.

\paragraph*{{\em 2354$+$159}}.  Both \citet{vos07} and \citet{bea99} give $T_{\rm eff}\approx
24500$ K -- assuming no hydrogen -- for this DBZ star.  However, both authors give alternative 
effective temperatures near 22500 K for a nominal hydrogen abundance.  The higher temperature
overpredicts the {\em GALEX} fluxes for this star, and the fit shown in Figure \ref{fig52} employs a 
temperature of 21500 K to match all the photometry; possibly indicating some hydrogen is 
present and closer to the 19000 K value given by \citet{koe05a}.

This white dwarf is another exemplary case of a 2MASS excess not being corroborated by 
targeted $JHK$ photometry; the lack of infrared excess is also confirmed by {\em Spitzer} 
IRAC observations \citep{far09}.

\section{DISCUSSION}

\subsection{Near-Infrared Excesses Predicted by 2MASS}

\begin{figure}
\includegraphics[width=85mm]{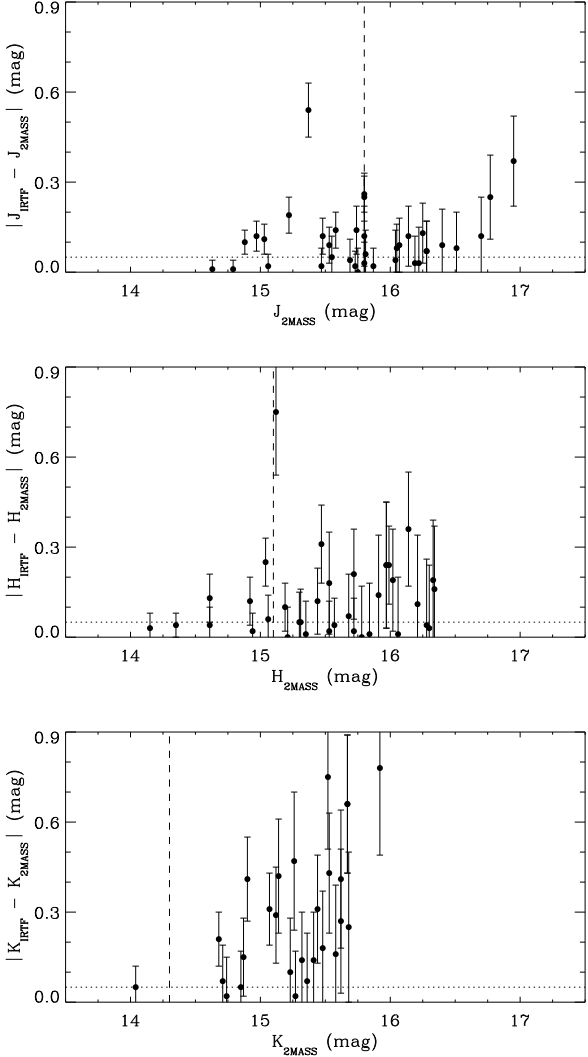}
\caption{The absolute value of the magnitude difference between the IRTF and 2MASS 
near-infrared photometry for all mutually observed target stars, plotted versus magnitude and 
uncorrected for distinct filter sets.  The displayed error bars are the 2MASS uncertainties, the
dashed line represents the 2MASS $10\sigma$ detection threshold, and the dotted line 
represents a 5\% error in the IRTF photometry.
\label{fig53}}
\end{figure}

The white dwarf fluxes in the 2MASS catalog were the prime motivation for obtaining follow 
up data, hence it is relevant to ask how well they predicted the IRTF $JHK$ photometry, and 
the presence of any near-infrared excess.  Figure \ref{fig53} compares the IRTF and 2MASS 
fluxes (in Vega magnitudes) for all 39 Table \ref{tbl2} white dwarf targets with both sets of 
photometry; transformations between filter sets were ignored as they are generally within 
a few percent \citep{car01}.  

The 2MASS $10\sigma$ minimum detection limit for the whole sky is $(J,H,K_s)=(15.8,15.1,
14.3)$ mag \citep{skr06}, and there is decent accord for measurements brighter than this limit.
The generally disagreeable behavior of the 2MASS data below the 10$\sigma$ limit in all three 
filters is not surprising, yet is most pronounced in the $K$ band.  Somewhat unexpectedly, the
$J$-band shows several discordant measures at the $1-2\sigma$ level at relatively bright stars.
In fact, of the co-observed white dwarfs at each bandpass, the $H$-band 2MASS data agree 
most frequently with the IRTF observations over all brightnesses.  At the $1\sigma$ level, the 
2MASS and IRTF photometry agree for 29 of 37 stars (78\%) at $H$, 25 of 38 stars (66\%) at 
$J$, and 14 of 29 stars (48\%) at $K$.

An infrared excess is revealed by the relative flux levels of three or more photometric fluxes,
at least two of which should be consistent with photospheric emission and at least one of which 
is significantly higher than expected for the photosphere alone.  Using this practical definition,
the 2MASS photometry suggest an $H$ and/or $K_s$-band excess -- relative to 2MASS $J$
or ultraviolet/optical photometry -- for 27 white dwarfs in Table \ref{tbl2}.  The potential excess 
fluxes are most readily discerned in the Figures, and where confirmed by the IRTF data, these 
excesses are often of a significantly milder or altogether different nature than suggested by 
the 2MASS or near-infrared data alone.  Hence, it is difficult to precisely quantify the number 
of IRTF confirmed versus potential 2MASS excesses.

\begin{figure}
\includegraphics[width=85mm]{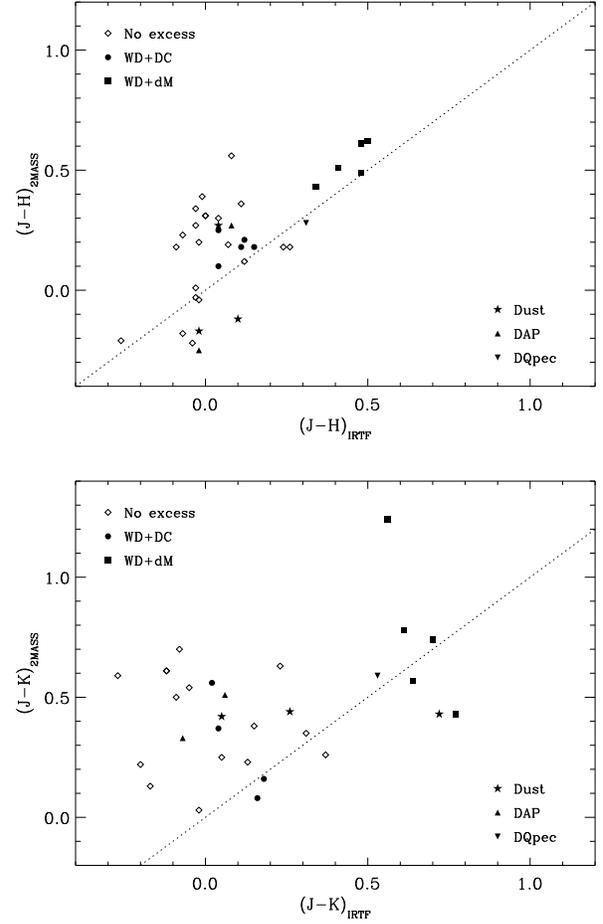}
\caption{IRTF versus 2MASS near-infrared colors.  The upper figure plots $J-H$ for all 37 
white dwarfs in common, while the lower figure plots $J-K$ for all 30 white dwarfs in common.
Objects with a comfirmed or likely near-infrared excess are shown as filled symbols; error bars
are not shown to avoid overcrowding in the figure.
\label{fig54}}
\end{figure}

\begin{figure*}
\includegraphics[width=140mm]{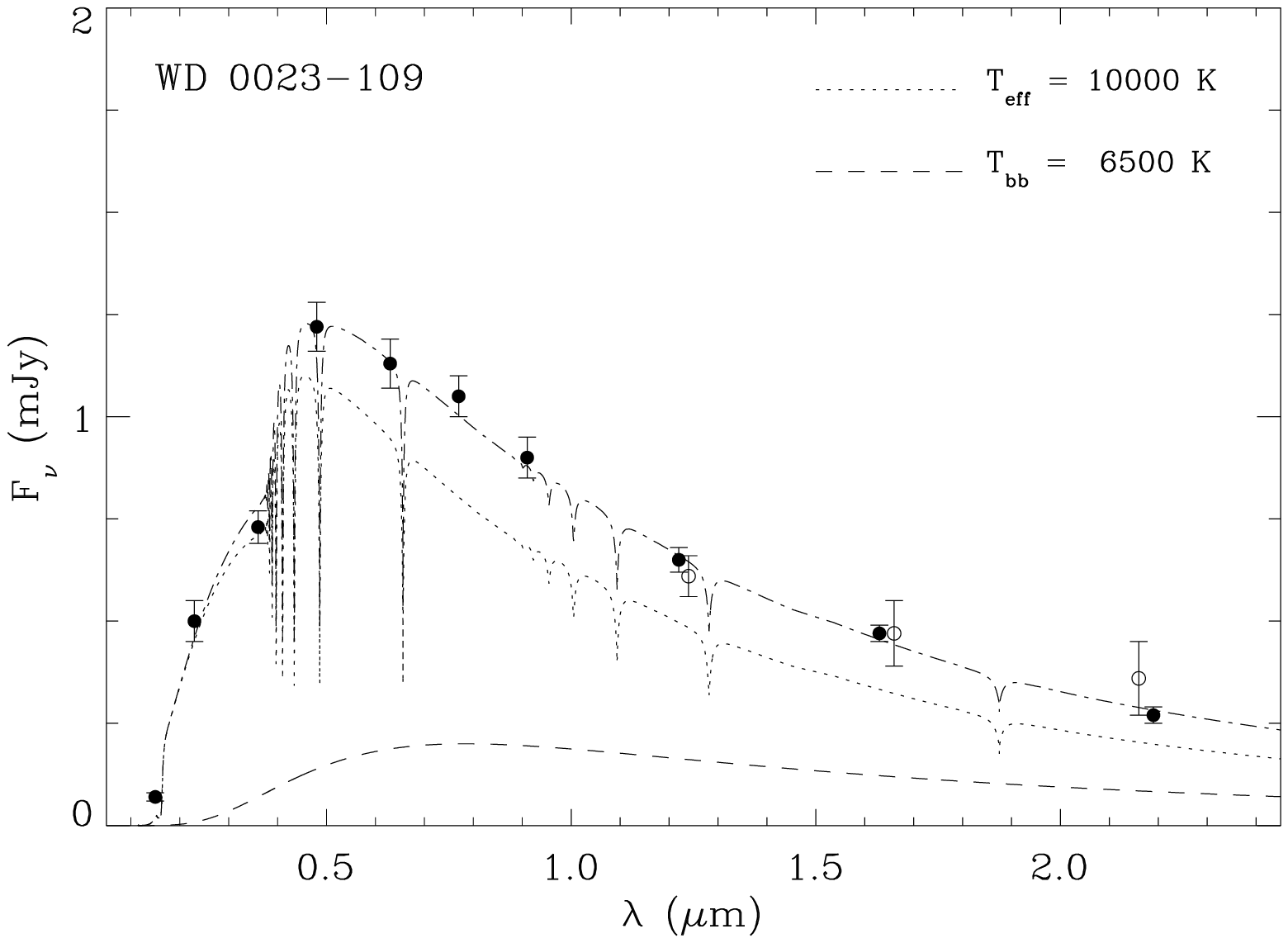}
\caption{Spectral energy distribution of G158-78.  The solid circles are {\em GALEX} far- and 
near-ultraviolet, SDSS $ugriz$, and IRTF $JHK$ photometry, while the open circles are 2MASS 
$JHK_s$ photometry.  The star is a likely double degenerate.
\label{fig1}}
\end{figure*}

\begin{figure*}
\includegraphics[width=140mm]{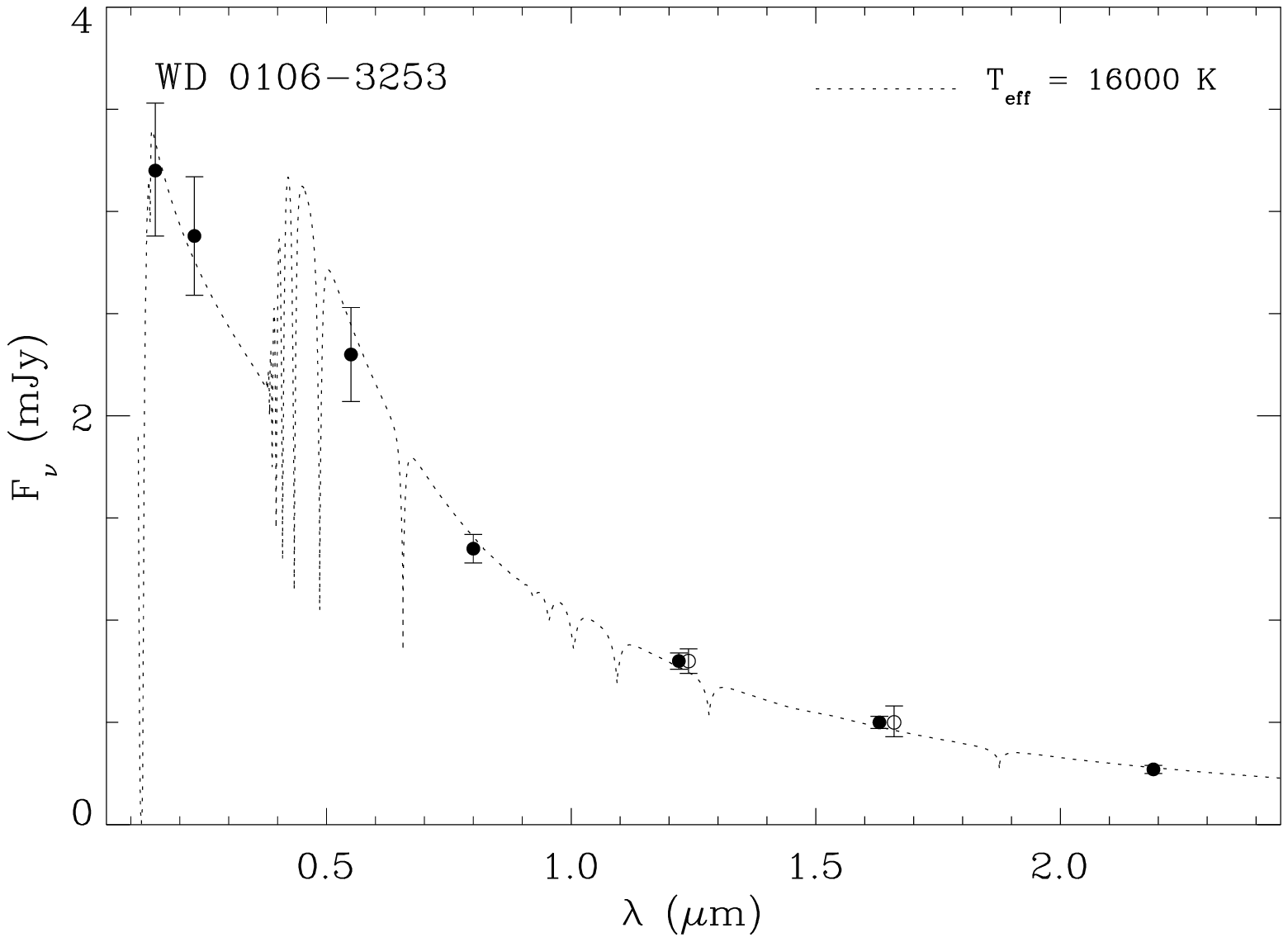}
\caption{Spectral energy distribution of HE 0106$-$3253.  The solid circles are {\em GALEX} far- and 
near-ultraviolet, optical $VI$, and IRTF $JHK$ photometry, while the open circles are 2MASS 
$JH$ photometry.
\label{fig2}}
\end{figure*}

\begin{figure*}
\includegraphics[width=140mm]{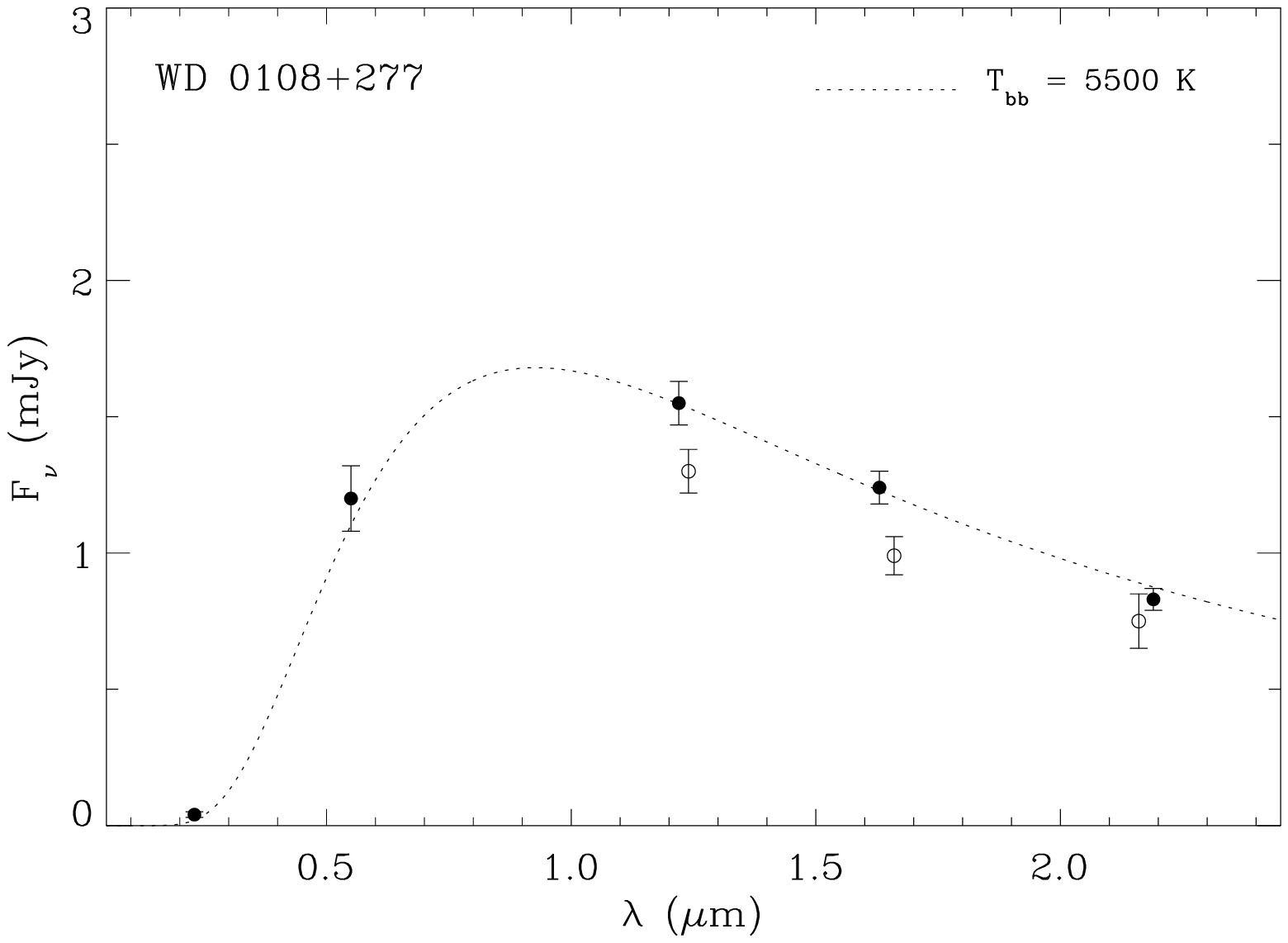}
\caption{Spectral energy distribution of NLTT 3915.  The solid circles are {\em GALEX} near-ultraviolet, 
optical $V$, and IRTF $JHK$ photometry, while the open circles are 2MASS $JHK_s$ photometry.
\label{fig3}}
\end{figure*}

\begin{figure*}
\includegraphics[width=140mm]{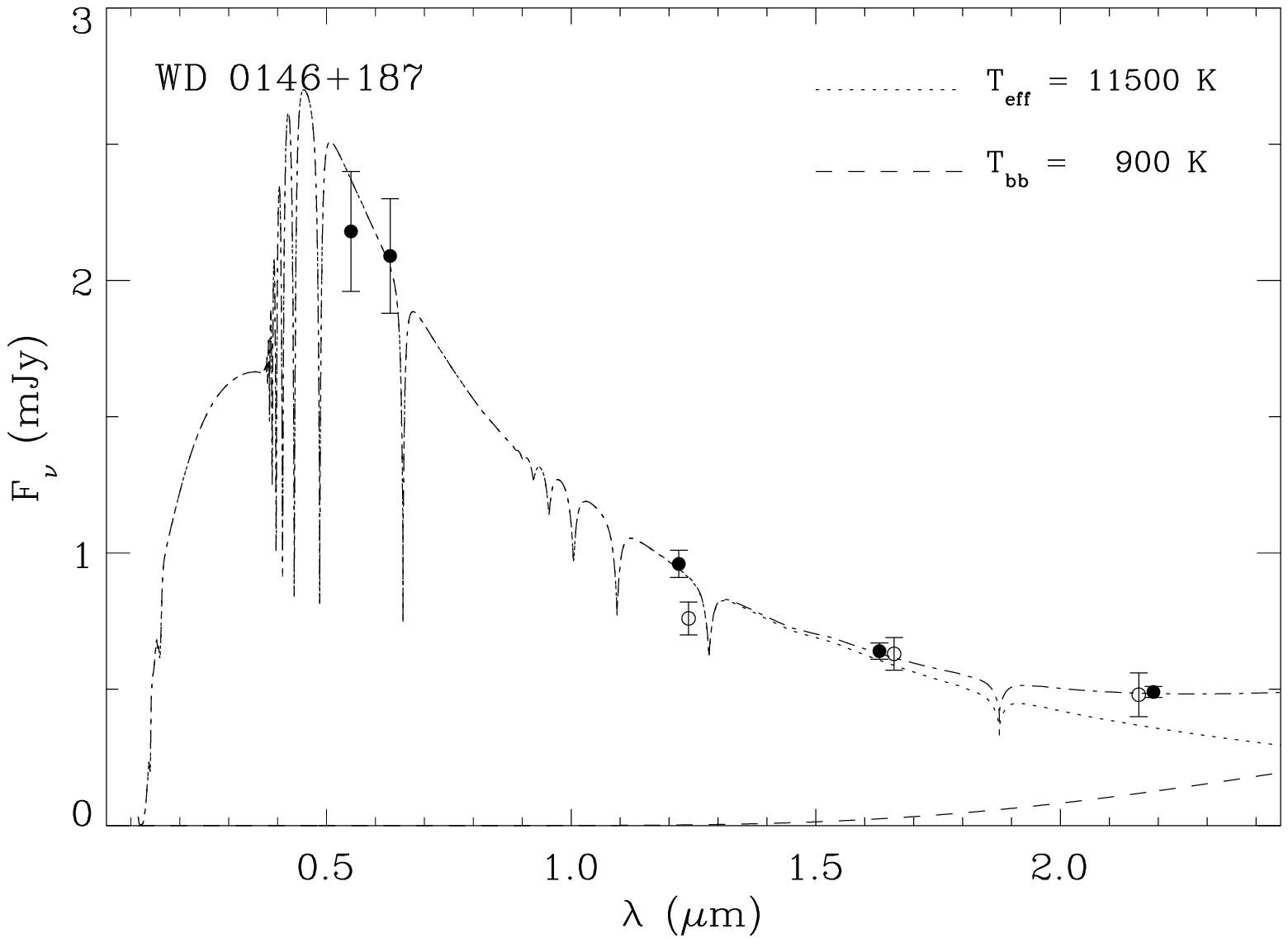}
\caption{Spectral energy distribution of GD 16.  The solid circles are optical $Vr$, and IRTF $JHK$ 
photometry, while the open circles are 2MASS $JHK_s$ photometry.  There are no {\em GALEX} data 
available for this star.
\label{fig4}}
\end{figure*}

\begin{figure*}
\includegraphics[width=140mm]{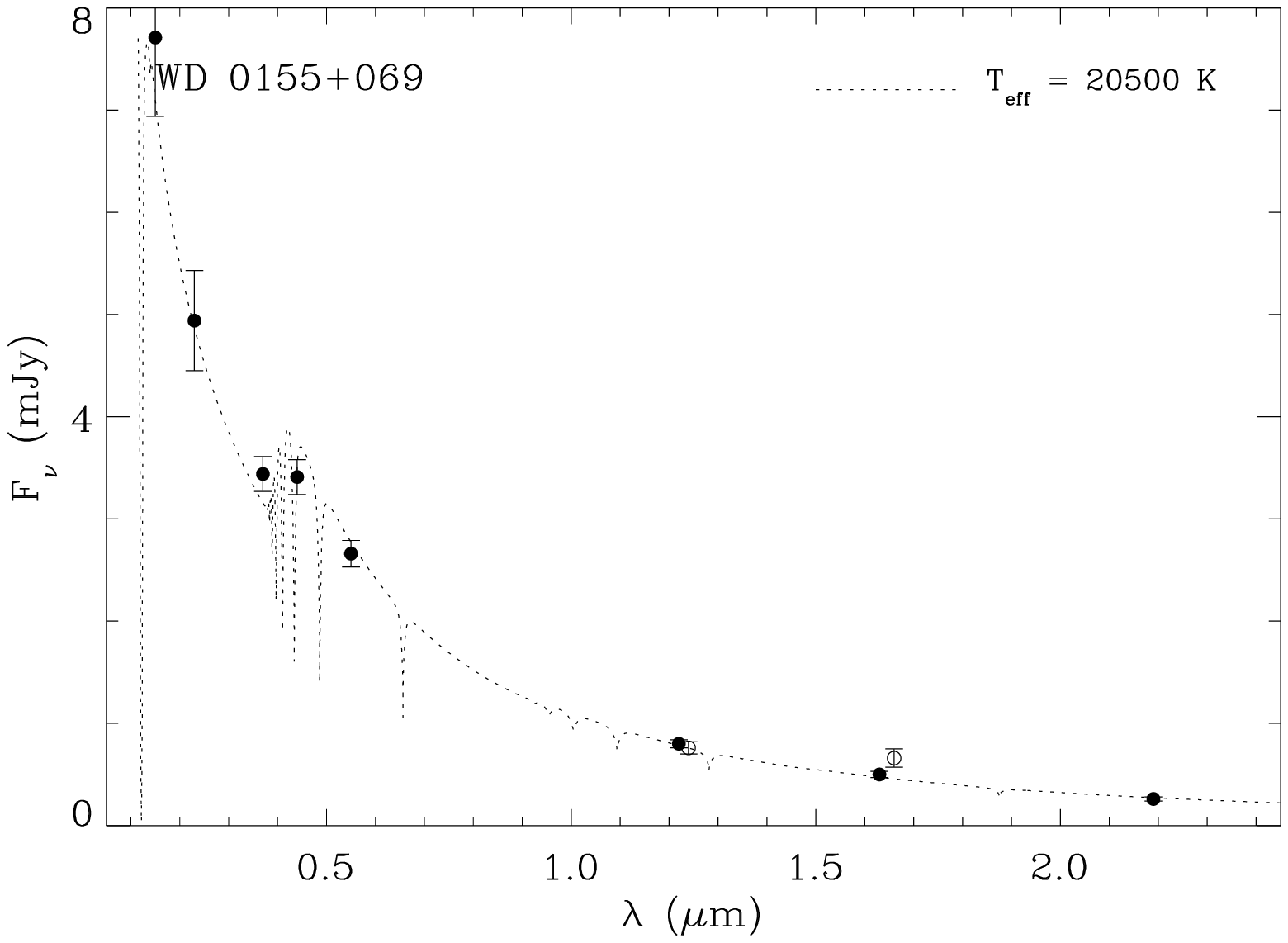}
\caption{Spectral energy distribution of GD 20.  The solid circles are {\em GALEX} far- and 
near-ultraviolet, optical $UBV$, and IRTF $JHK$ photometry, while the open circles are 2MASS 
$JH$ photometry.
\label{fig5}}
\end{figure*}

\begin{figure*}
\includegraphics[width=140mm]{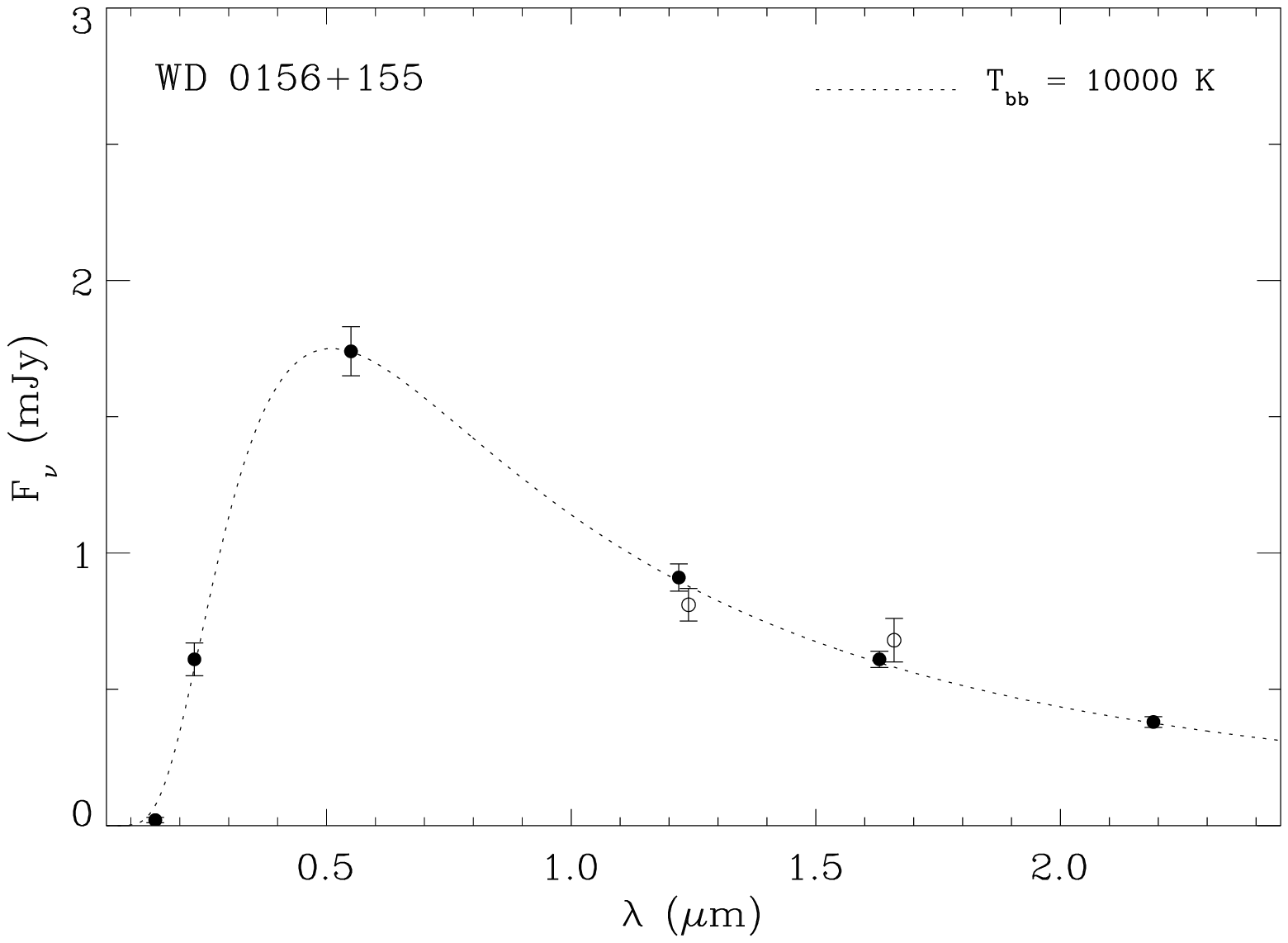}
\caption{Spectral energy distribution of PG 0156$+$155.  The solid circles are {\em GALEX} far- and 
near-ultraviolet, optical $V$, and IRTF $JHK$ photometry, while the open circles are 2MASS 
$JH$ photometry.
\label{fig6}}
\end{figure*}

\begin{figure*}
\includegraphics[width=140mm]{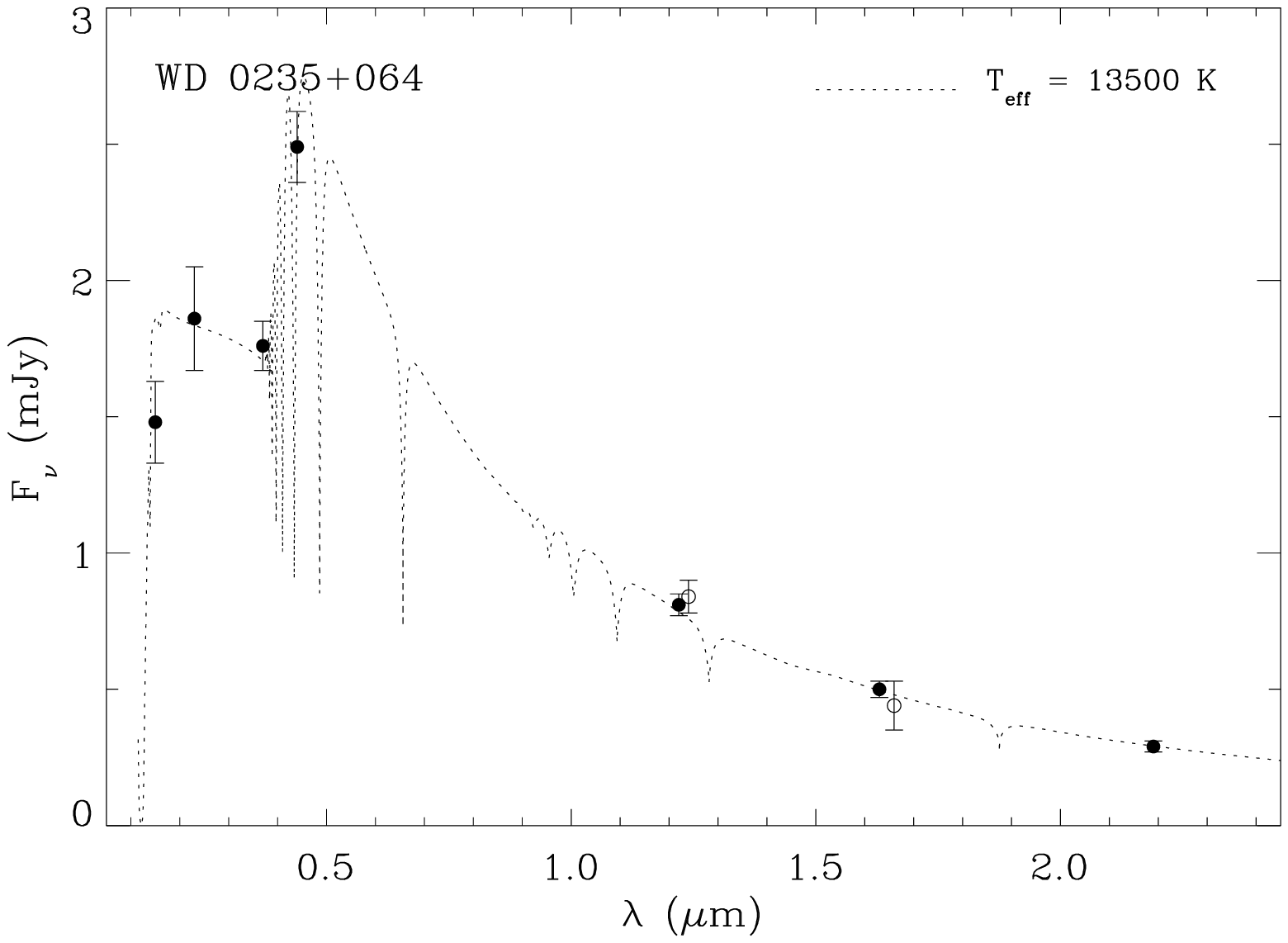}
\caption{Spectral energy distribution of PG 0235$+$064.  The solid circles are {\em GALEX} far- and 
near-ultraviolet, optical $UB$, and IRTF $JHK$ photometry, while the open circles are 2MASS 
$JH$ photometry.  The solid-circle photometry is uncontaminated by the nearby common-proper 
motion companion \citep{far08b}, indicating a new, higher effective temperature for this white 
dwarf.
\label{fig7}}
\end{figure*}

\begin{figure*}
\includegraphics[width=140mm]{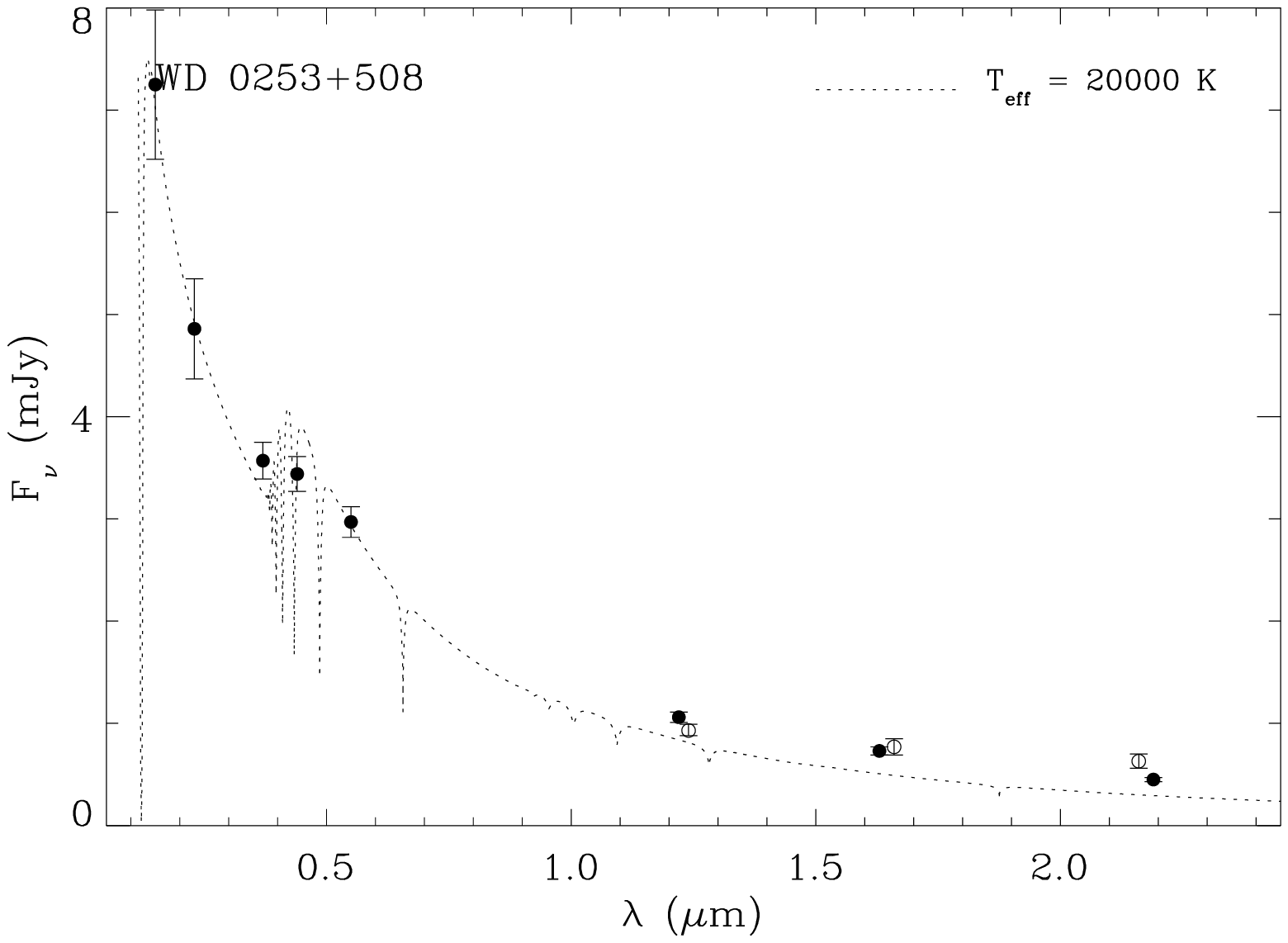}
\caption{Spectral energy distribution of KPD 0253$+$508.  The solid circles are {\em GALEX} far- and 
near-ultraviolet, optical $UBV$, and IRTF $JHK$ photometry, while the open circles are 2MASS 
$JHK_s$ photometry.  A single effective temperature does not fit this magnetic white dwarf; it has
clear excess (relative to non-magnetic DA models) emission at $JHK$.
\label{fig8}}
\end{figure*}

\begin{figure*}
\includegraphics[width=140mm]{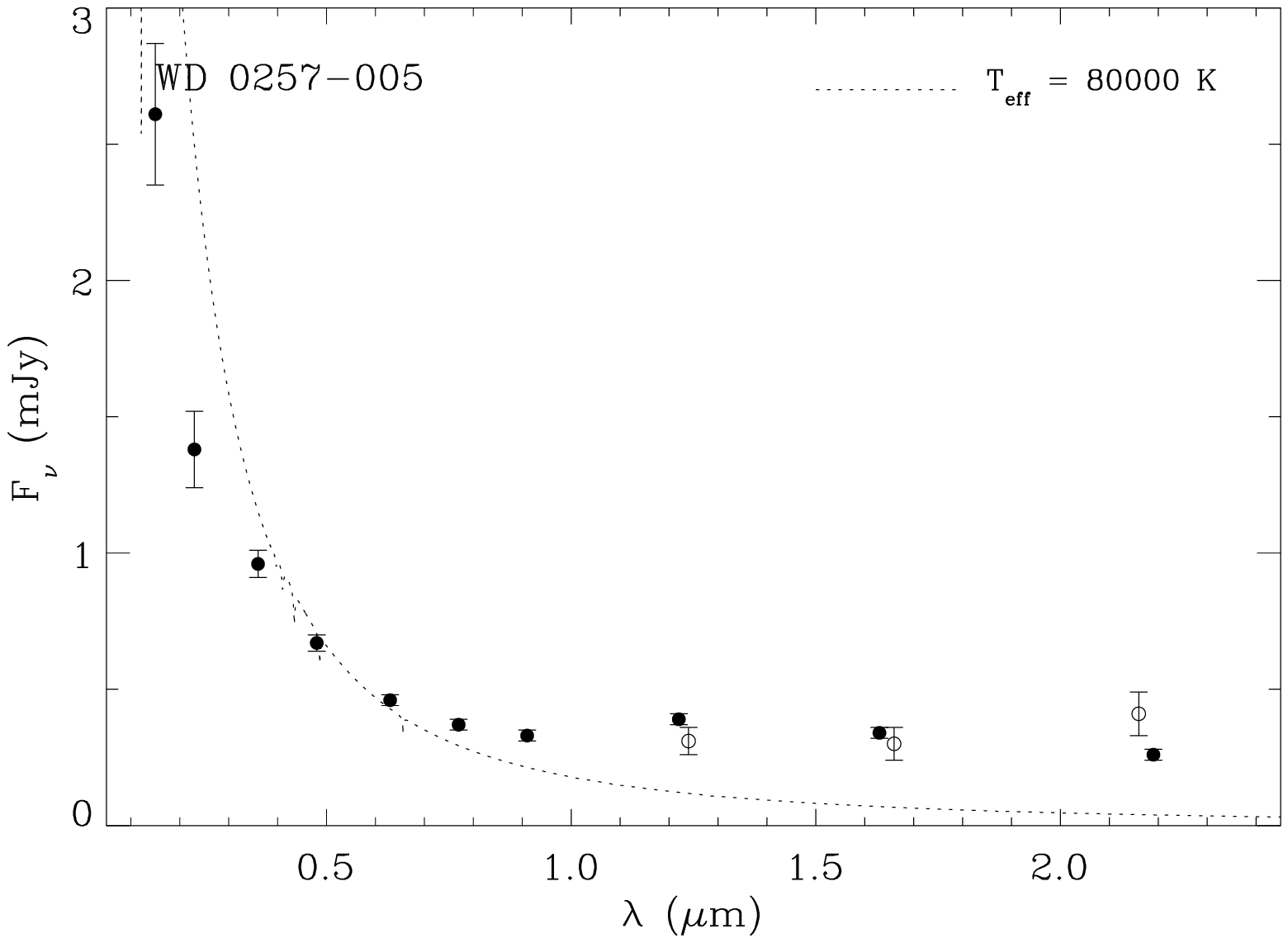}
\caption{Spectral energy distribution of KUV 0257$-$005.  The solid circles are {\em GALEX} far- and 
near-ultraviolet, SDSS $ugriz$, and IRTF $JHK$ photometry, while the open circles are 2MASS 
$JHK_s$ photometry.  There is clear evidence for extinction at ultraviolet and optical wavelengths.
\label{fig9}}
\end{figure*}

\begin{figure*}
\includegraphics[width=140mm]{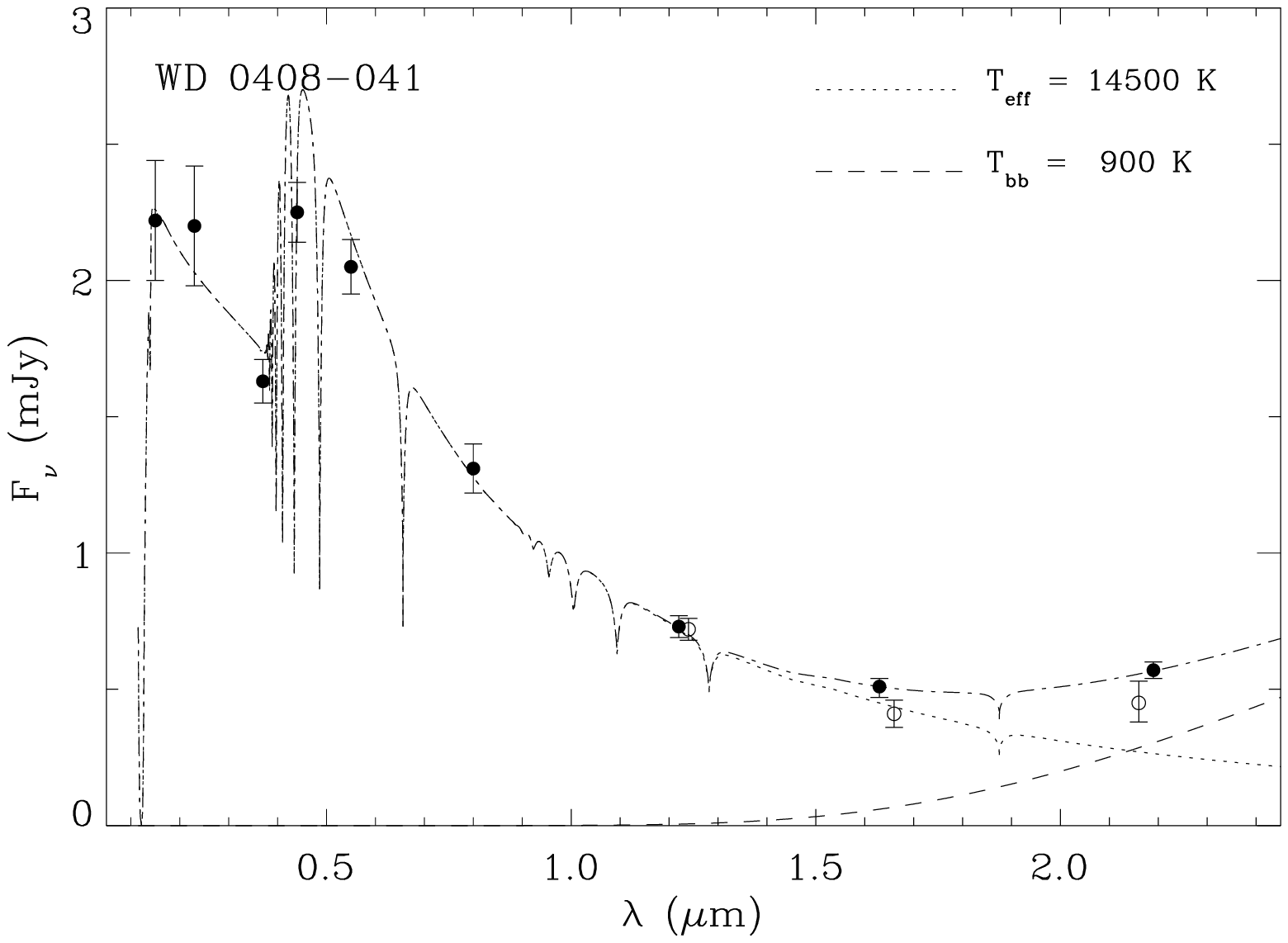}
\caption{Spectral energy distribution of GD 56.  The solid circles are {\em GALEX} far- and 
near-ultraviolet, optical $UBVI$, and IRTF $JHK$ photometry, while the open circles are 2MASS 
$JHK_s$ photometry.
\label{fig10}}
\end{figure*}

\begin{figure*}
\includegraphics[width=140mm]{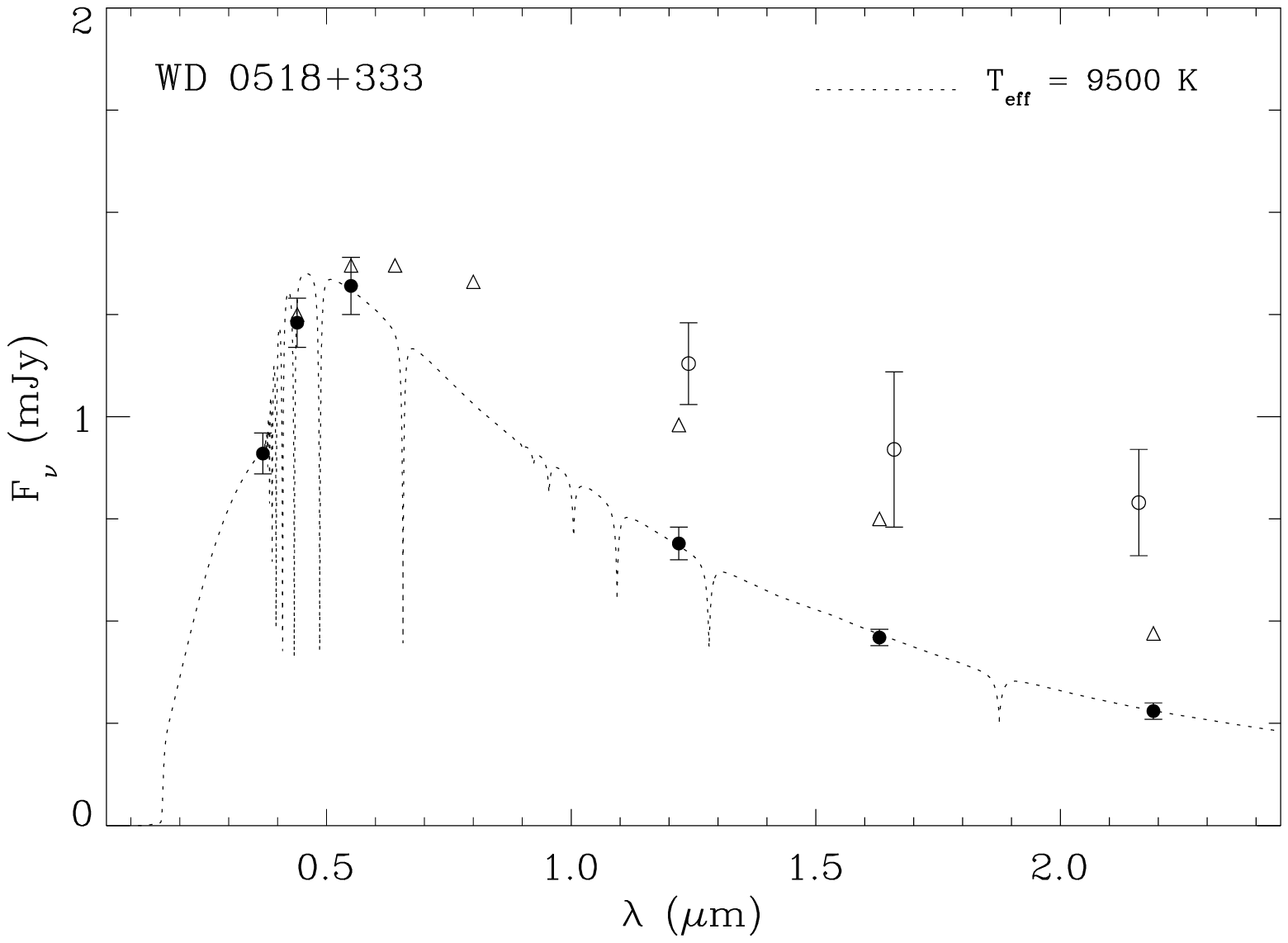}
\caption{Spectral energy distribution of EG 43.  The solid circles are optical $UBVI$, and 
IRTF $JHK$ photometry, while the open circles are 2MASS $JHK_s$ photometry, clearly 
contaminated by the light from the visual M dwarf companion.  The open triangles are 
$BVRIJHK$ from \citet{ber01}, also apparently contaminated at optical wavelengths.
\label{fig11}}
\end{figure*}

\begin{figure*}
\includegraphics[width=140mm]{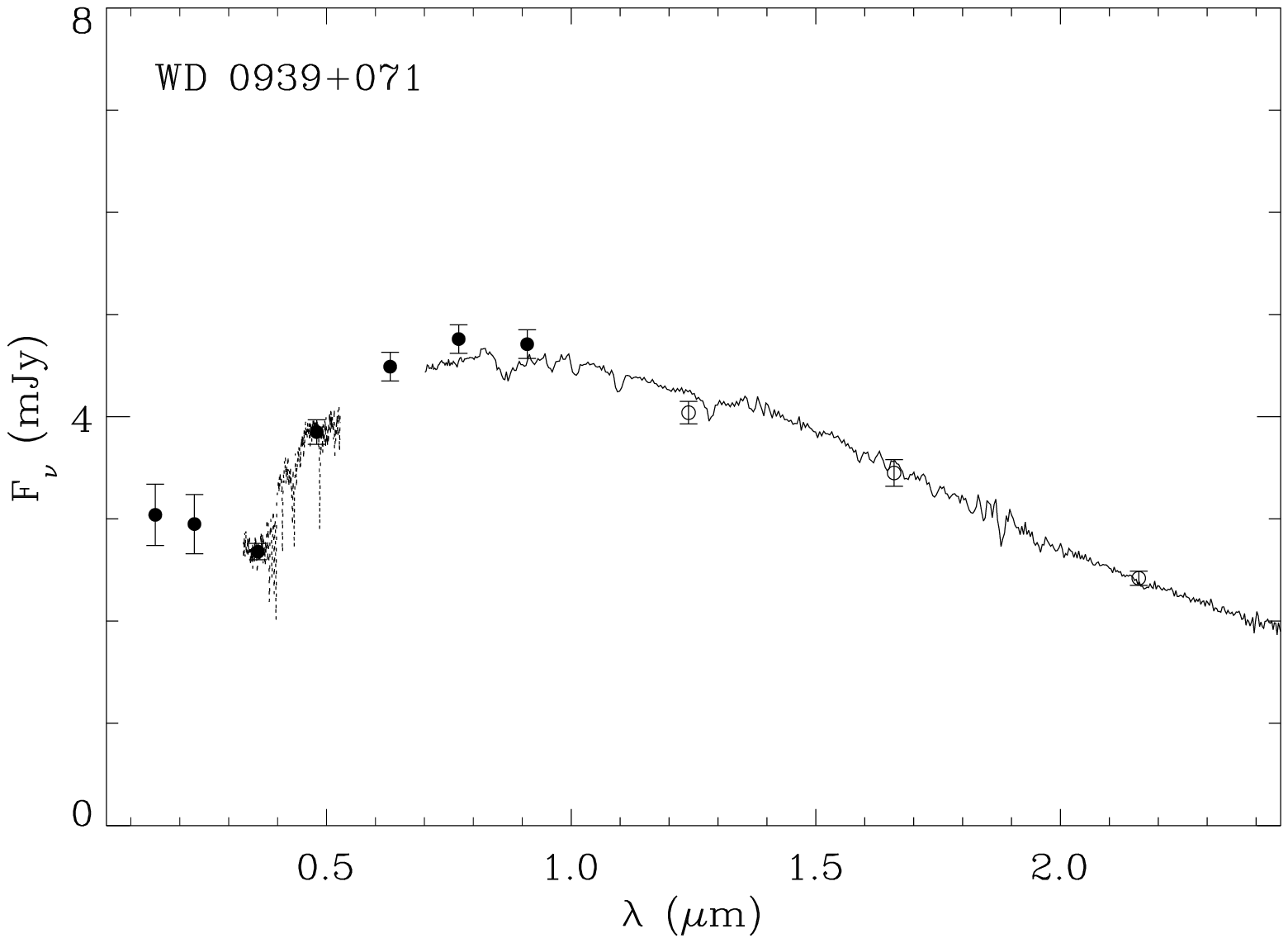}
\caption{Spectral energy distribution of PG 0939$+$071.  The solid circles are {\em GALEX} far- and 
near-ultraviolet, and SDSS $ugriz$ photometry, while the open circles are 2MASS $JHK_s$ 
photometry.  The $0.7-2.5$ $\mu$m spectrum was taken with SpeX, and the blue optical 
spectrum is also shown (see Figure \ref{fig12}).
\label{fig12}}
\end{figure*}

\begin{figure*}
\includegraphics[width=140mm]{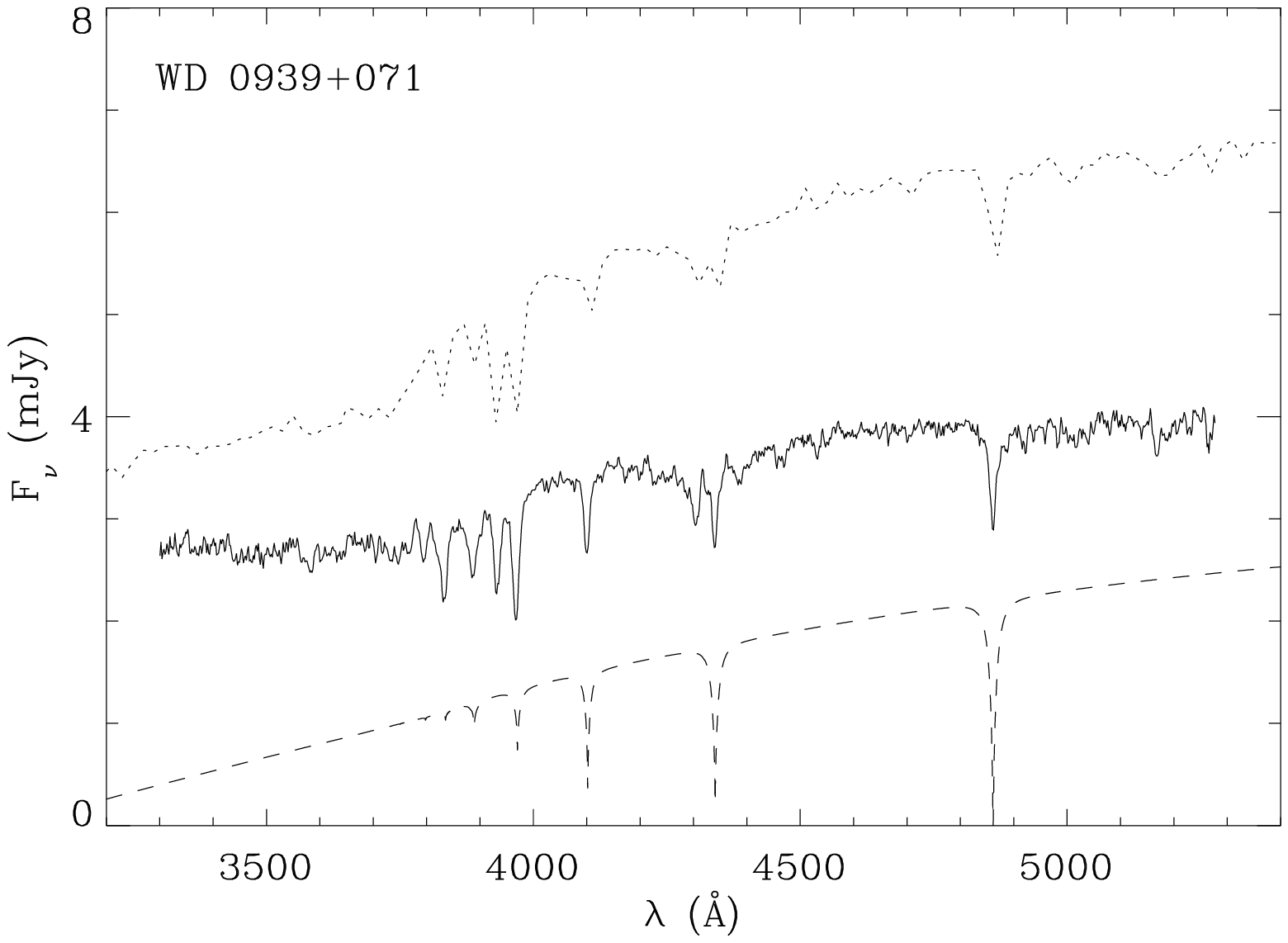}
\caption{The blue optical spectrum of PG 0939$+$071 (A. Gianninas 2008, private 
communication), shown together with a 7000 K DA white dwarf model (dashed line), and a low
resolution F2V stellar model (dotted line).  Note the blue continuum in the actual data is not seen 
in either model.
\label{fig13}}
\end{figure*}

\begin{figure*}
\includegraphics[width=140mm]{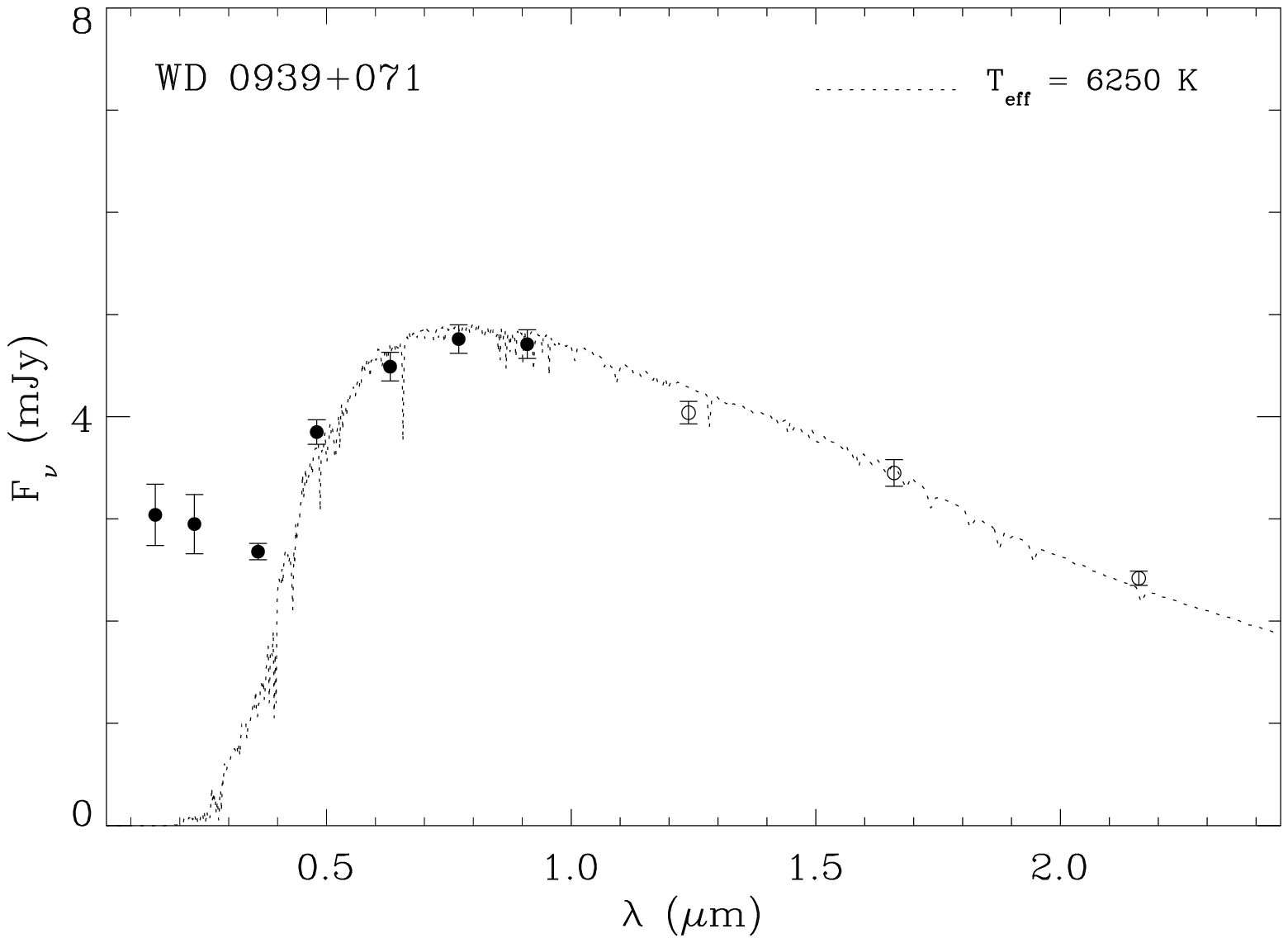}
\caption{An F8V stellar model fitted to the optical and near-infrared photometry of PG 0939$+$071;
revealing a clear ultraviolet flux excess.
\label{fig14}}
\end{figure*}

\begin{figure*}
\includegraphics[width=140mm]{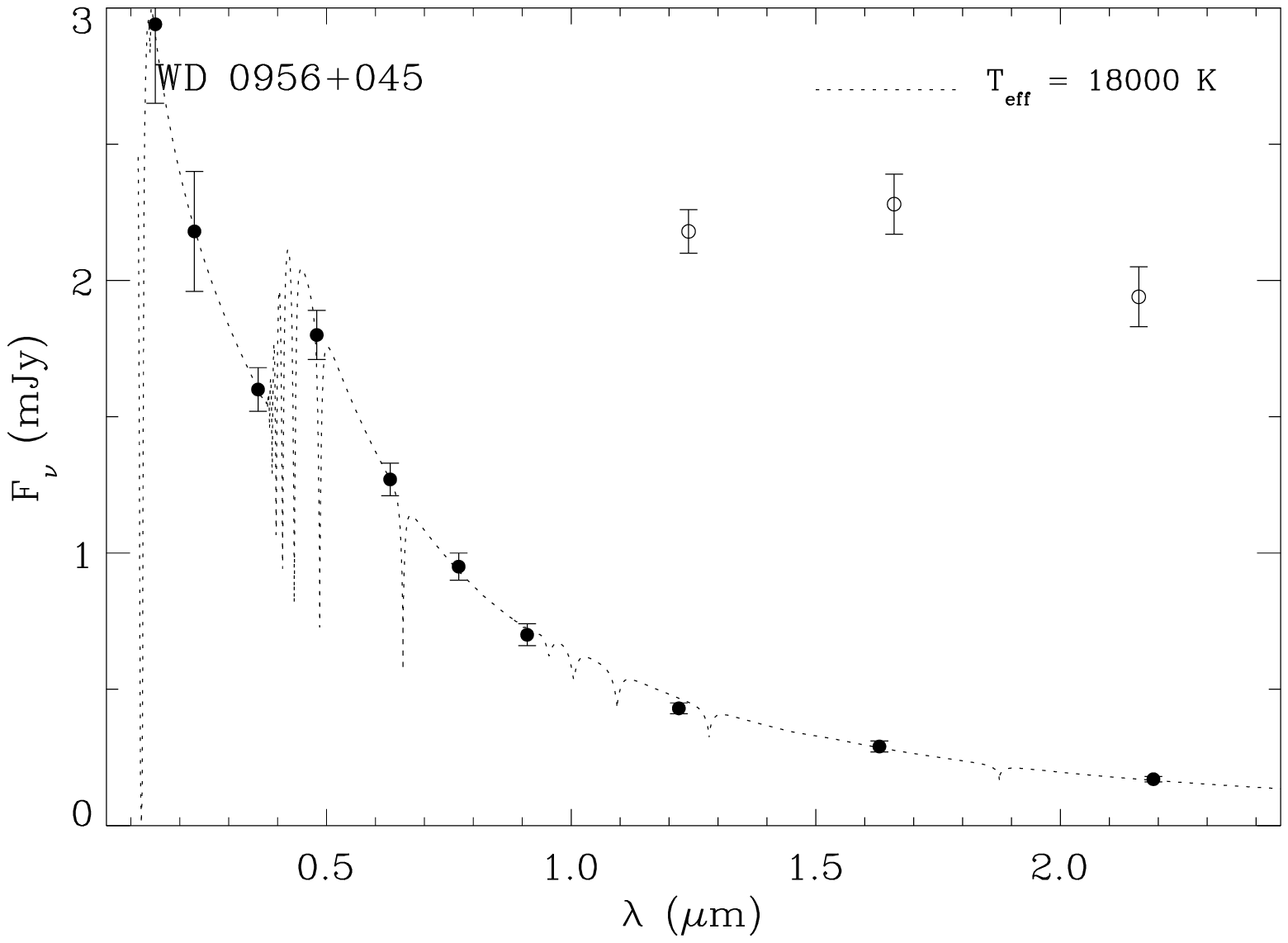}
\caption{Spectral energy distribution of PG 0956$+$045A.  The solid circles are {\em GALEX} far- and
near-ultraviolet, SDSS $ugriz$, and IRTF $JHK$ photometry.  The open circles are 2MASS $JHK_s$ 
photometry, which are primarily due to the flux from PG 0956$+$045B.
\label{fig15}}
\end{figure*}

\begin{figure*}
\includegraphics[width=140mm]{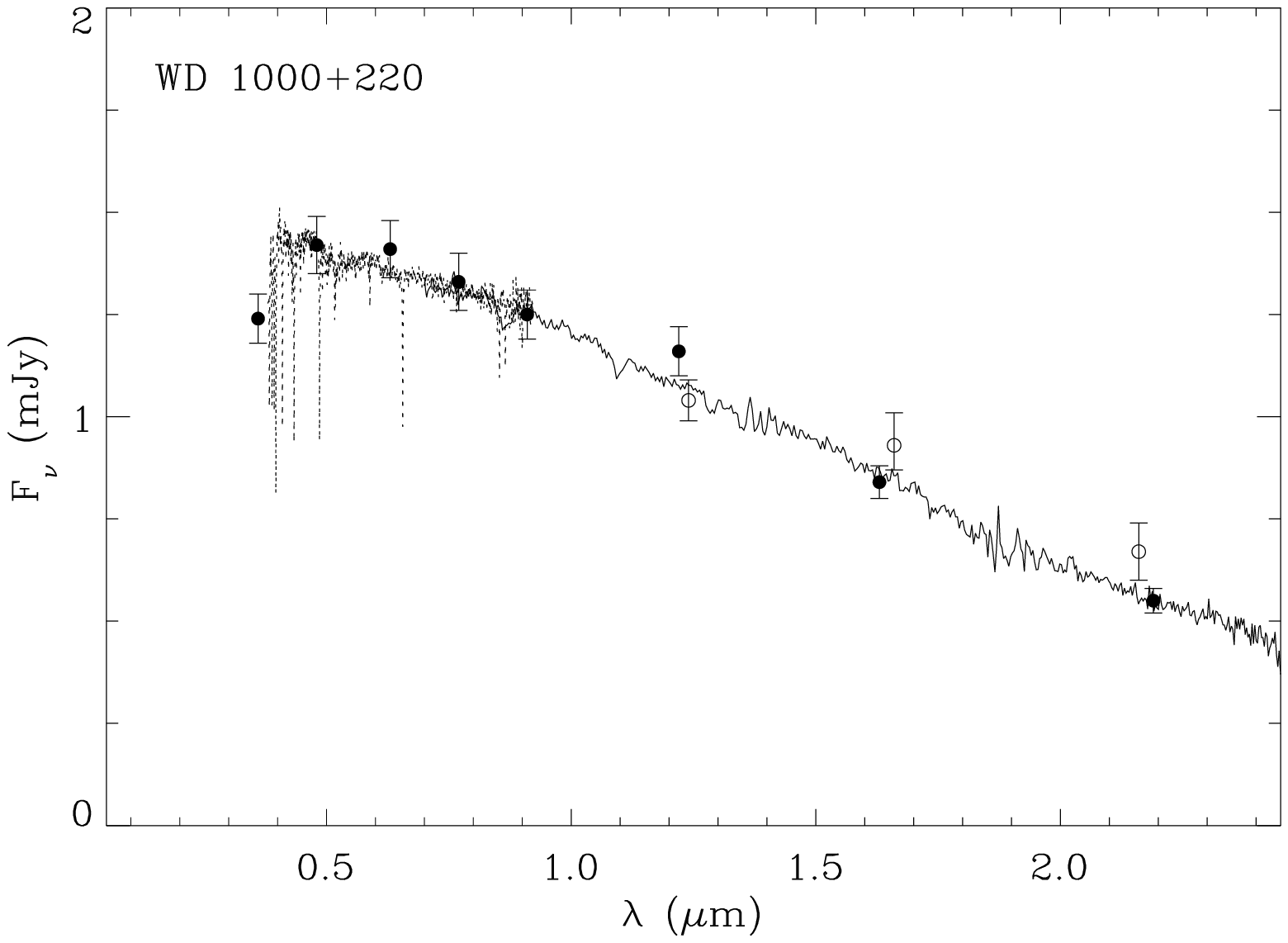}
\caption{Spectral energy distribution of TON 1145.  The solid circles are SDSS $ugriz$, and IRTF 
$JHK$ photometry, while the open circles are 2MASS $JHK_s$ photometry.  The $0.7-2.5$ $\mu$m 
spectrum was taken with SpeX, and the optical SDSS spectrum is also shown (see Figure \ref{fig17}).
\label{fig16}}
\end{figure*}

\begin{figure*}
\includegraphics[width=140mm]{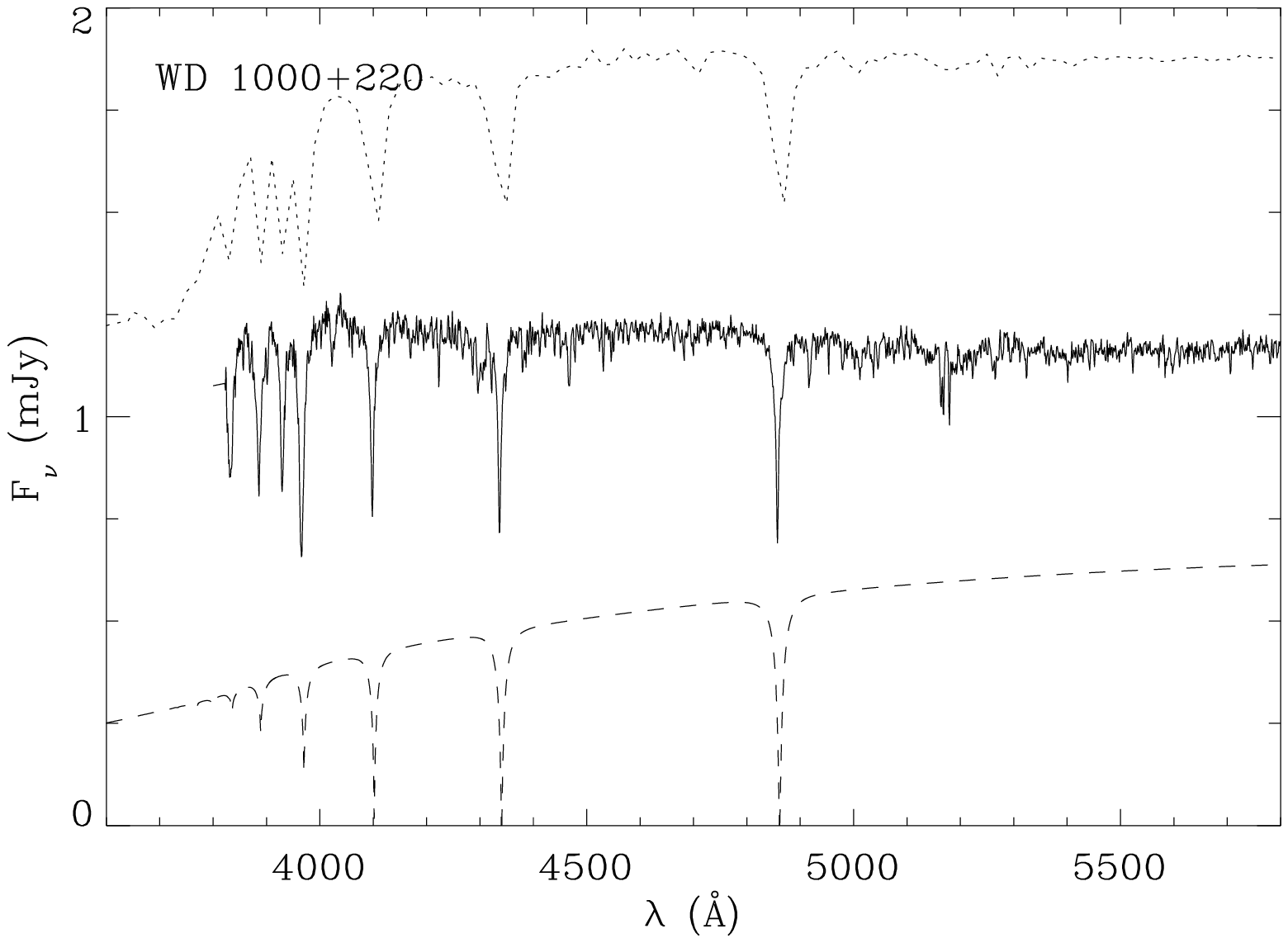}
\caption{The SDSS blue optical spectrum of TON 1145, shown together with a 7500 K DA white 
dwarf model (dashed line), and a low resolution A8V stellar model (dotted line).  Note the flat
psuedo-continuum below 4500\AA \ in the actual data is not seen in either model.
\label{fig17}}
\end{figure*}

\begin{figure*}
\includegraphics[width=140mm]{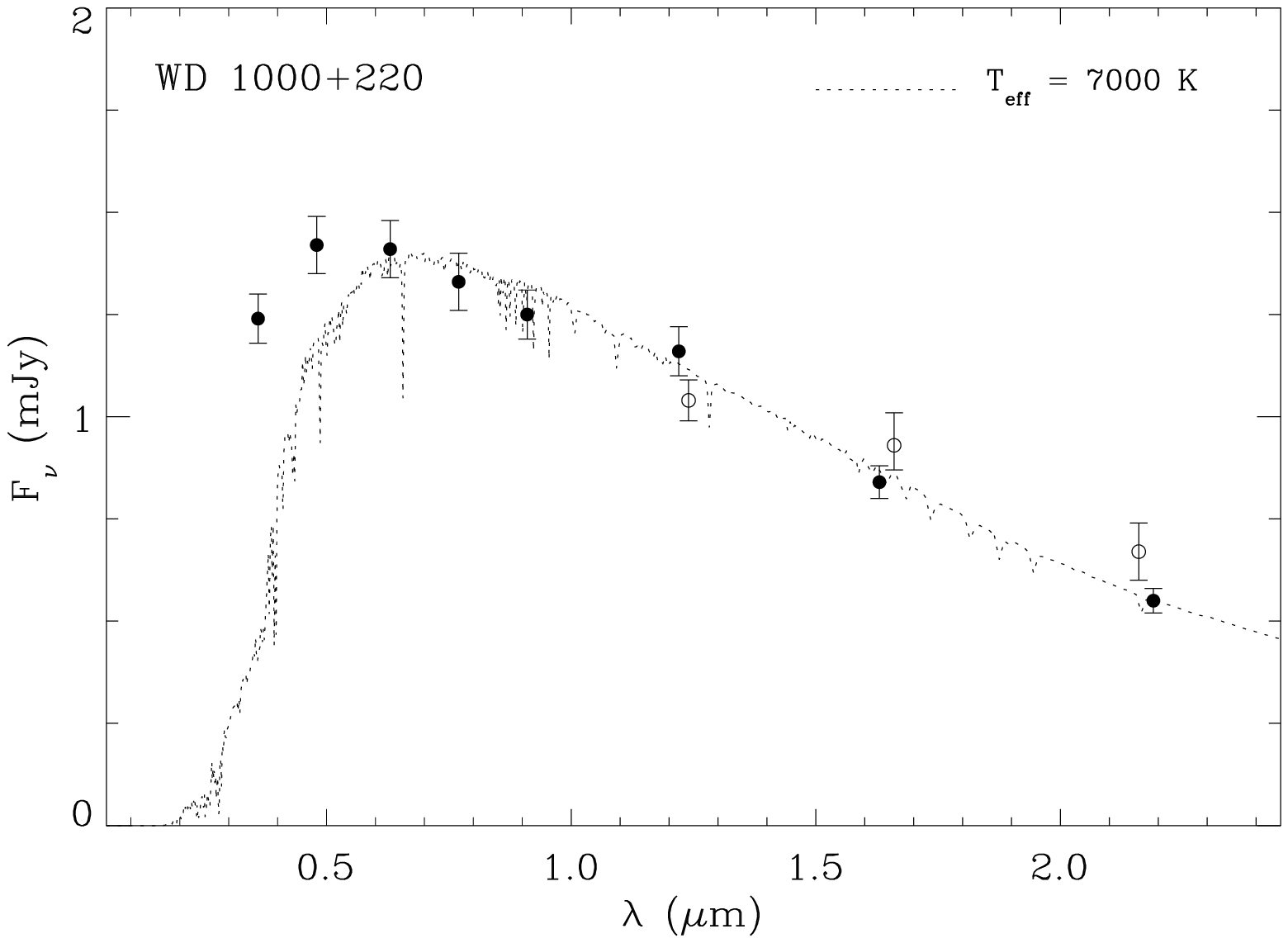}
\caption{An F2V stellar model fitted to the optical and near-infrared photometry of TON 1145.
While there are no {\em GALEX} data for this star, it appears possible it also has an ultraviolet excess.
\label{fig18}}
\end{figure*}

\begin{figure*}
\includegraphics[width=140mm]{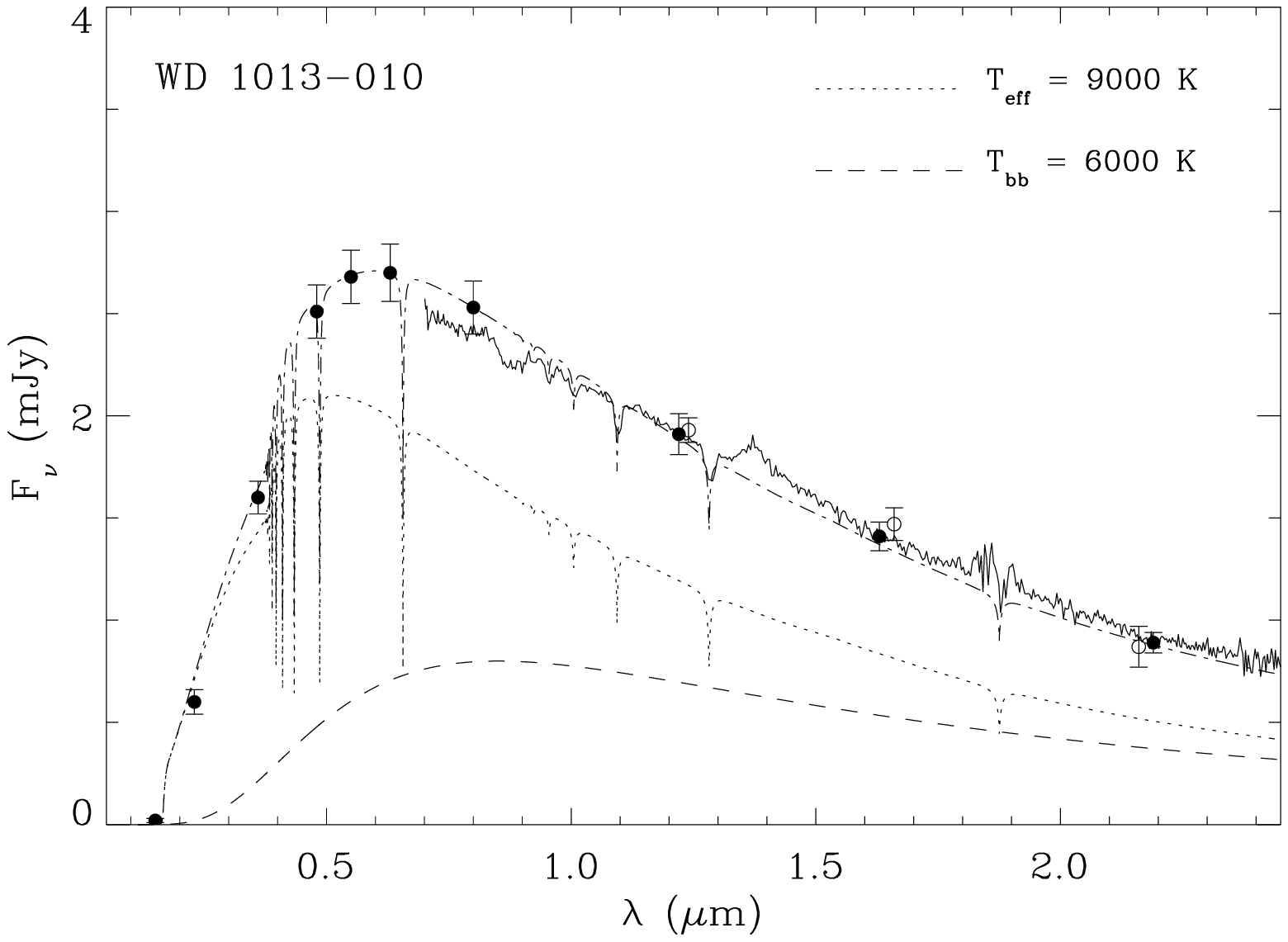}
\caption{Spectral energy distribution of G53-38.  The solid circles are {\em GALEX} far- and
near-ultraviolet, SDSS $ugr$, optical $VI$, and IRTF $JHK$ photometry, while the open circles 
are 2MASS $JHK_s$ photometry.  The $0.7-2.5$ $\mu$m spectrum was taken with SpeX.  There 
appears to be photometric evidence of the known, hidden white dwarf companion \citep{nel05}.
\label{fig19}}
\end{figure*}

\begin{figure*}
\includegraphics[width=140mm]{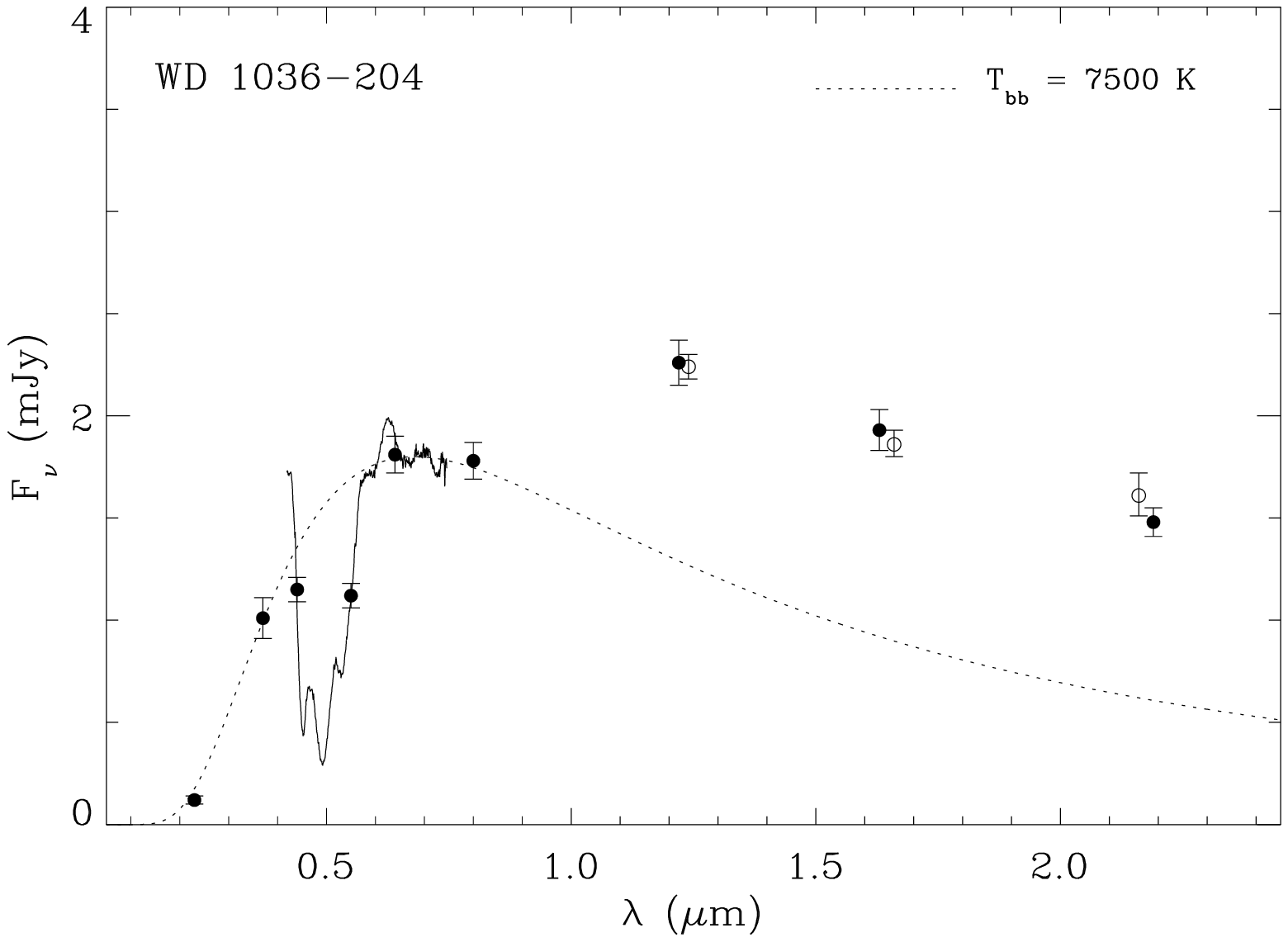}
\caption{Spectral energy distribution of LHS 2293.  The solid circles are {\em GALEX} near-ultraviolet,
optical $UBVRI$, and IRTF $JHK$ photometry, while the open circles are 2MASS $JHK_s$ 
photometry.  Also shown is the optical spectrum from \citet{sch95}.  The ultraviolet and optical 
photometry seem to indicate an effective temperature near 7500 K \citep{lie78}.
\label{fig20}}
\end{figure*}

\begin{figure*}
\includegraphics[width=140mm]{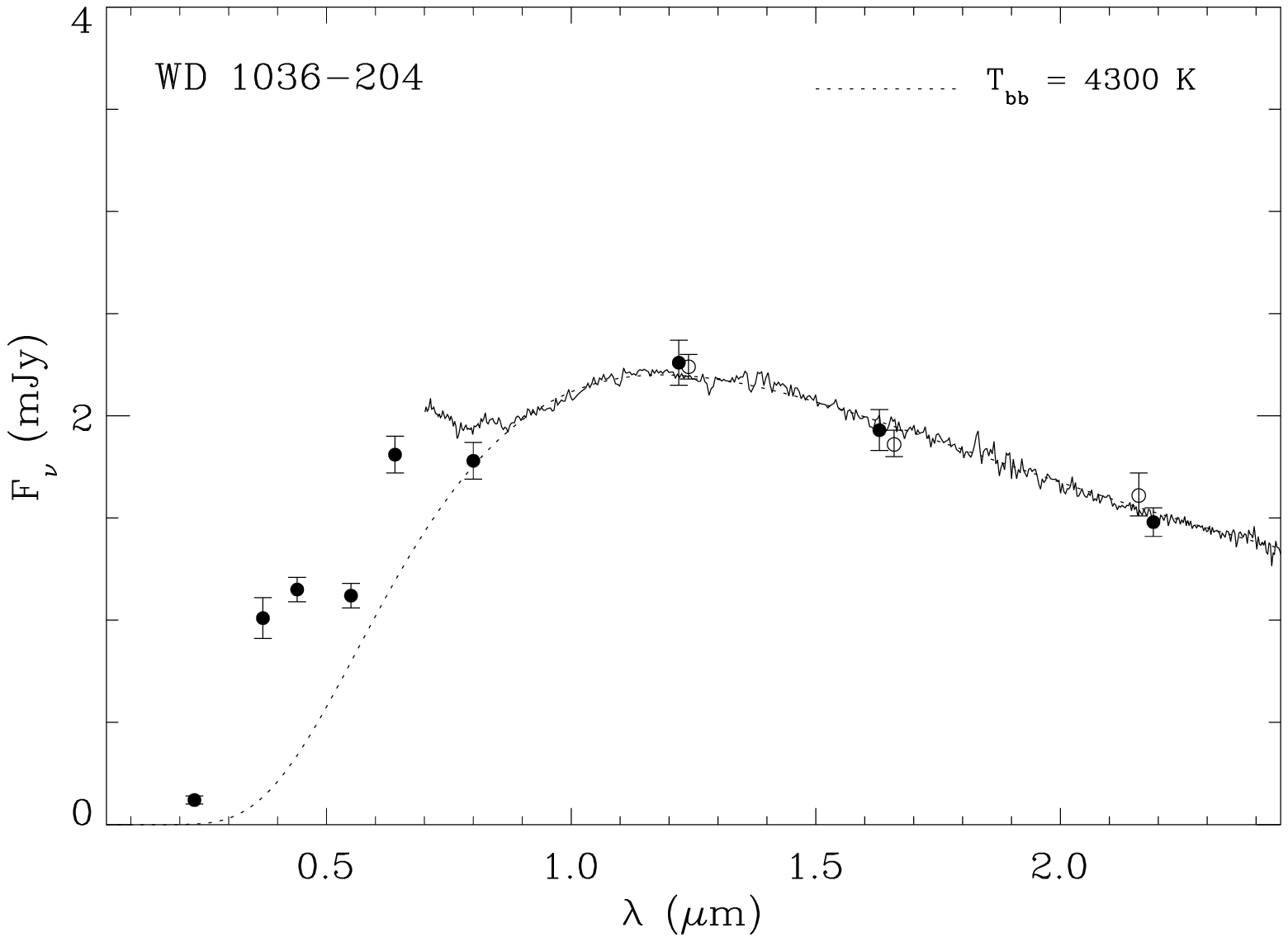}
\caption{The $0.7-2.5$ $\mu$m spectrum taken with SpeX indicates a significantly lower 
effective temperature of 4300 K.  The region below 0.9 $\mu$m may suffer from poor correction 
to the instrument response.
\label{fig21}}
\end{figure*}

\begin{figure*}
\includegraphics[width=140mm]{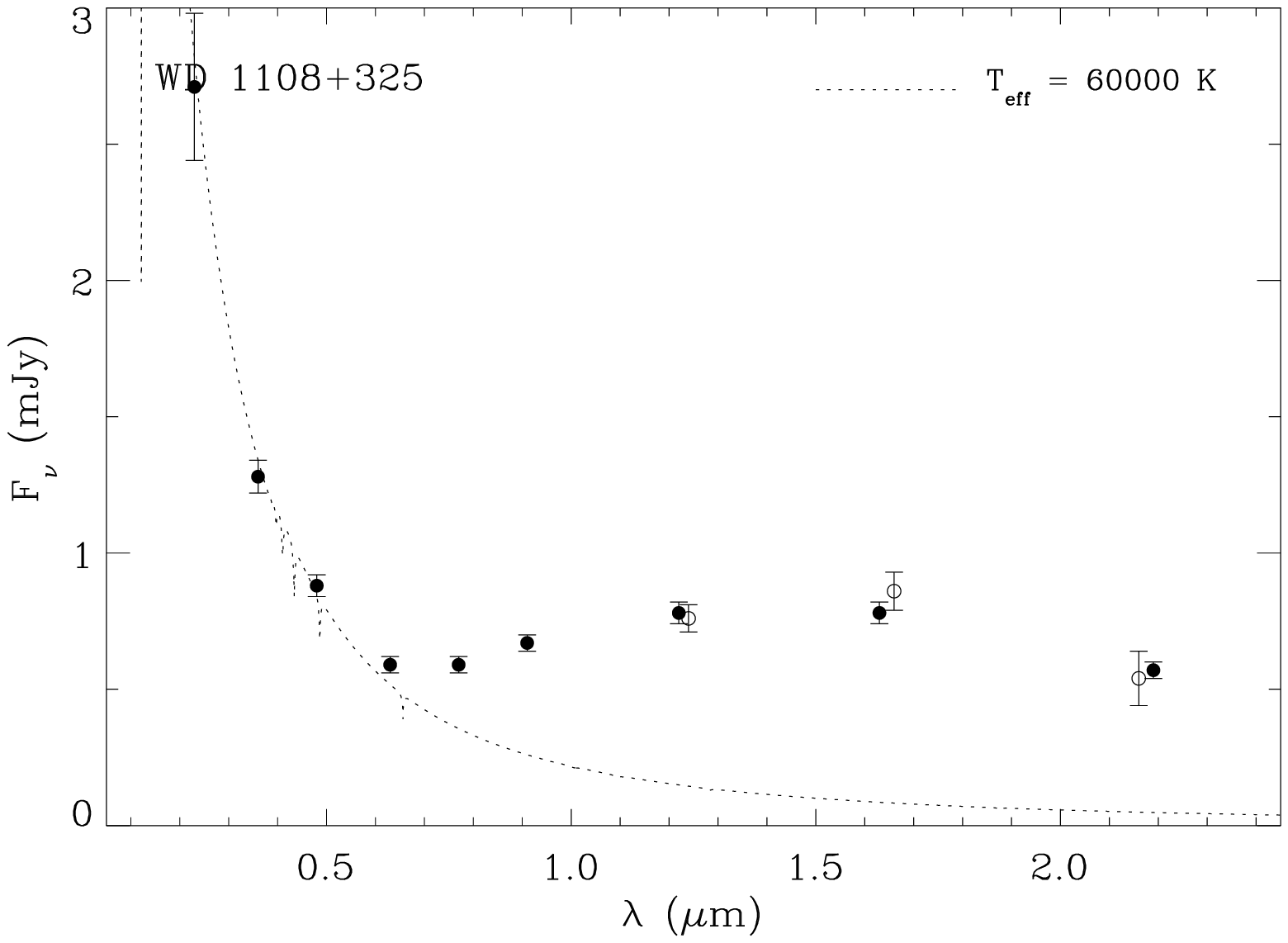}
\caption{Spectral energy distribution of TON 60.  The solid circles are {\em GALEX} 
near-ultraviolet, SDSS $ugriz$, and IRTF $JHK$ photometry, while the open circles are 2MASS 
$JHK_s$ photometry.  The {\em GALEX} far-ultraviolet flux of $4.9\pm0.5$ mJy is not shown for scaling 
purposes.  The light of TON 60B dominates at wavelengths beyond 0.8 $\mu$m.
\label{fig22}}
\end{figure*}

\begin{figure*}
\includegraphics[width=140mm]{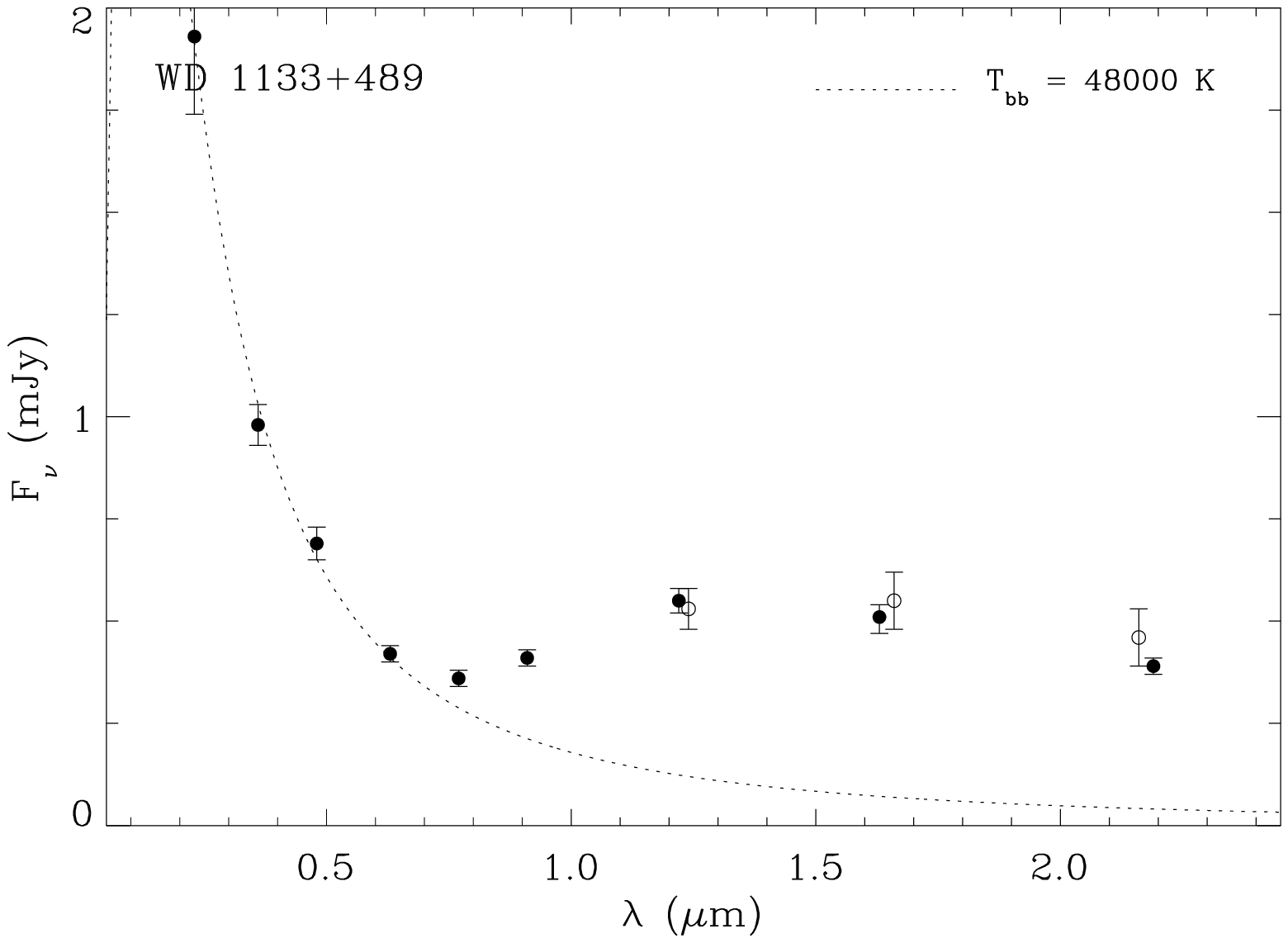}
\caption{Spectral energy distribution of PG 1133$+$489.  The solid circles are {\em GALEX} 
near-ultraviolet, SDSS $ugriz$, and IRTF $JHK$ photometry, while the open circles are 2MASS 
$JHK_s$ photometry.  The {\em GALEX} far-ultraviolet flux of $3.1\pm0.3$ mJy is not shown for scaling 
purposes.  The companion flux dominates at wavelengths beyond 0.8 $\mu$m.
\label{fig23}}
\end{figure*}

\begin{figure*}
\includegraphics[width=140mm]{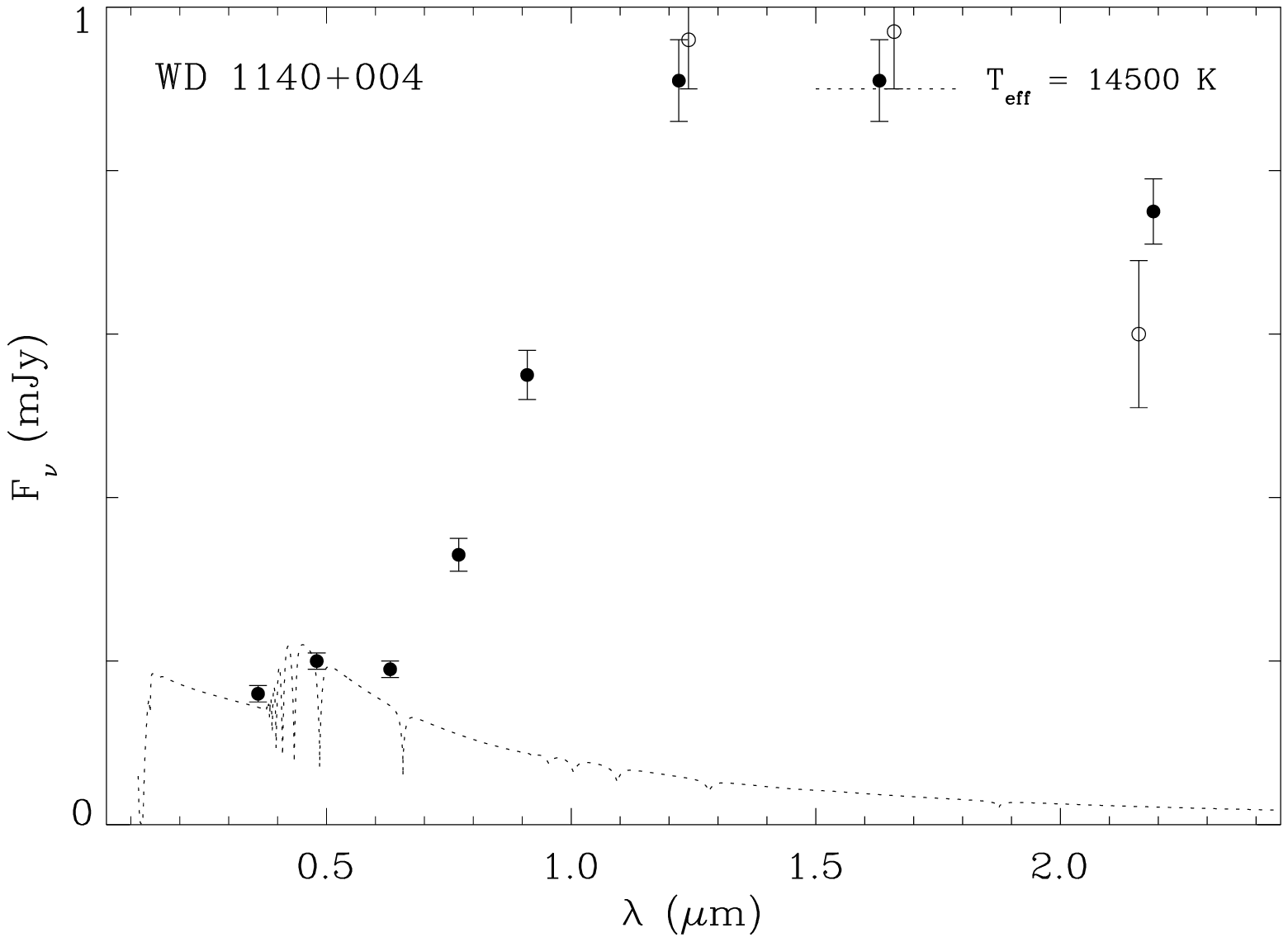}
\caption{Spectral energy distribution of SDSS J114312.57$+$000926.5.  The solid circles 
are SDSS $ugriz$, and IRTF $JHK$ photometry, while the open circles are 2MASS $JHK_s$ 
photometry.  The {\em GALEX} data for this source appear unreliable due to a nearby, bright star.
The companion flux strongly dominates the system at red optical wavelengths.
\label{fig24}}
\end{figure*}

\begin{figure*}
\includegraphics[width=140mm]{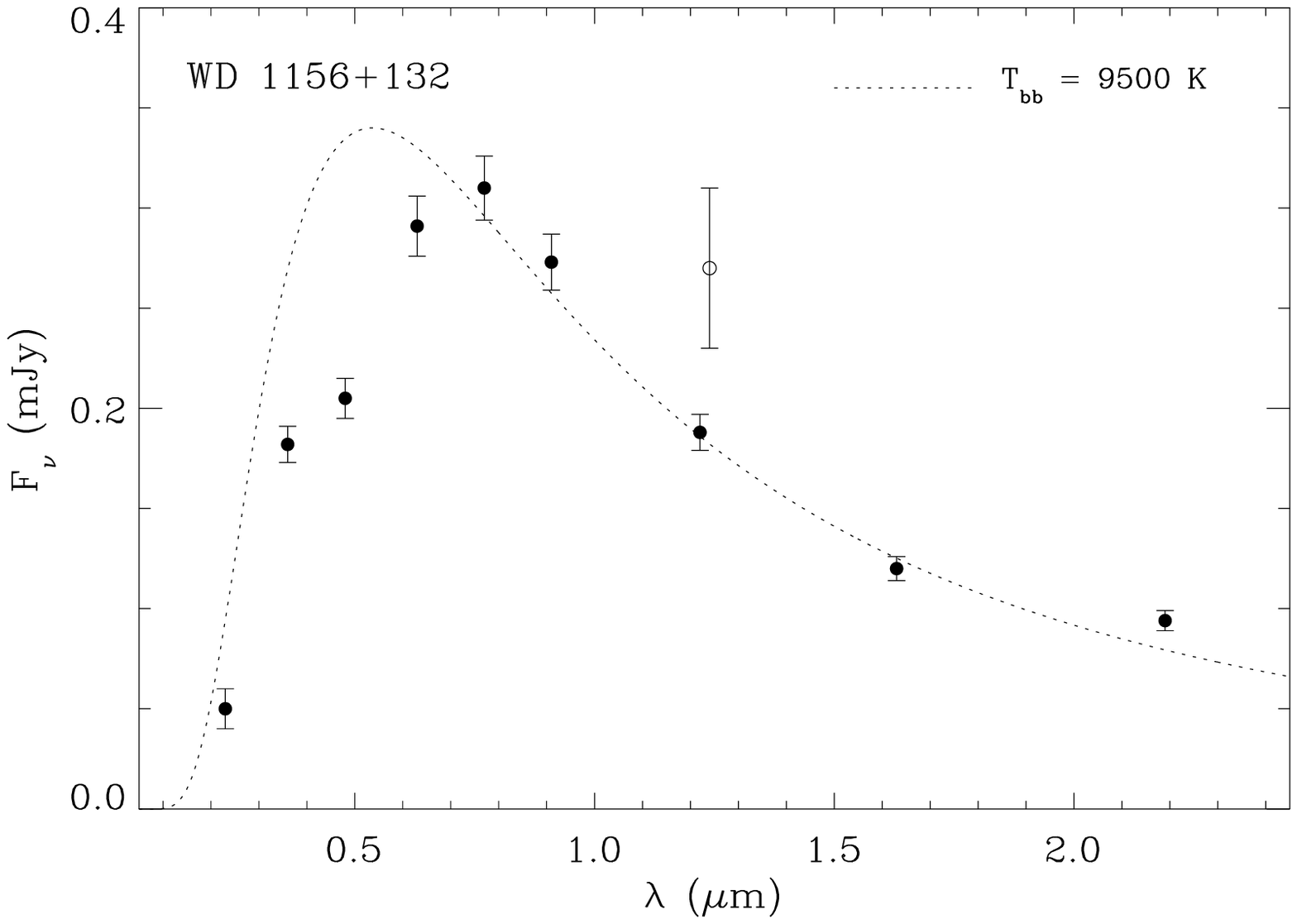}
\caption{Spectral energy distribution of LP 494-12.  The solid circles are {\em GALEX} 
near-ultraviolet, SDSS $ugriz$, and IRTF $JHK$ photometry, while the open circle is 2MASS $J$ 
photometry.  The $ugr$ flux points were ignored in the fit as large absorption bands affect those 
bandpasses, while the $iz$ photometry is unaffected by features in the SDSS spectrum.
\label{fig25}}
\end{figure*}

\begin{figure*}
\includegraphics[width=140mm]{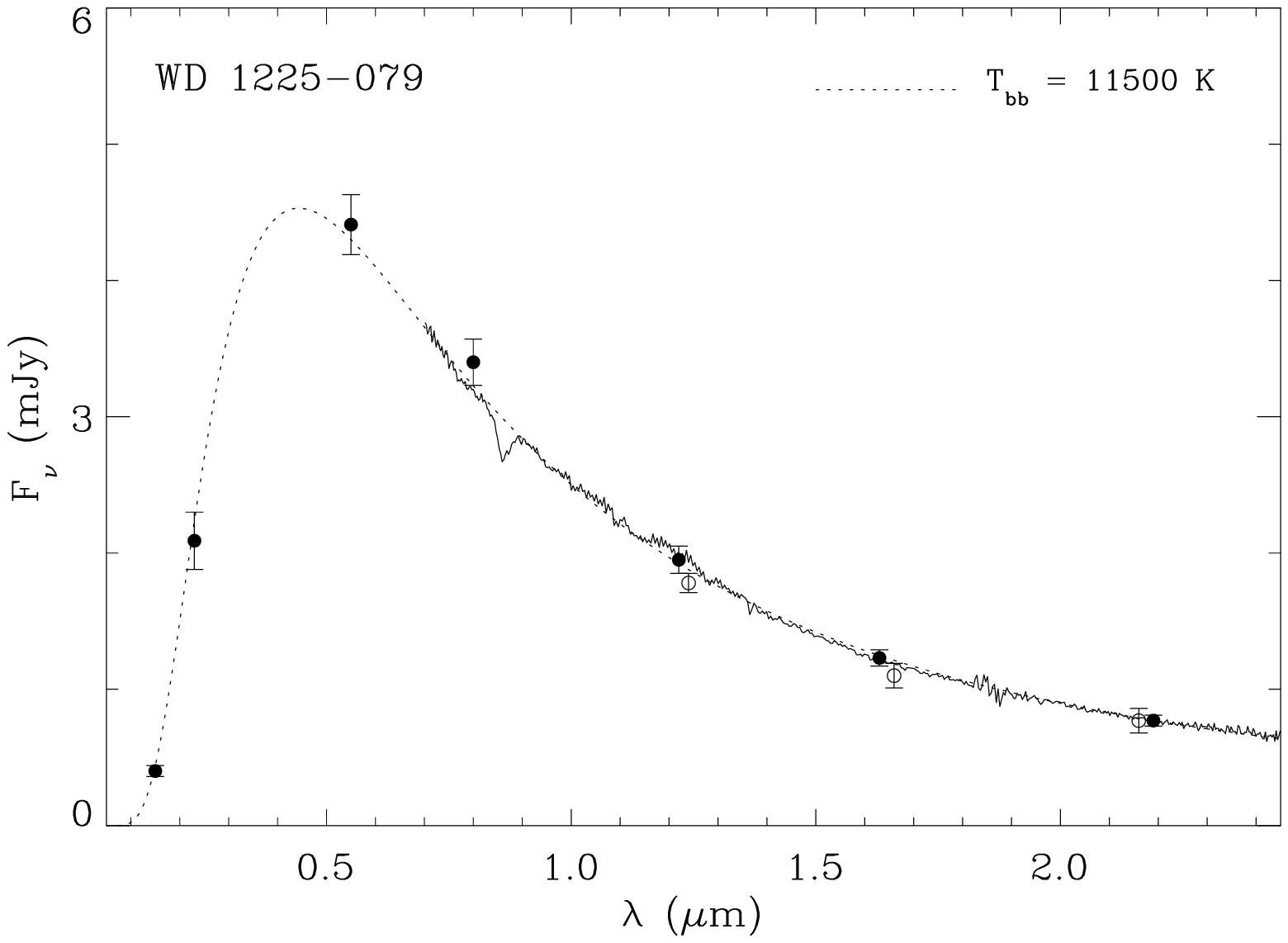}
\caption{Spectral energy distribution of PG 1225$-$079.  The solid circles are {\em GALEX} far- and 
near-ultraviolet, optical $V$, and IRTF $JHK$ photometry, while the open circles are 2MASS 
$JHK_s$ photometry.   The $0.7-2.5$ $\mu$m spectrum was taken with SpeX; the calcium 
absorption triplet centered near 8600 \AA \ is clearly detected.
\label{fig26}}
\end{figure*}

\begin{figure*}
\includegraphics[width=140mm]{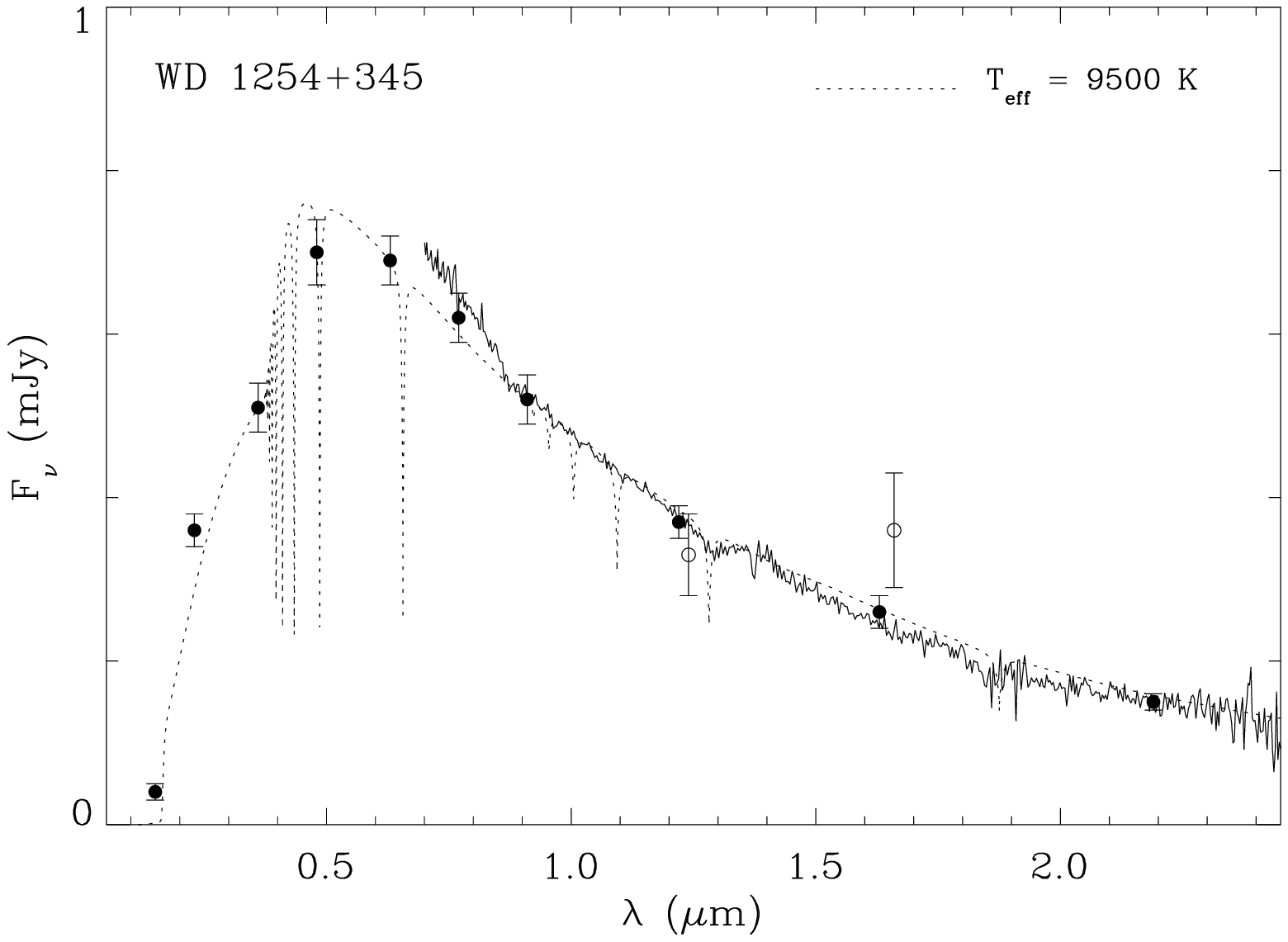}
\caption{Spectral energy distribution of HS 1254$+$345.  The solid circles are {\em GALEX} far- and 
near-ultraviolet, SDSS $ugriz$, and IRTF $JHK$ photometry, while the open circles are 2MASS 
$JH$ photometry.  The $0.7-2.5$ $\mu$m spectrum was taken with SpeX; the absence of 
hydrogen lines is consistent with the presence of a high magnetic field.  The region below 
0.9 $\mu$m may suffer from poor correction to the instrument response.
\label{fig27}}
\end{figure*}

\begin{figure*}
\includegraphics[width=140mm]{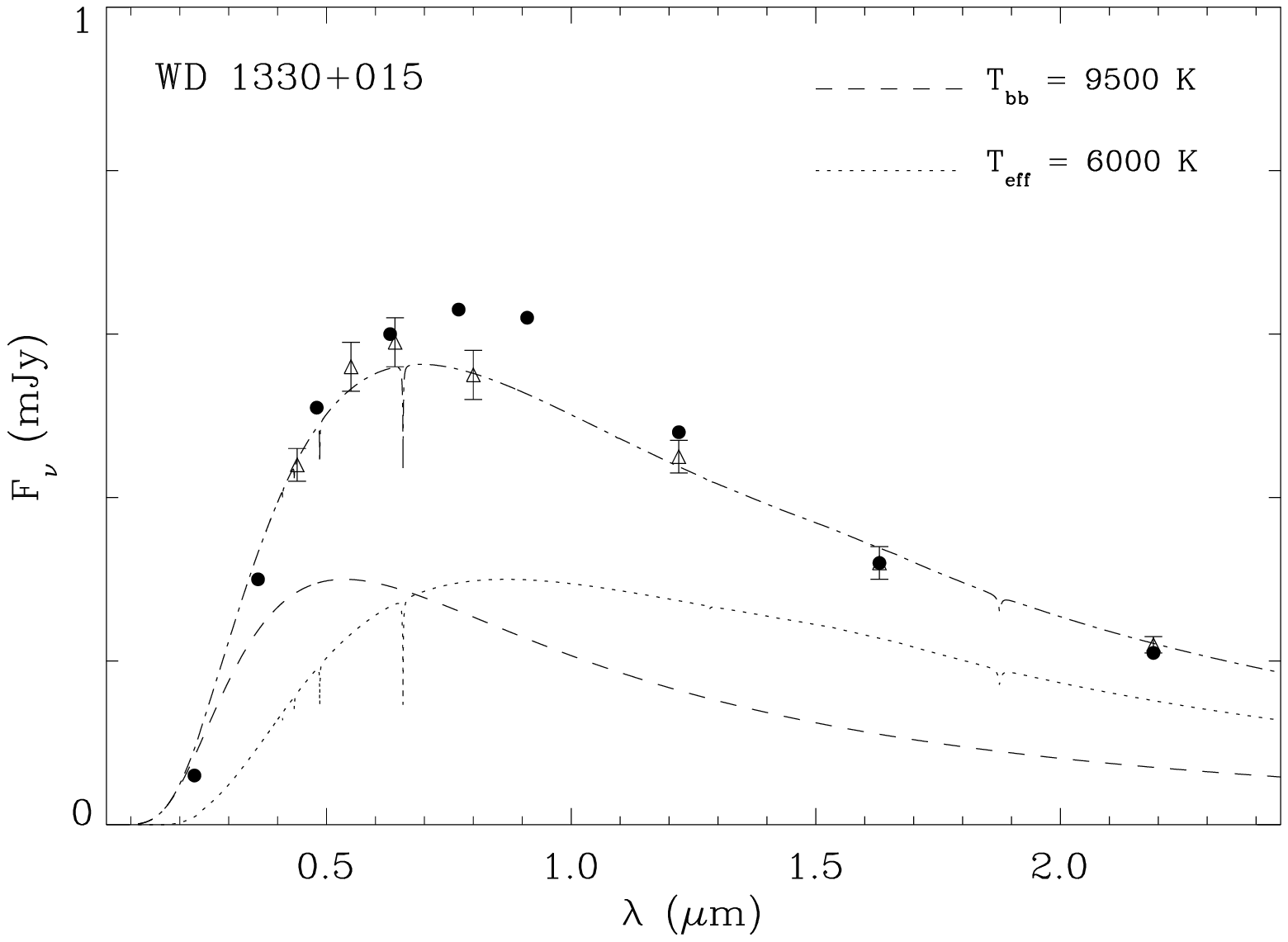}
\caption{Spectral energy distribution of G62-46.  The solid circles are {\em GALEX} near-ultraviolet, 
SDSS $ugriz$, and IRTF $JHK$ photometry, while the open circles are 2MASS $JHK_s$ 
photometry.  The open triangles are $BVRIJHK$ from \citet{ber97} and the models are fitted 
to their data.  The $I$-band flux predicts a hotter DC component at odds with the SDSS $iz$ 
photometry.
\label{fig28}}
\end{figure*}

\begin{figure*}
\includegraphics[width=140mm]{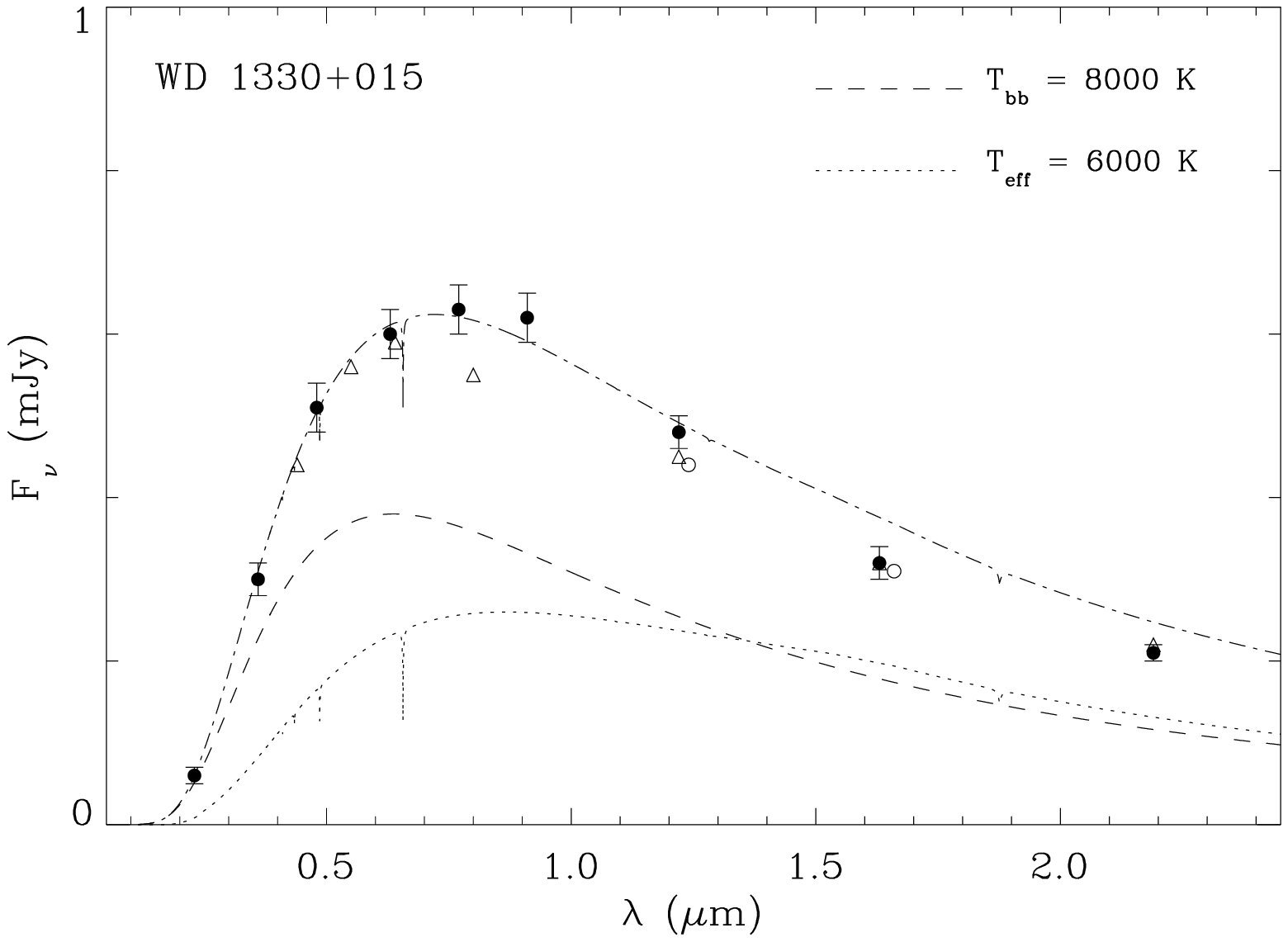}
\caption{Same as Figure \ref{fig28} but now the models are fitted to the $ugriz$ photometry.
While a cooler DC component reproduces the optical data, the full spectral energy distribution
cannot be fitted simultaneously.  Variability appears possible in this likely binary (see \S4).
\label{fig29}}
\end{figure*}

\begin{figure*}
\includegraphics[width=140mm]{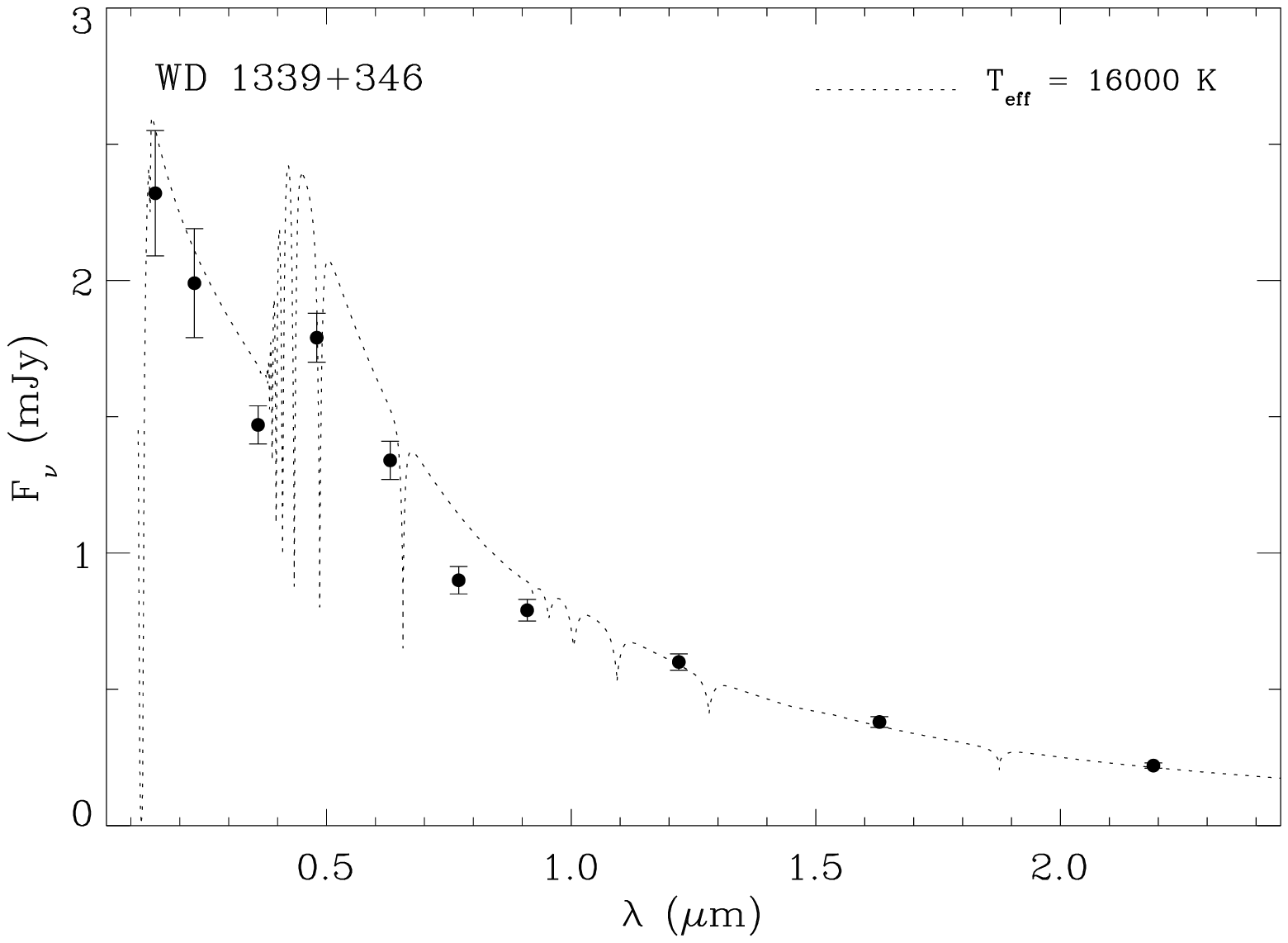}
\caption{Spectral energy distribution of PG 1339$+$346.  The solid circles are {\em GALEX} far- and 
near-ultraviolet, SDSS $ugriz$, and IRTF $JHK$ photometry.  The white dwarf was spatially and 
photometrically resolved from its apparent (i.e. non-physical) companion $2''.6$ away (see \S4) 
during the IRTF observations, but the reliability of the SDSS photometry is uncertain.  It appears 
that the $ugriz$ photometry was adversely affected by the background star.
\label{fig30}}
\end{figure*}

\begin{figure*}
\includegraphics[width=140mm]{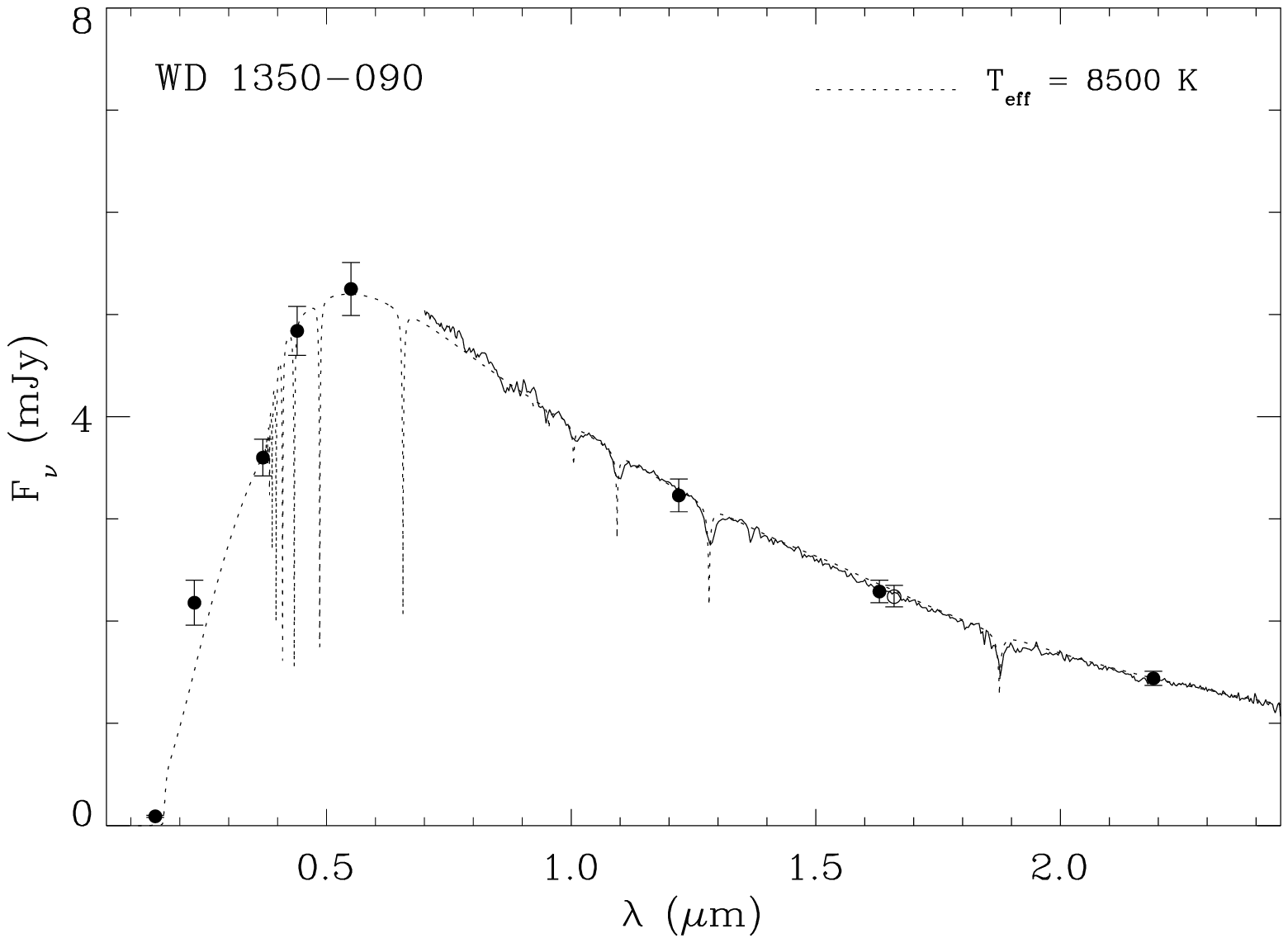}
\caption{Spectral energy distribution of LP907-37.  The solid circles are {\em GALEX} far- and 
near-ultraviolet, optical $UBV$, and IRTF $JHK$ photometry, while the open circle is 2MASS 
$H$ photometry.  The $0.7-2.5$ $\mu$m spectrum was taken with SpeX.
\label{fig31}}
\end{figure*}

\begin{figure*}
\includegraphics[width=140mm]{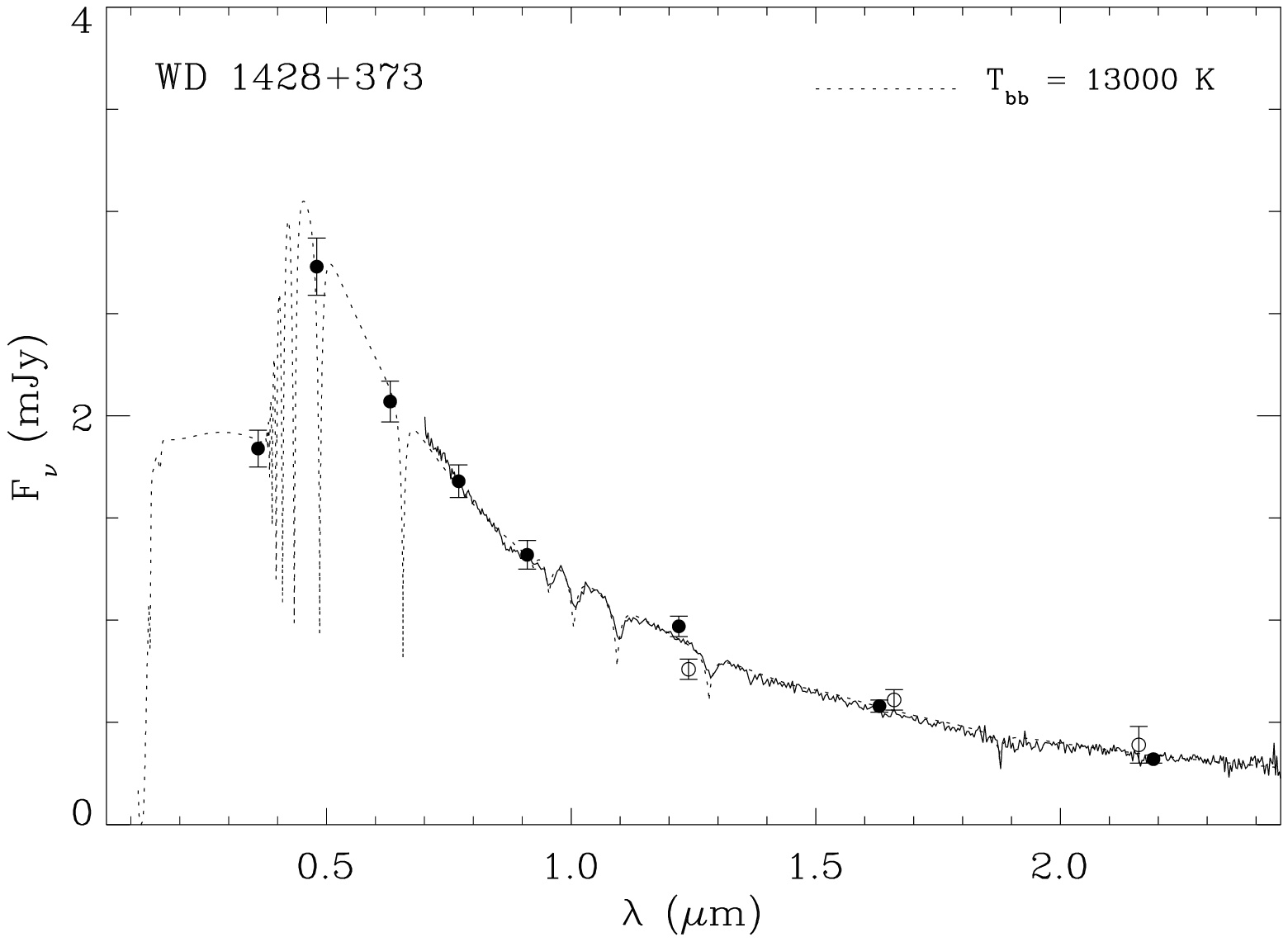}
\caption{Spectral energy distribution of PG 1428$+$373.  The solid circles are SDSS $ugriz$, 
and IRTF $JHK$ photometry, while the open circles are 2MASS $JHK_s$ photometry.  The 
$0.7-2.5$ $\mu$m spectrum was taken with SpeX.  There are no {\em GALEX} data available for this 
star, and there is no photometric evidence of the known white dwarf companion \citep{mor05}.
\label{fig32}}
\end{figure*}

\begin{figure*}
\includegraphics[width=140mm]{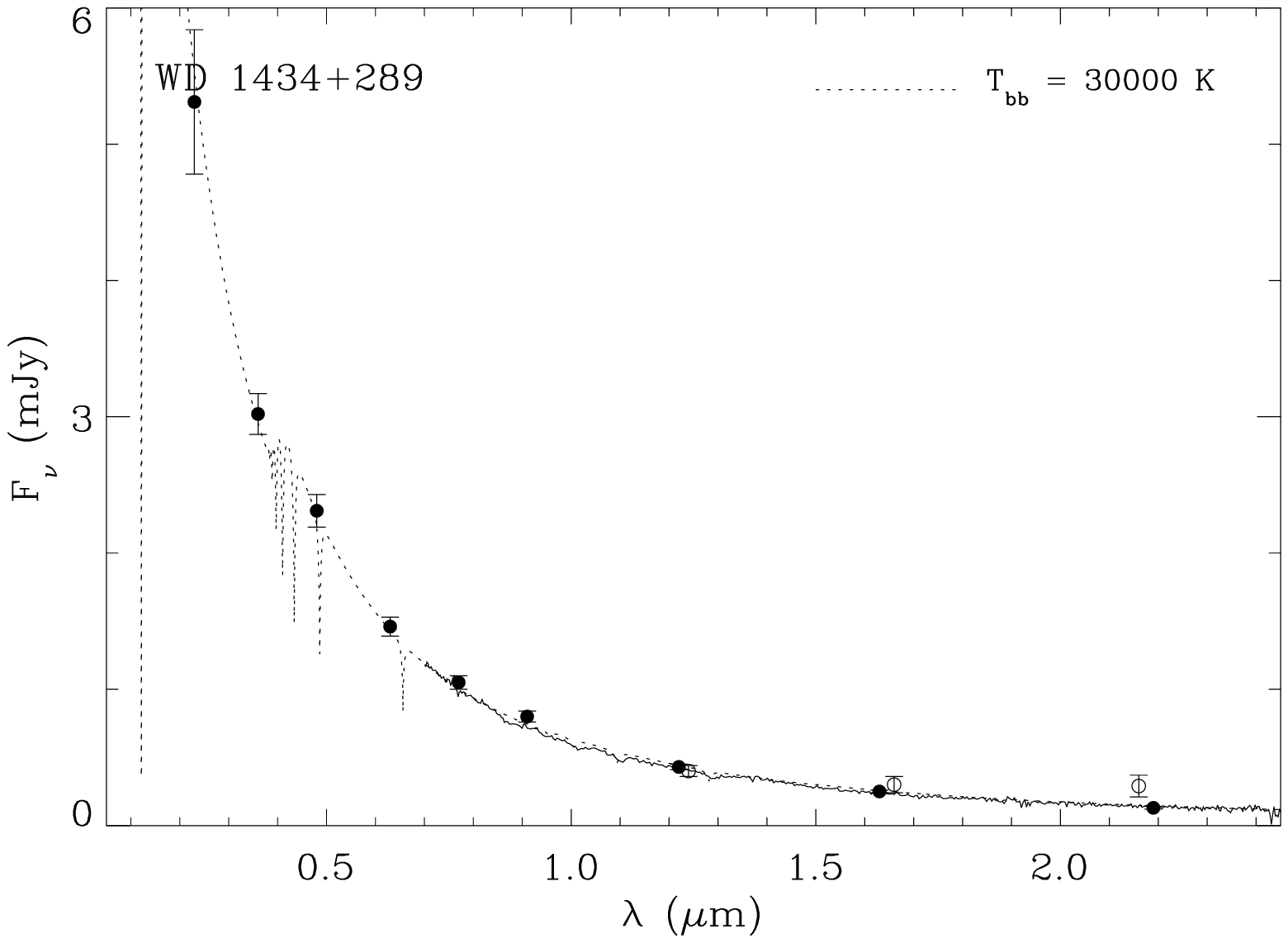}
\caption{Spectral energy distribution of TON 210.  The solid circles are {\em GALEX} far- and 
near-ultraviolet, SDSS $ugriz$, and IRTF $JHK$ photometry, while the open circles are 2MASS 
$JHK_s$ photometry.  The {\em GALEX} far-ultraviolet flux of $7.8\pm0.8$ mJy is not shown for scaling 
purposes.  The $0.7-2.5$ $\mu$m spectrum was taken with SpeX.  
\label{fig33}}
\end{figure*}

\begin{figure*}
\includegraphics[width=140mm]{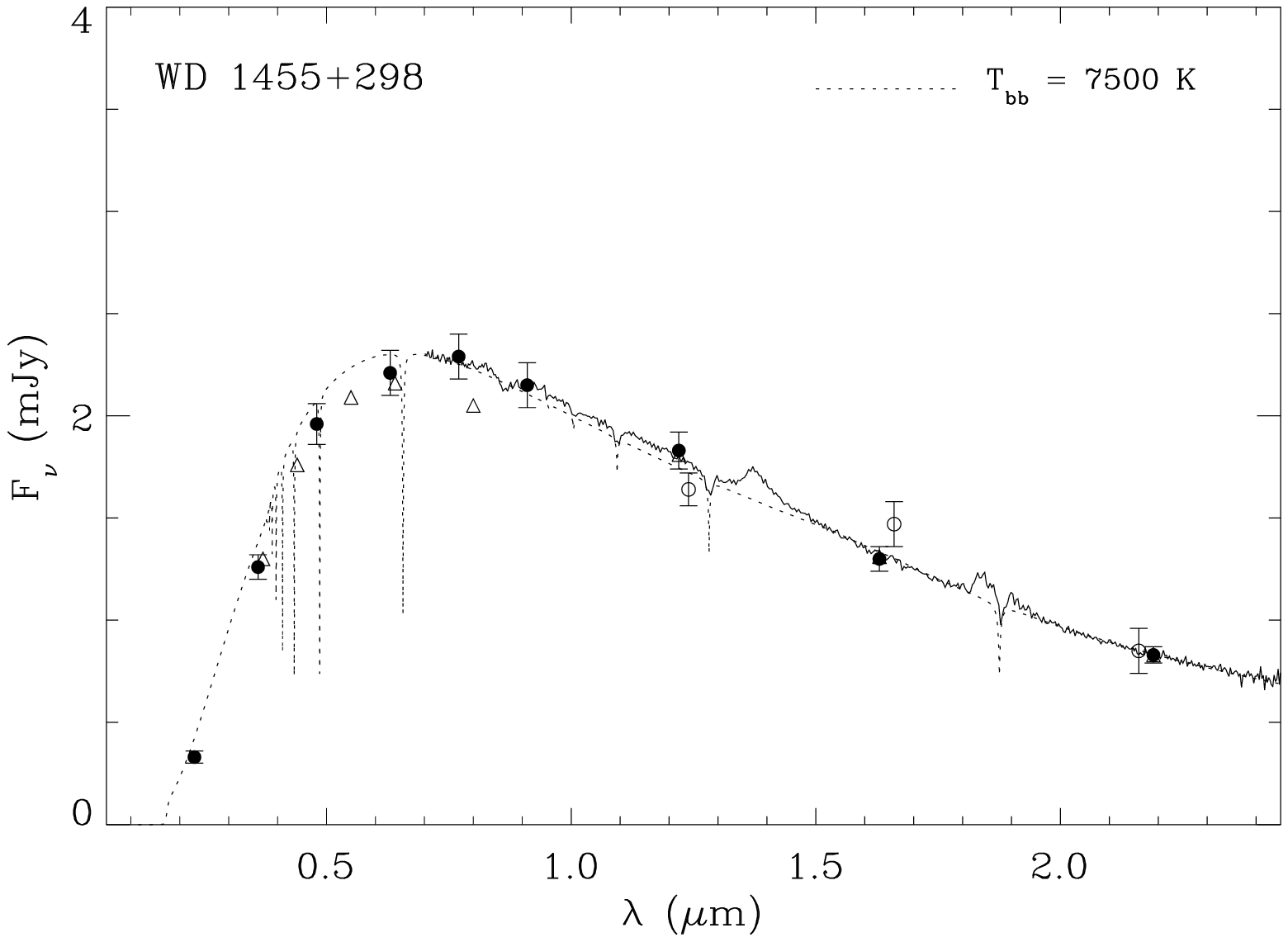}
\caption{Spectral energy distribution of G166-58.  The solid circles are {\em GALEX} 
near-ultraviolet, SDSS $ugriz$, and IRTF $JHK$ photometry, while the open circles are 
2MASS $JHK_s$ photometry.  The open triangles are $UBVRIJHK$ photometry from \citet{ber01,
mcc99}, and the $0.7-2.5$ $\mu$m spectrum was taken with SpeX, which suffers from poor telluric 
correction.
\label{fig34}}
\end{figure*}

\begin{figure*}
\includegraphics[width=140mm]{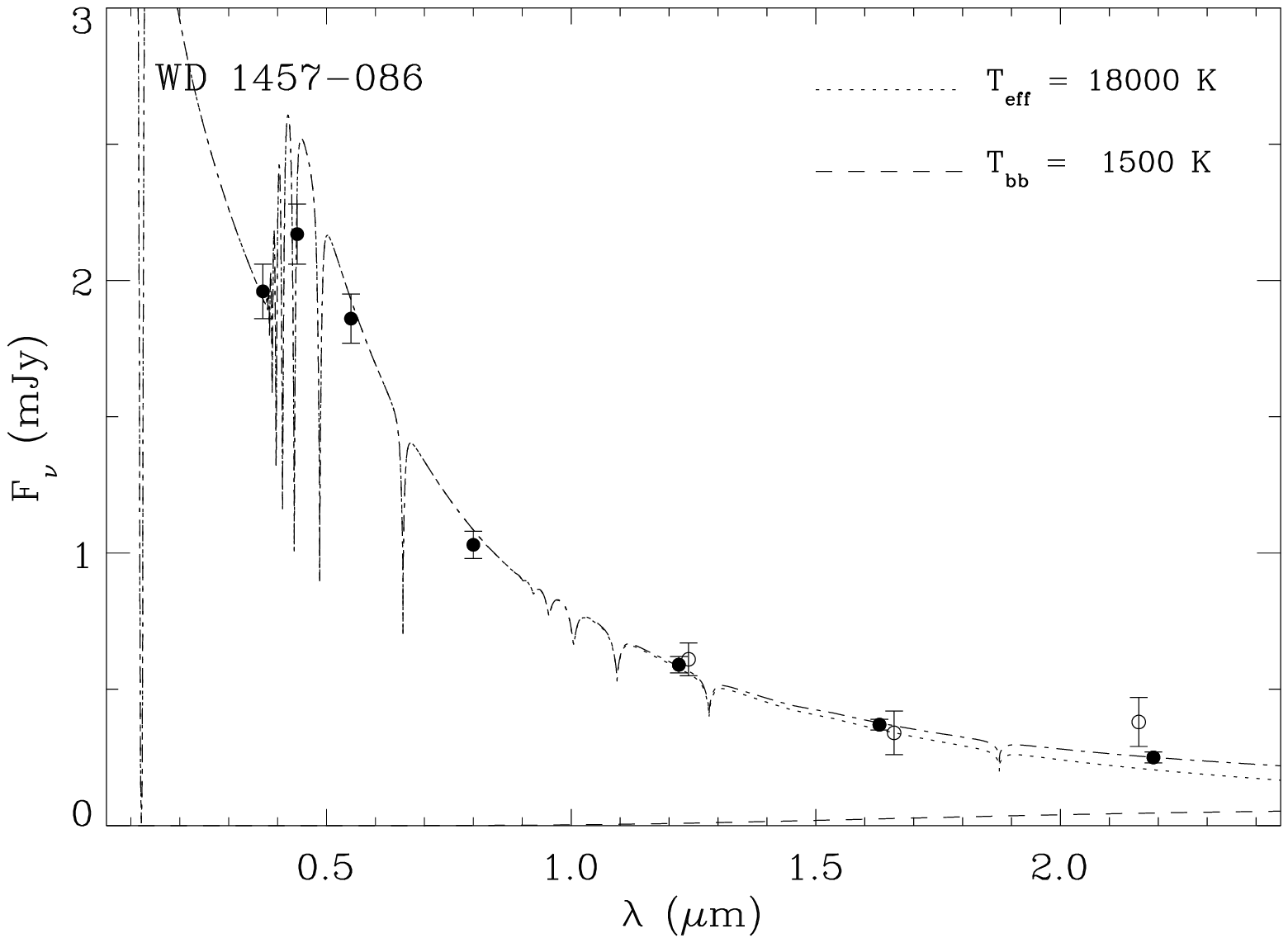}
\caption{Spectral energy distribution of PG 1457$-$086.  The solid circles are optical 
$UBVI$ and IRTF $JHK$ photometry, while the open circles are 2MASS $JHK_s$ photometry.  
There are no {\em GALEX} data available for this star.
\label{fig35}}
\end{figure*}

\begin{figure*}
\includegraphics[width=140mm]{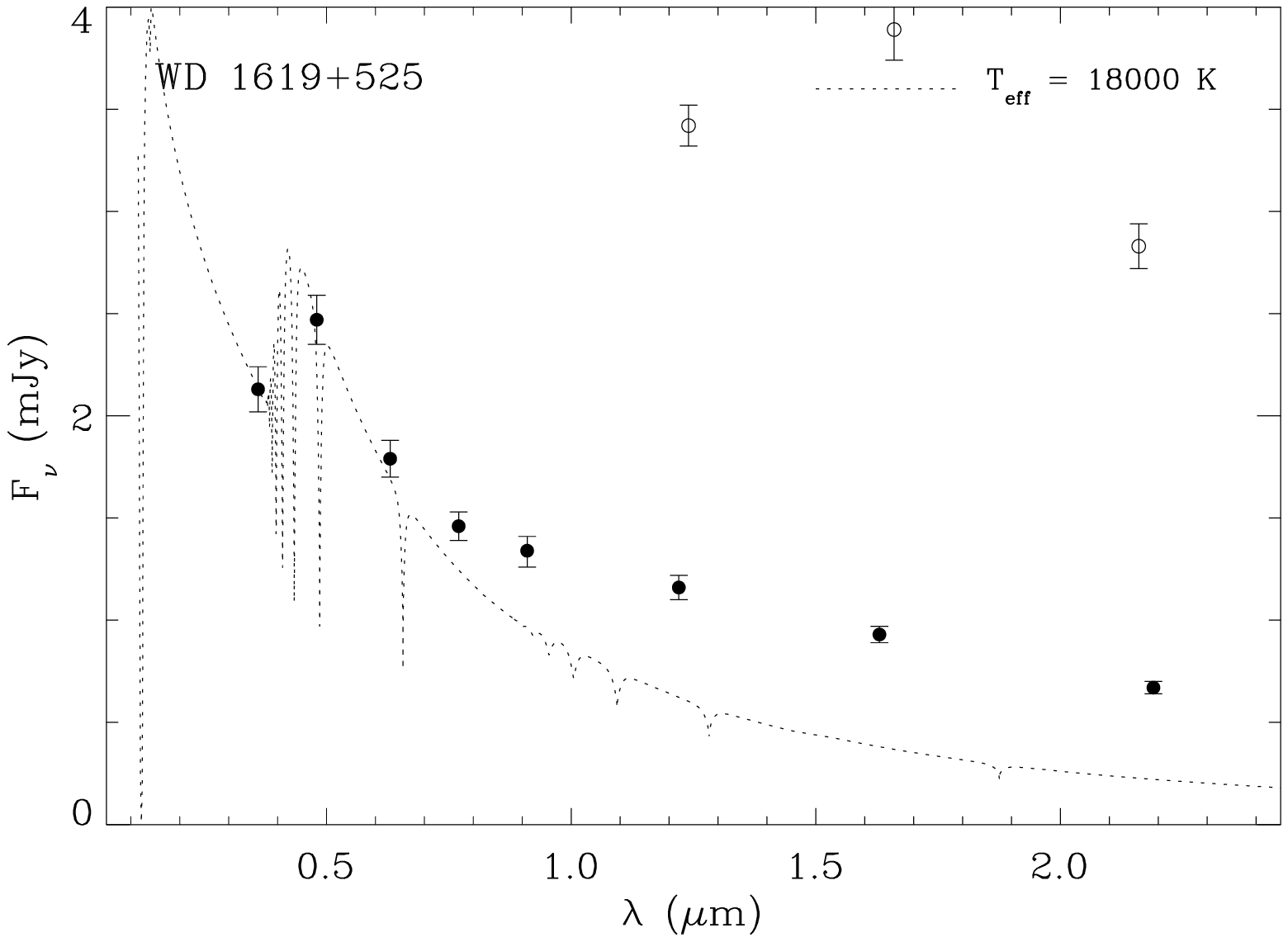}
\caption{Spectral energy distribution of PG 1619$+$525AB.  The solid circles are SDSS 
$ugriz$, and IRTF $JHK$ photometry.  The open circles are 2MASS $JHK_s$ fluxes, primarily
due to the light of PG 1619$+$525C, which does not contaminate the IRTF photometry.  The 
secondary star is spatially resolved only in {\em HST} / ACS observations \citep{far06}.  There 
are no {\em GALEX} data available for this star.
\label{fig36}}
\end{figure*}

\begin{figure*}
\includegraphics[width=140mm]{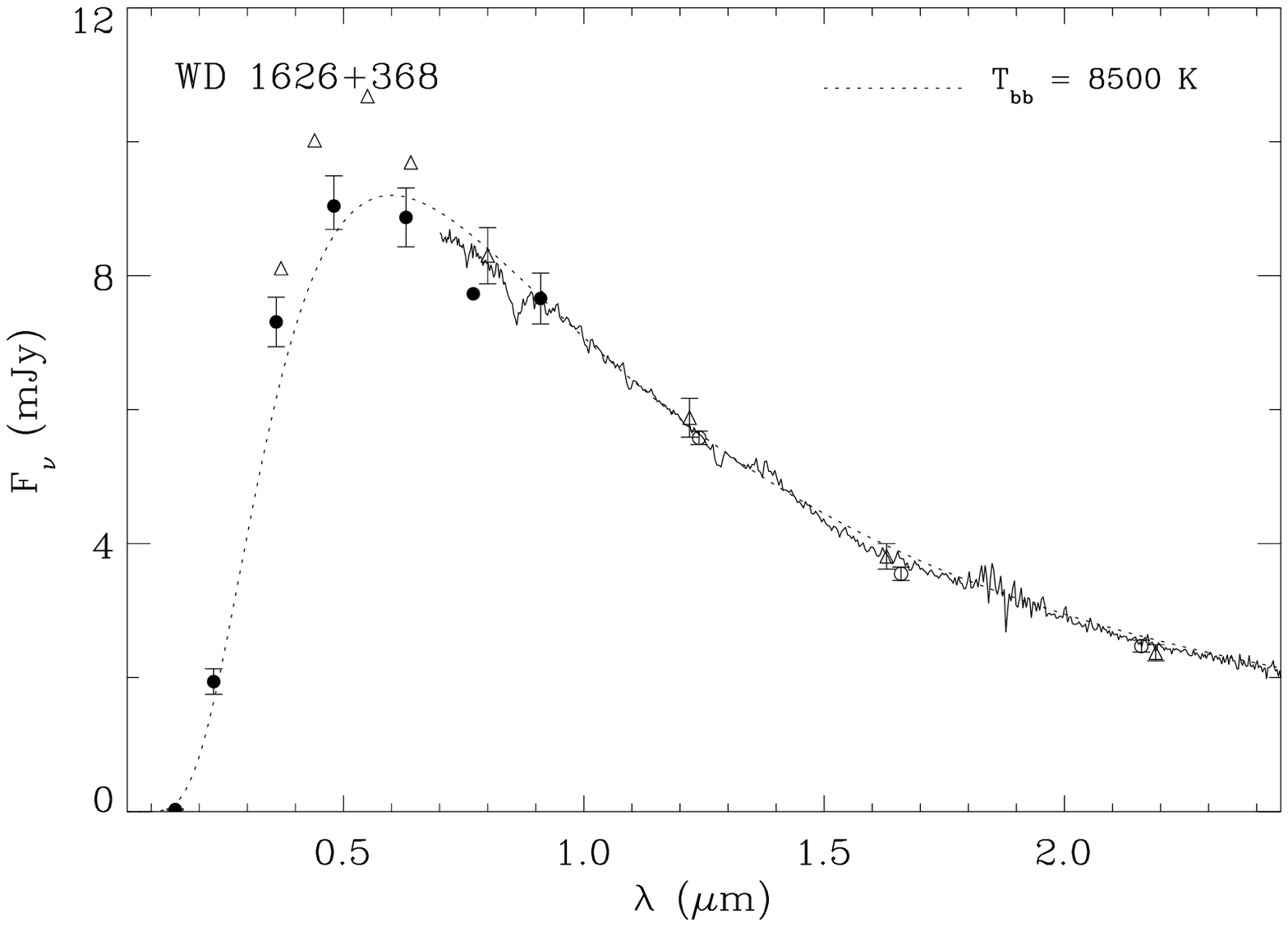}
\caption{Spectral energy distribution of G180-57.  The solid circles are {\em GALEX} far- and
near-ultraviolet, SDSS $ugriz$, and IRTF $JHK$ photometry, while the open circles are 
2MASS $JHK_s$ photometry.  The open triangles are $UBVRIJHK$ photometry from
\citet{ber01,mcc99}, and the $0.7-2.5$ $\mu$m spectrum was taken with SpeX; the calcium 
absorption triplet centered near 8600 \AA  \ is clearly detected.
\label{fig37}}
\end{figure*}

\begin{figure*}
\includegraphics[width=140mm]{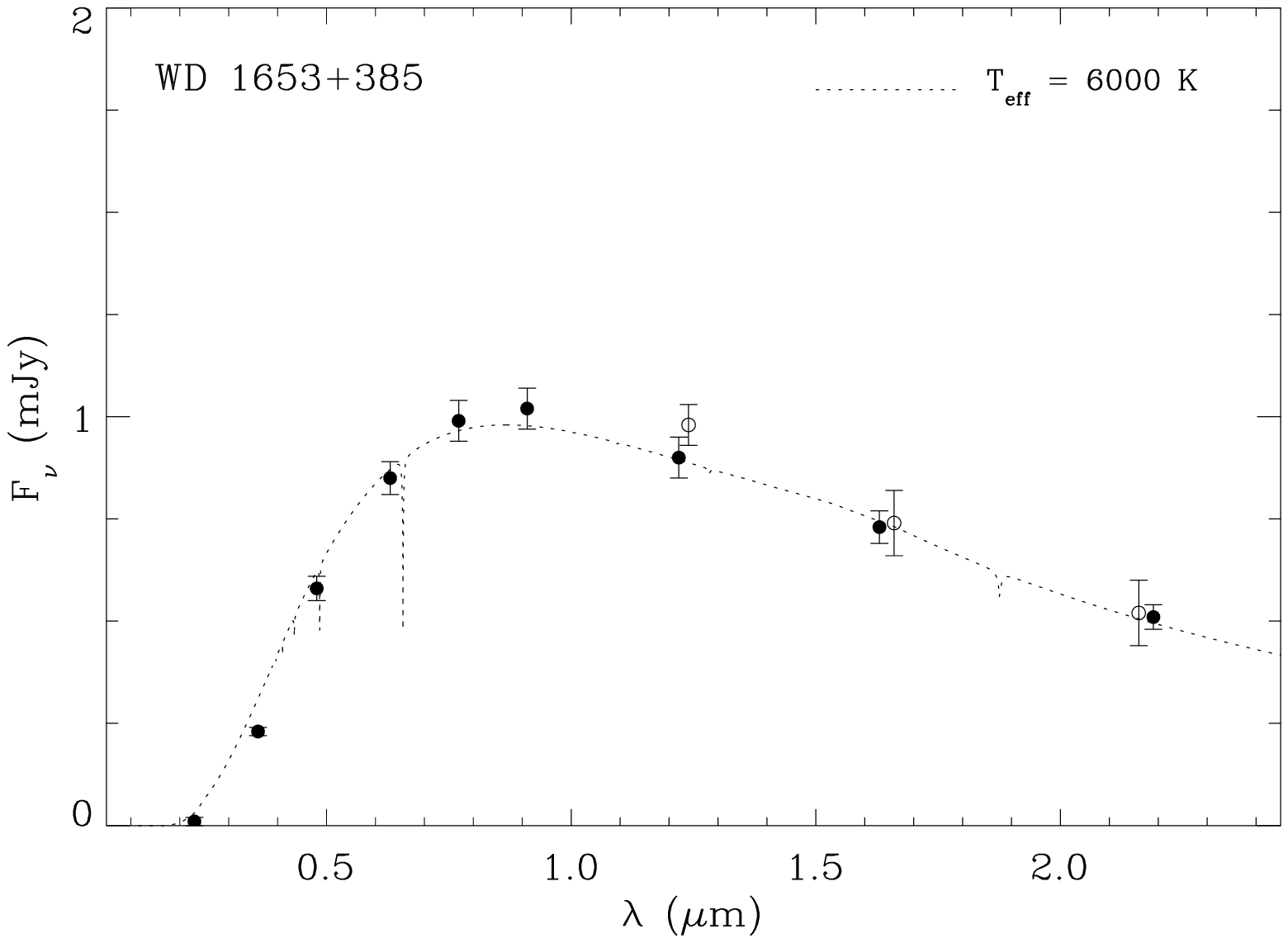}
\caption{Spectral energy distribution of NLTT 43806.  The solid circles are {\em GALEX} 
near-ultraviolet, SDSS $ugriz$, and IRTF $JHK$ photometry, while the open circles are 2MASS 
$JHK_s$ photometry.
\label{fig38}}
\end{figure*}

\begin{figure*}
\includegraphics[width=140mm]{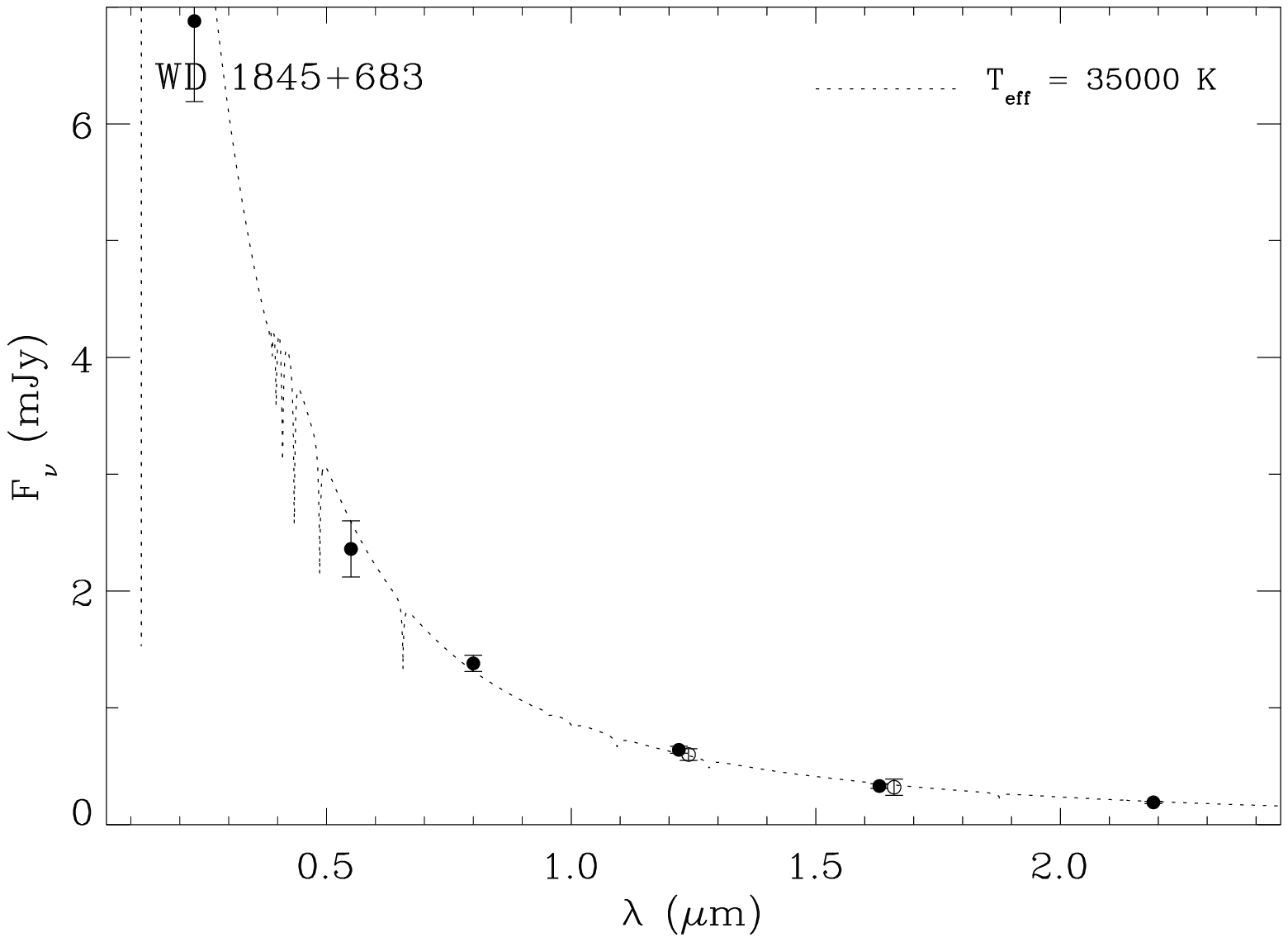}
\caption{Spectral energy distribution of KUV 1845$+$683.  The solid circles are {\em GALEX} 
near-ultraviolet, optical $VI$, and IRTF $JHK$ photometry, while the open circles are 2MASS 
$JHK_s$ photometry.  The {\em GALEX} far-ultraviolet flux of $14.1\pm0.7$ mJy is not shown for 
scaling purposes.
\label{fig39}}
\end{figure*}

\begin{figure*}
\includegraphics[width=140mm]{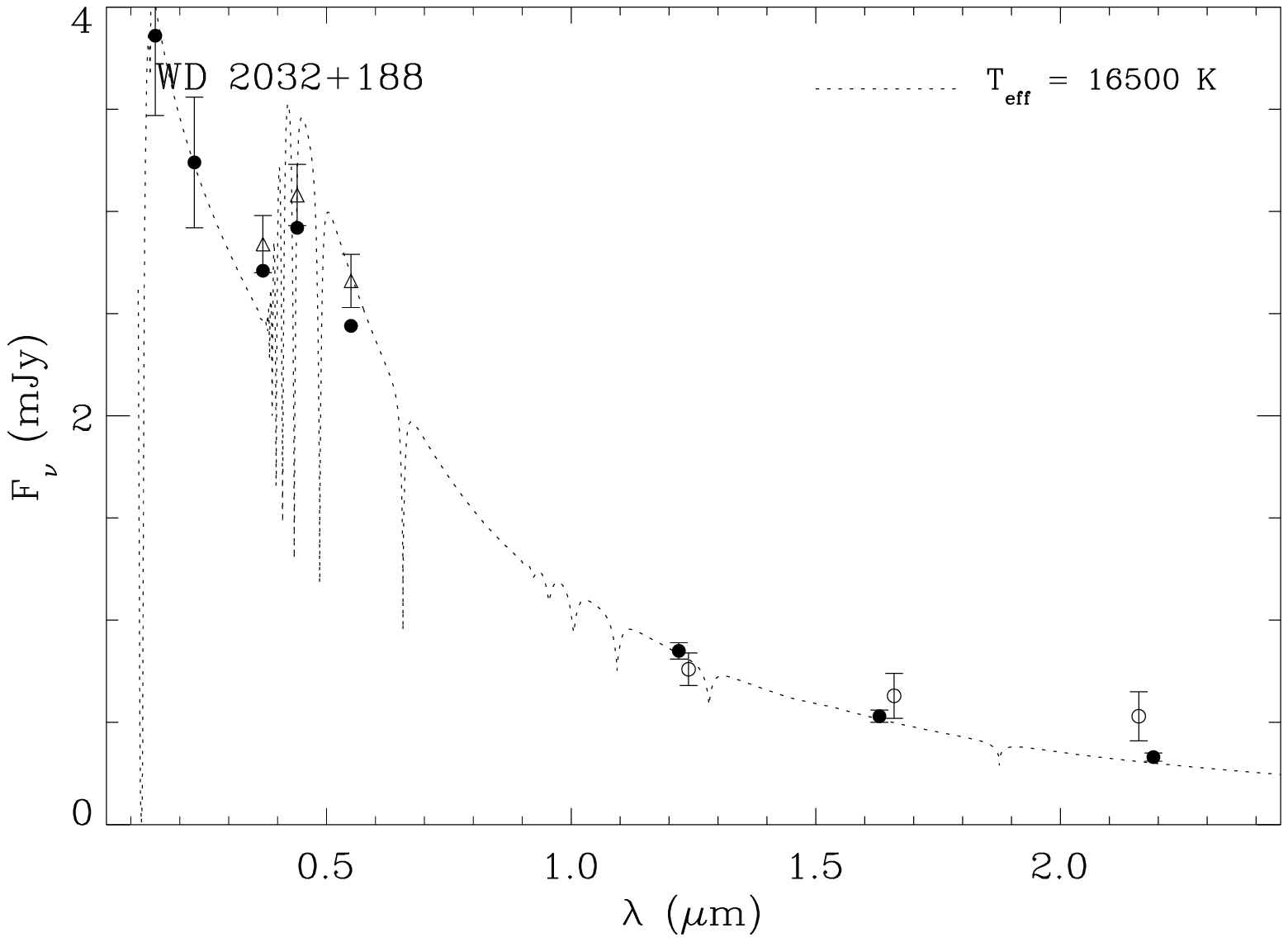}
\caption{Spectral energy distribution of GD 231.  The solid circles are {\em GALEX} far- and 
near-ultraviolet, optical $UBV$ (average values from \citealt{mcc99}), and IRTF $JHK$ 
photometry, while the open circles are 2MASS $JHK_s$ photometry.  The open triangles are 
$UBV$ from \citet{egg68}.
\label{fig40}}
\end{figure*}

\begin{figure*}
\includegraphics[width=140mm]{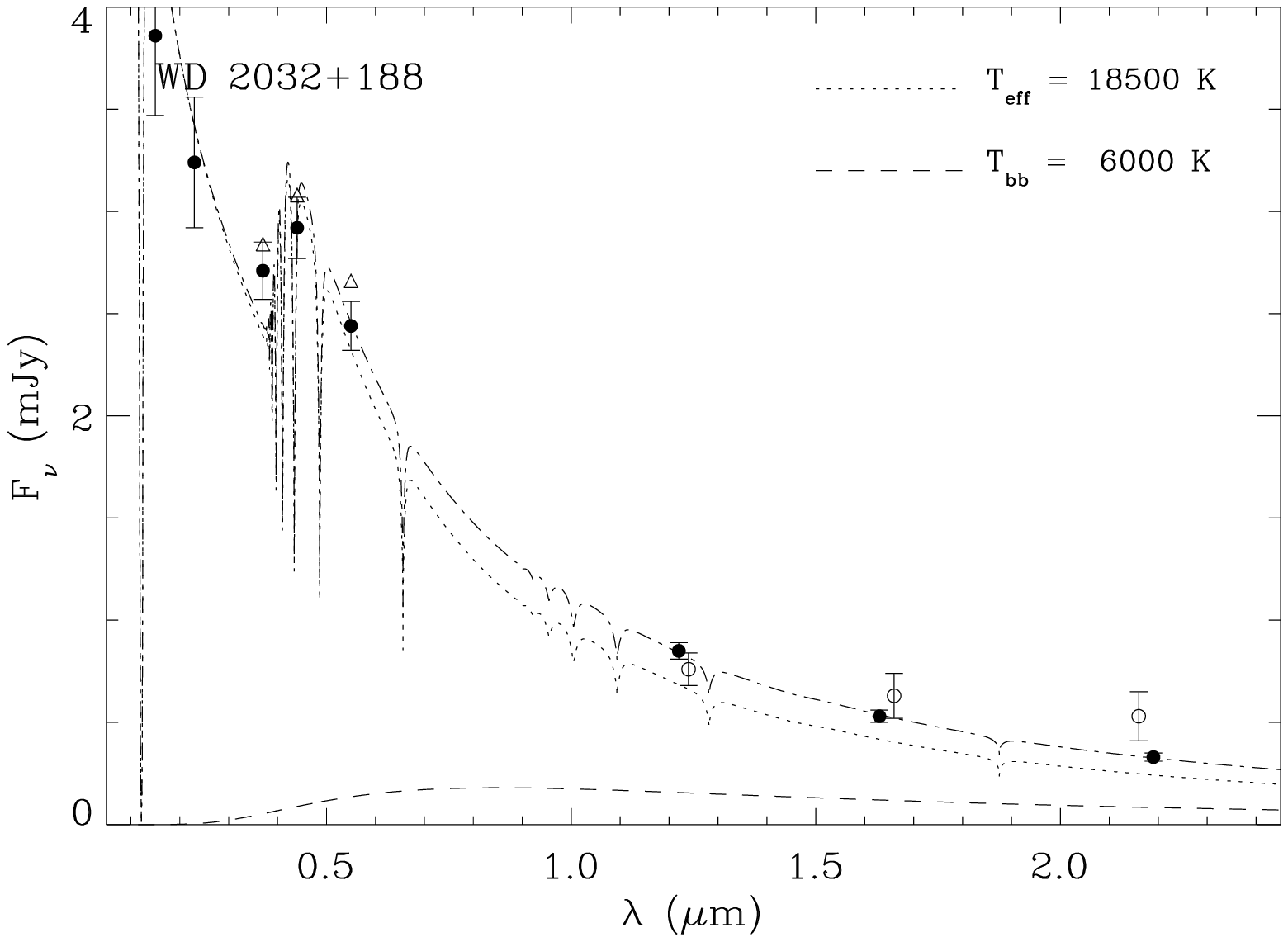}
\caption{Same as Figure \ref{fig40} but fitted with the effective temperature of \citet{ber92}.  In 
this case the near-infrared data reveal a mild photometric excess consistent with a cool white
dwarf companion.
\label{fig41}}
\end{figure*}

\begin{figure*}
\includegraphics[width=140mm]{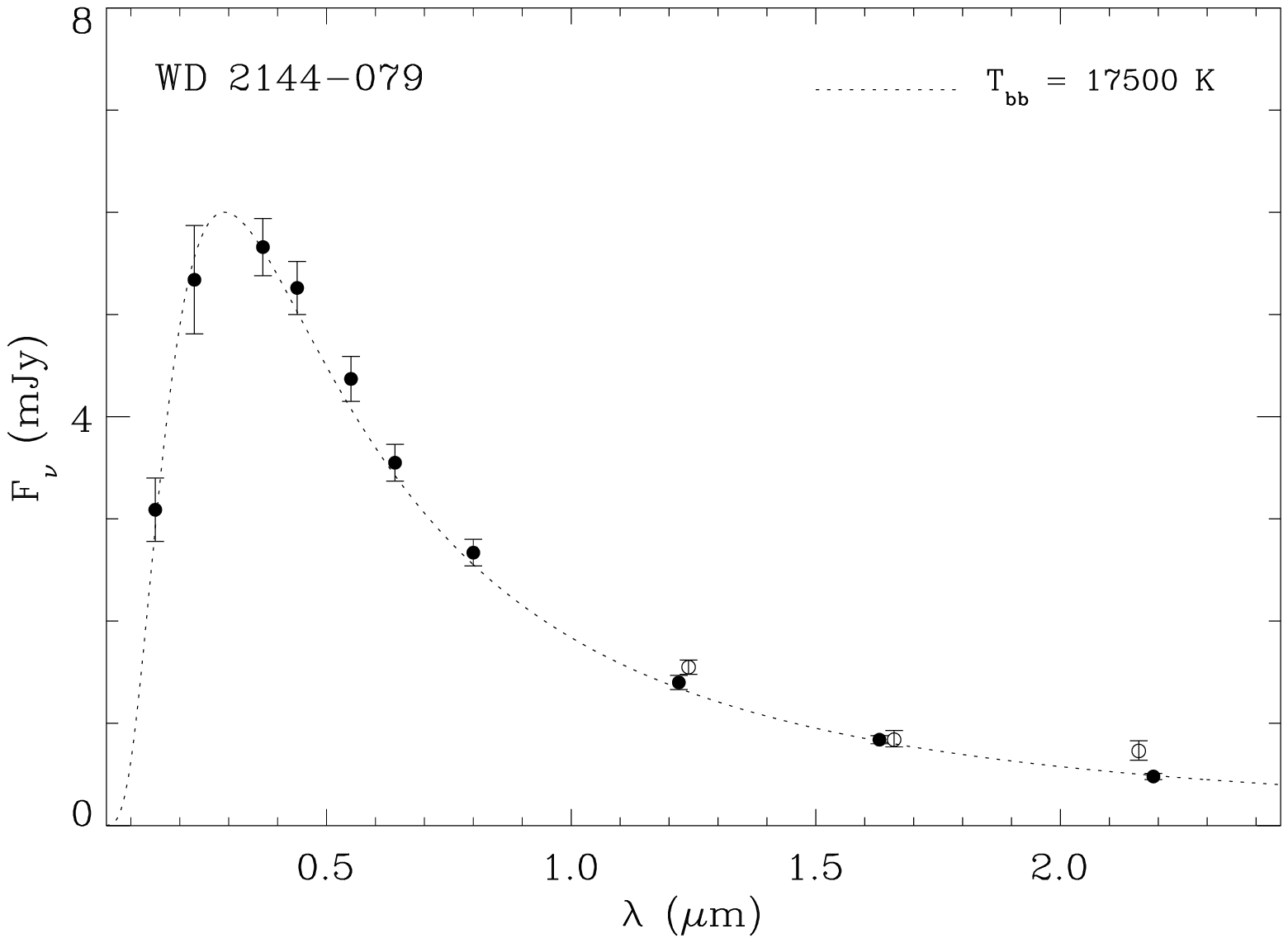}
\caption{Spectral energy distribution of G26-31.  The solid circles are {\em GALEX} far- and 
near-ultraviolet, optical $UBVRI$, and IRTF $JHK$ photometry, while the open circles are 
2MASS $JHK_s$ photometry.
\label{fig42}}
\end{figure*}

\begin{figure*}
\includegraphics[width=140mm]{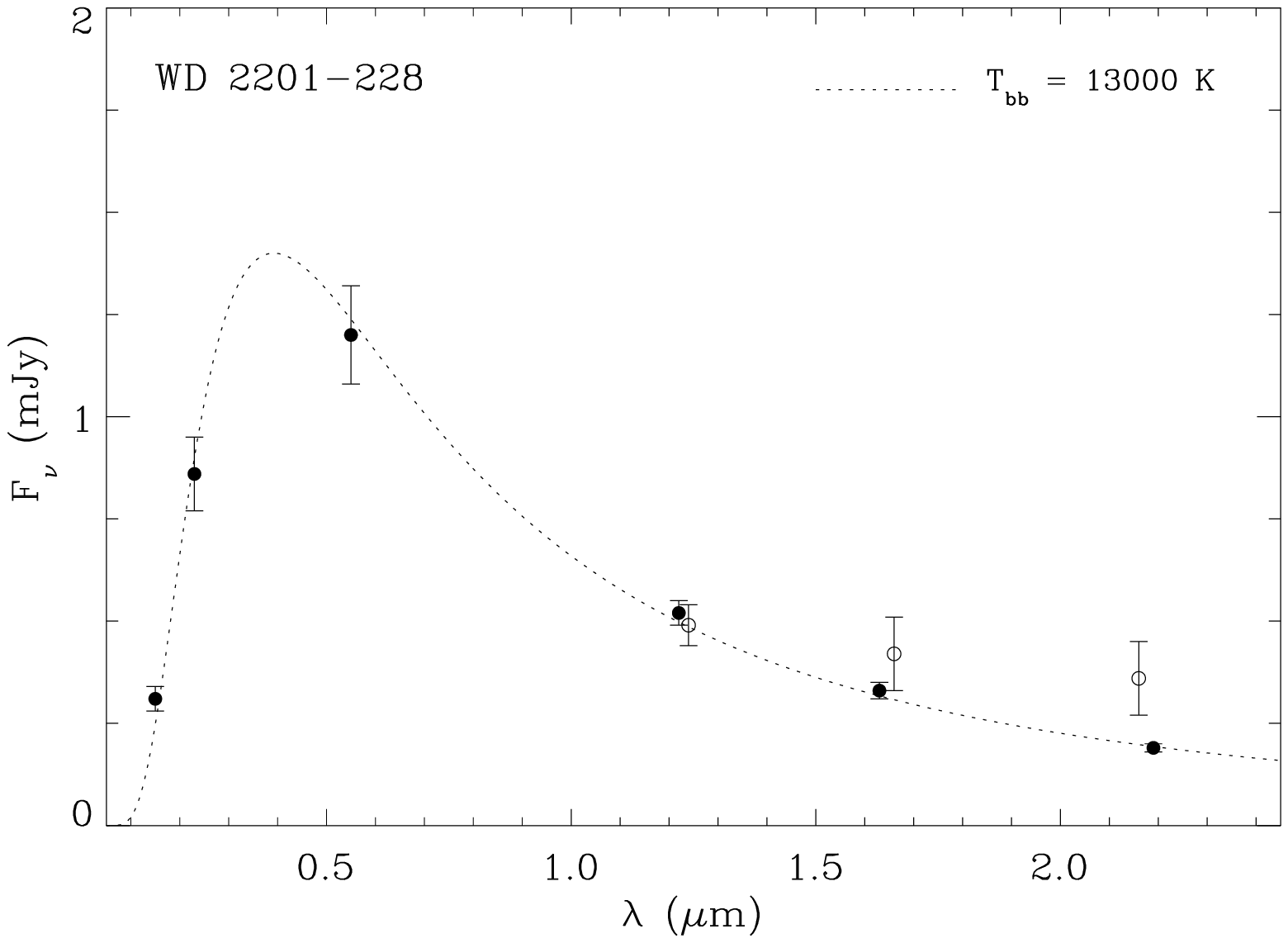}
\caption{Spectral energy distribution of HE 2201$-$228.  The solid circles are {\em GALEX} far- and 
near-ultraviolet, estimated optical $V$, and IRTF $JHK$ photometry, while the open circles are 
2MASS $JHK_s$ photometry.
\label{fig43}}
\end{figure*}

\begin{figure*}
\includegraphics[width=140mm]{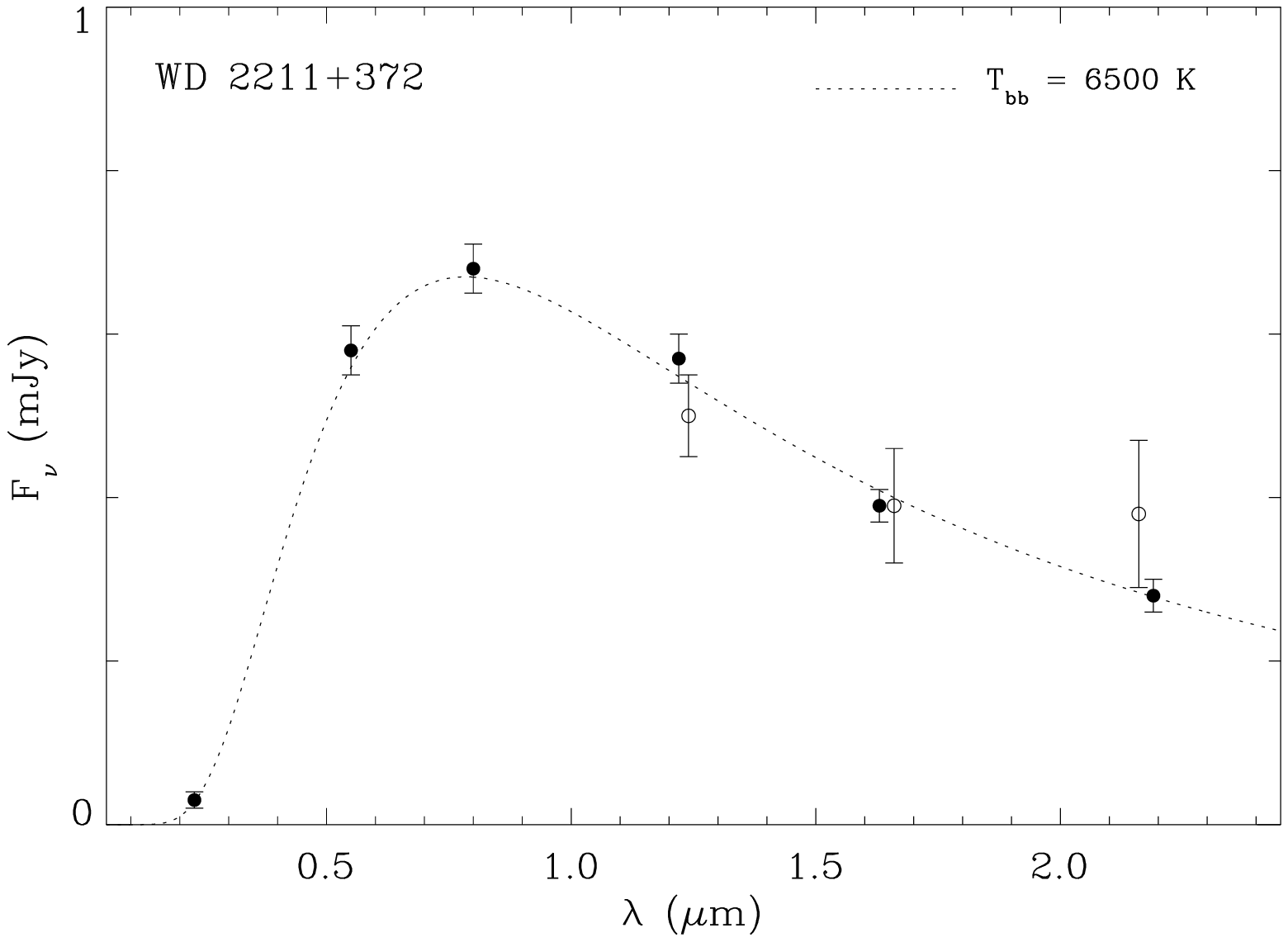}
\caption{Spectral energy distribution of LHS 3779.  The solid circles are {\em GALEX} 
near-ultraviolet, optical $VI$, and IRTF $JHK$ photometry, while the open circles are 2MASS 
$JHK_s$ photometry.
\label{fig44}}
\end{figure*}

\begin{figure*}
\includegraphics[width=140mm]{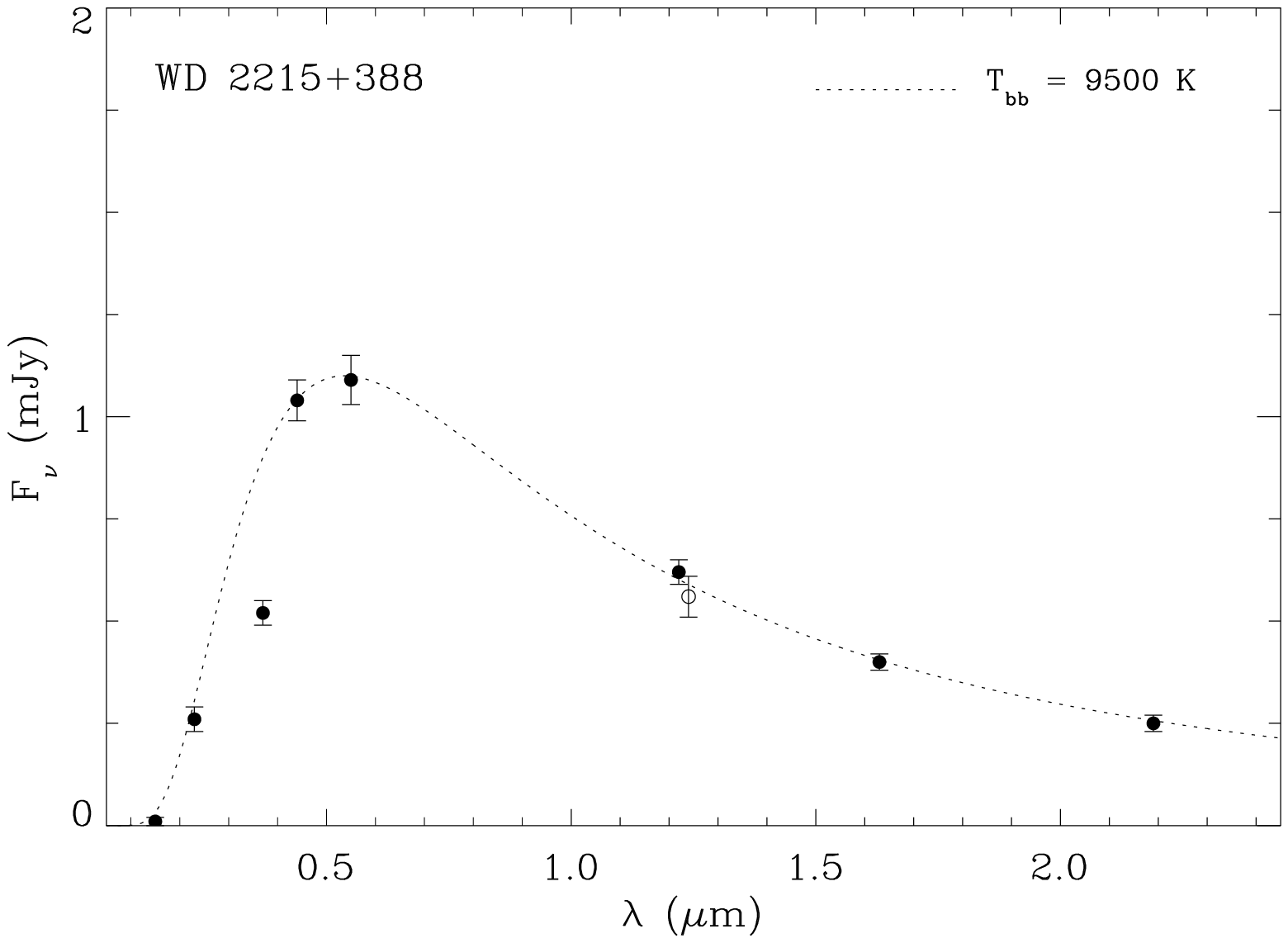}
\caption{Spectral energy distribution of GD 401.  The solid circles are {\em GALEX} far- and
near-ultraviolet, optical $UBV$, and IRTF $JHK$ photometry, while the open circles are 2MASS 
$JHK_s$ photometry.  This star has very strong Ca H and K absorption \citep{sio90}.
\label{fig45}}
\end{figure*}

\begin{figure*}
\includegraphics[width=140mm]{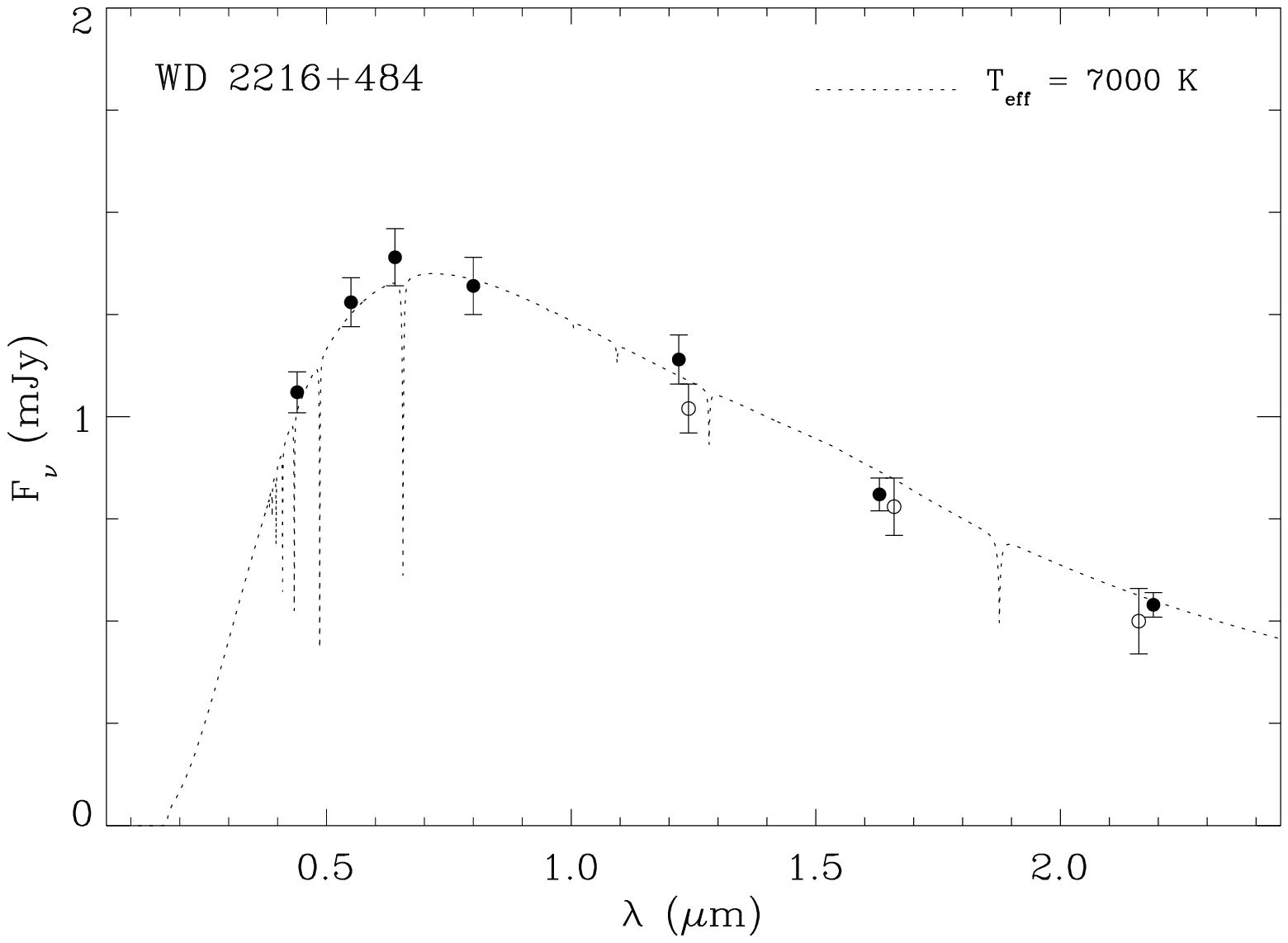}
\caption{Spectral energy distribution of GD 402.  The solid circles are optical $BVRI$, and IRTF 
$JHK$ photometry, while the open circles are 2MASS $JHK_s$ photometry.  It is possible the
full spectral energy distribution is better reproduced by two components \citep{ber90}.  There 
are no {\em GALEX} data available for this star.
\label{fig46}}
\end{figure*}

\begin{figure*}
\includegraphics[width=140mm]{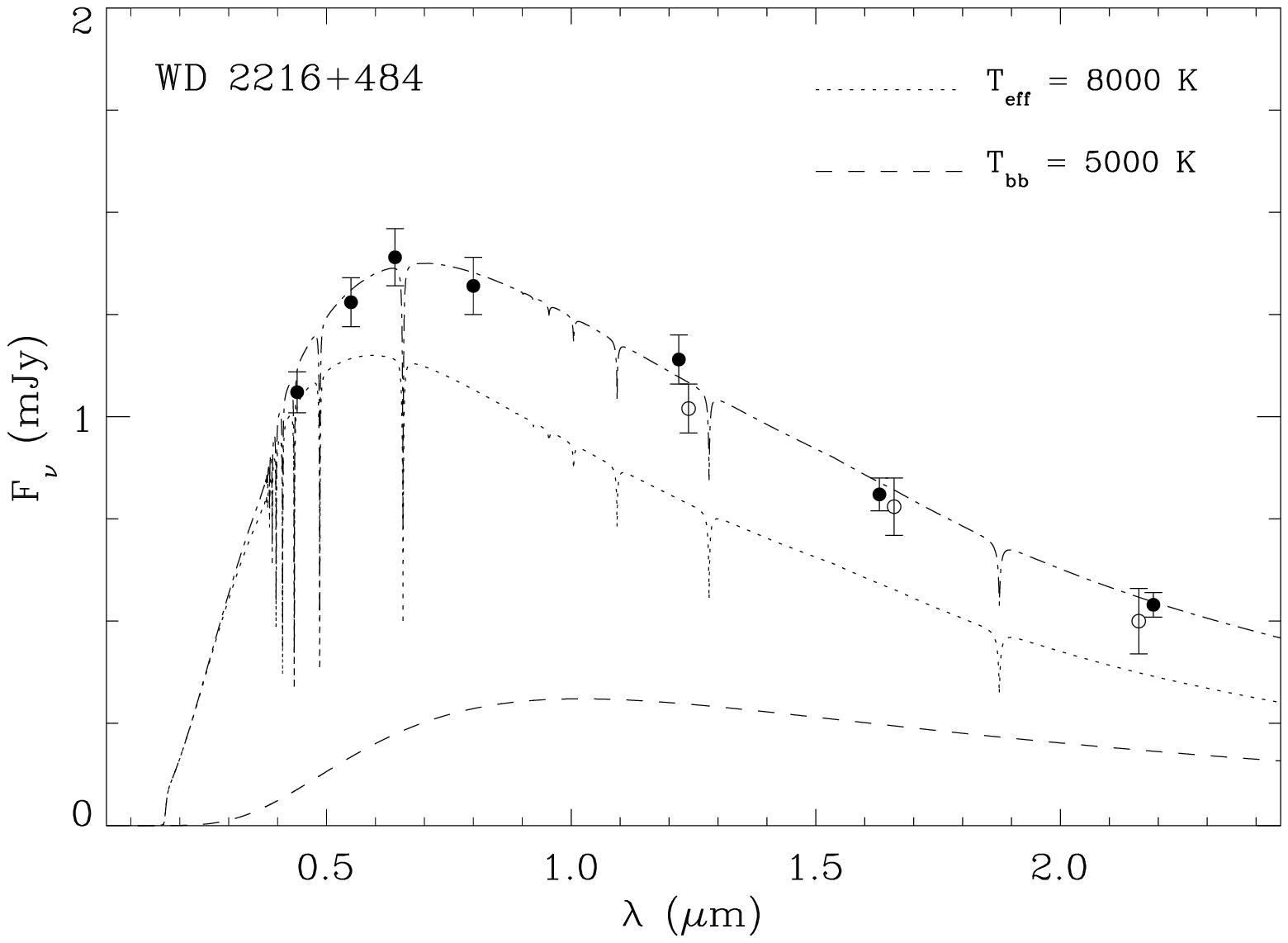}
\caption{Same as Figure \ref{fig46} but now fitted with two white dwarf components, similar to
the suggestion of \citet{ber90}.
\label{fig47}}
\end{figure*}

\begin{figure*}
\includegraphics[width=140mm]{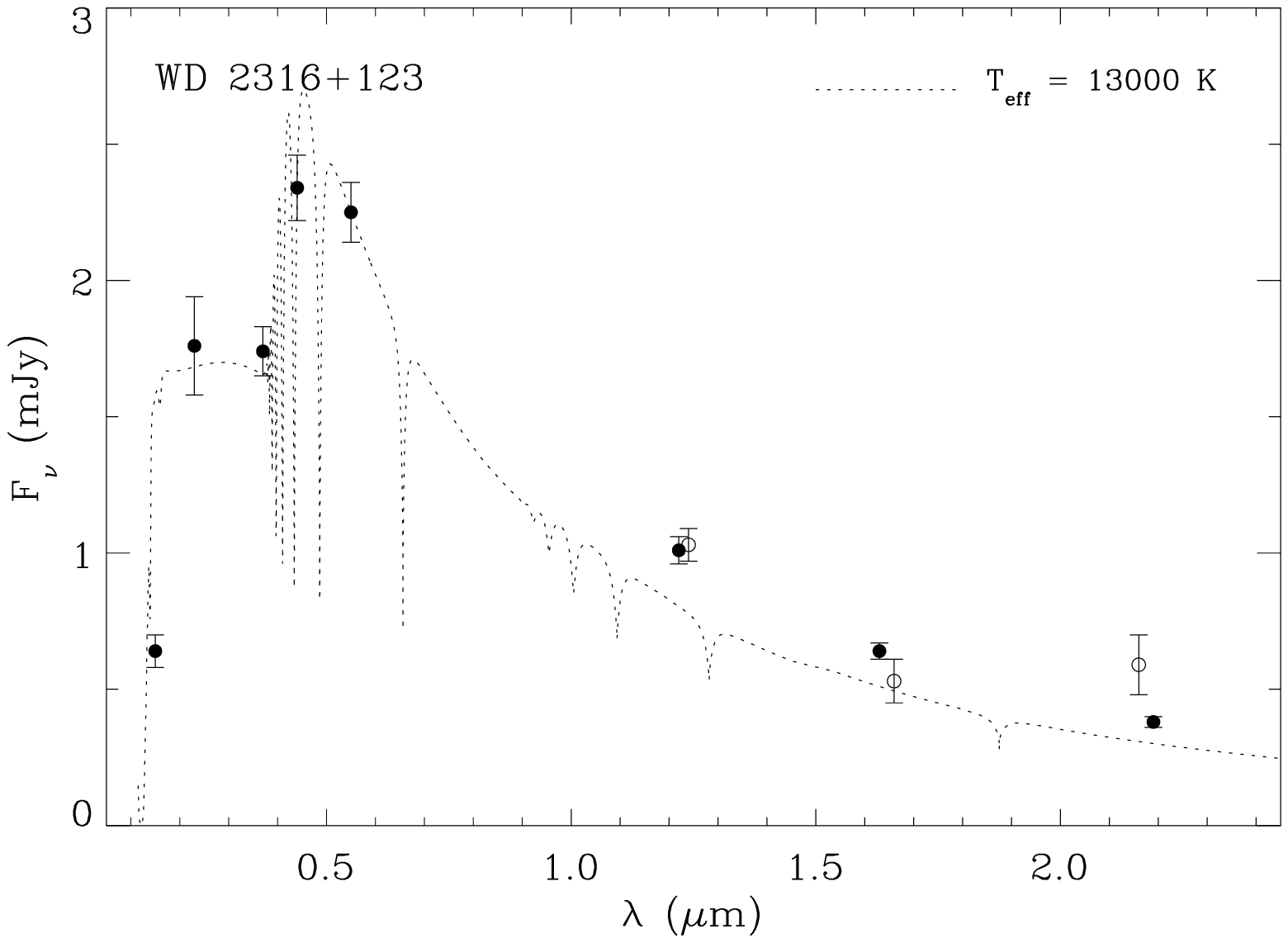}
\caption{Spectral energy distribution of KUV 2316$+$123.  The solid circles are {\em GALEX} far- and
near-ultraviolet, optical $UBV$, and IRTF $JHK$ photometry, while the open circles are 2MASS 
$JHK_s$ photometry.  The spectral energy distribution is not reproduced by a single stellar 
effective temperature, non-magnetic DA model.
\label{fig48}}
\end{figure*}

\begin{figure*}
\includegraphics[width=140mm]{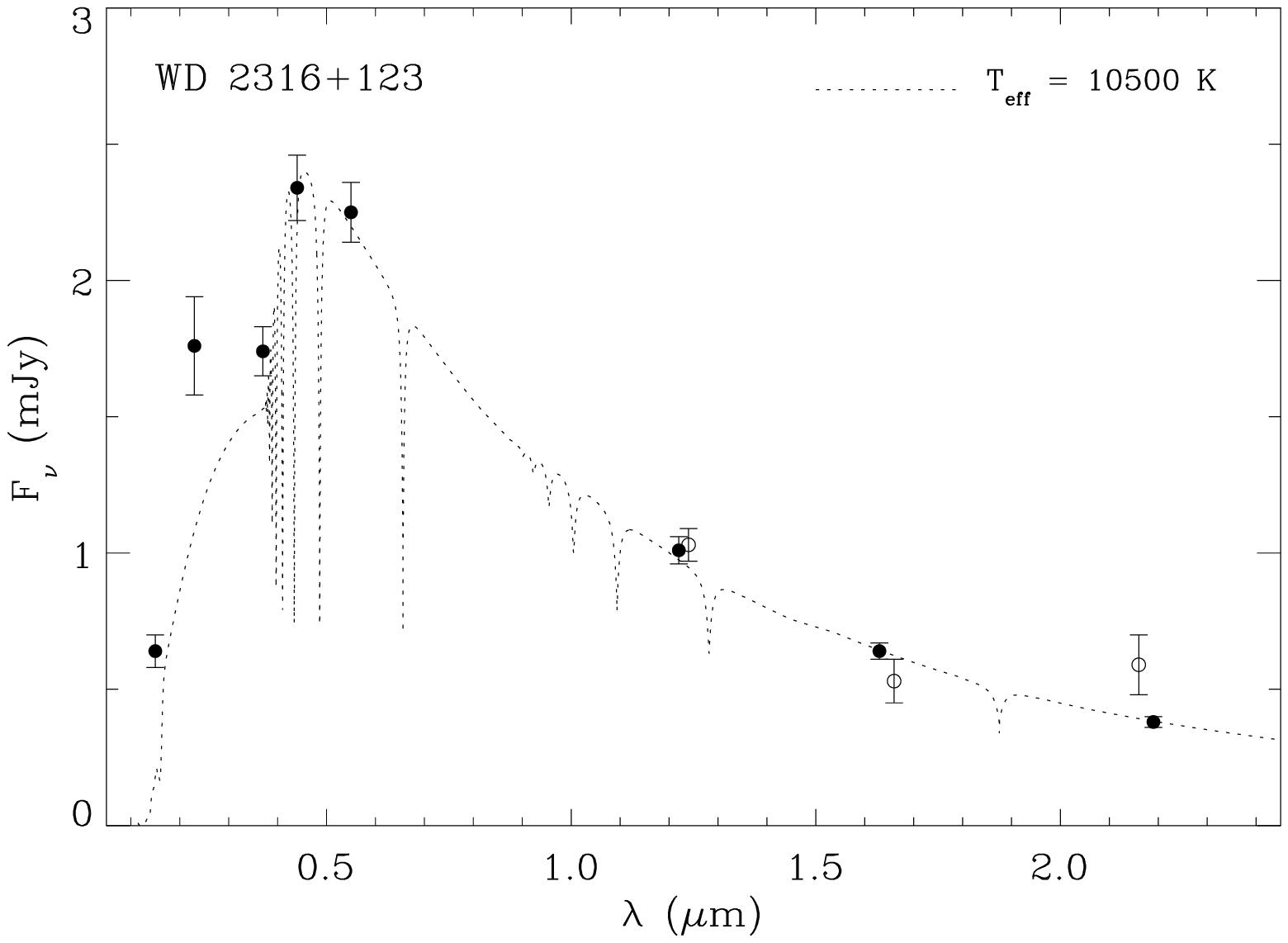}
\caption{Same as Figure \ref{fig48} but now fitted with the effective temperature estimate of
\citet{lie85}; this model underpredicts the ultraviolet fluxes.
\label{fig49}}
\end{figure*}

\begin{figure*}
\includegraphics[width=140mm]{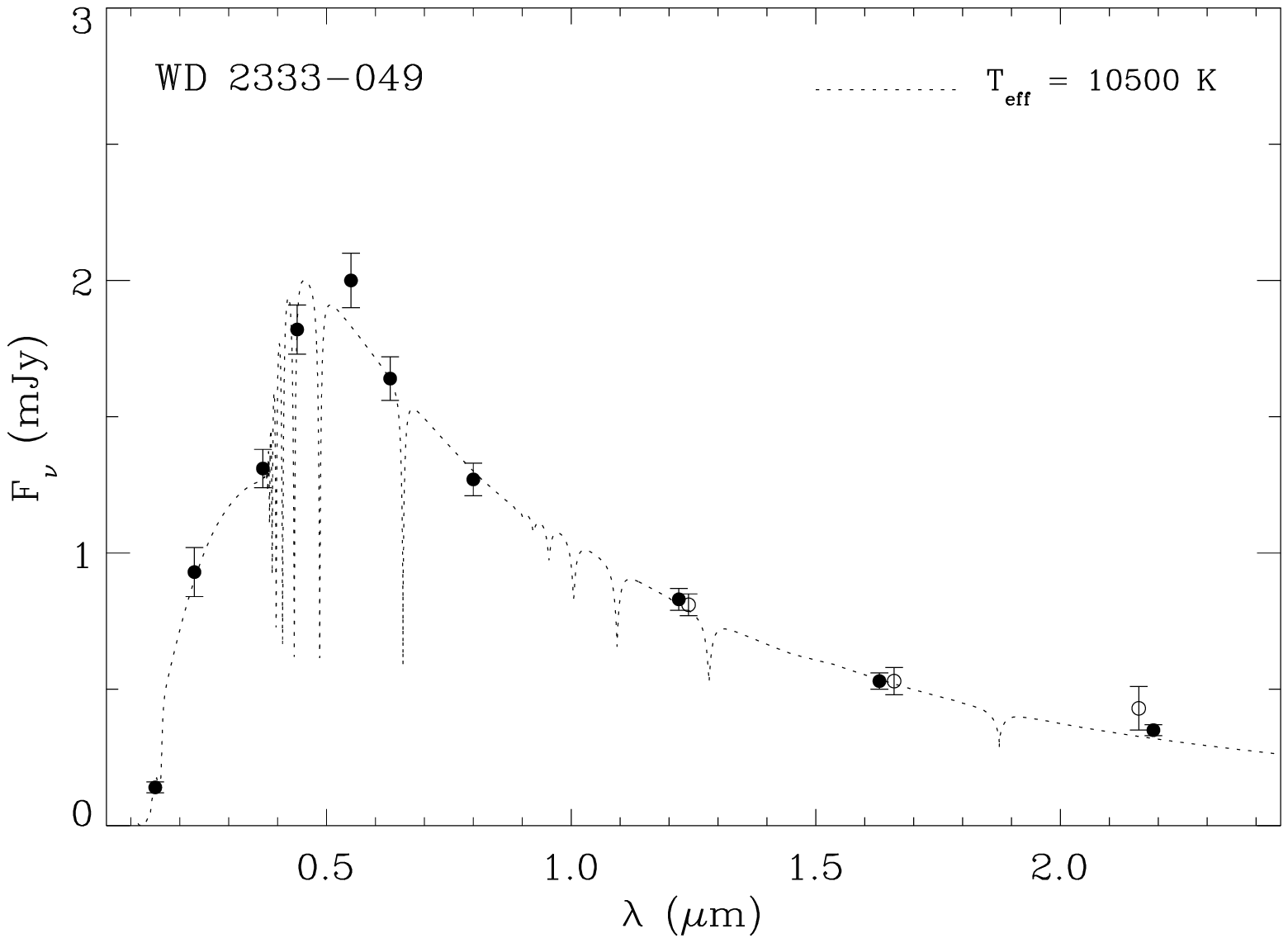}
\caption{Spectral energy distribution of G157-82.  The solid circles are {\em GALEX} far- and
near-ultraviolet, optical $UBVrI$, and IRTF $JHK$ photometry, while the open circles are 2MASS 
$JHK_s$ photometry.  This star may have a slight $K$-band excess.
\label{fig50}}
\end{figure*}

\begin{figure*}
\includegraphics[width=140mm]{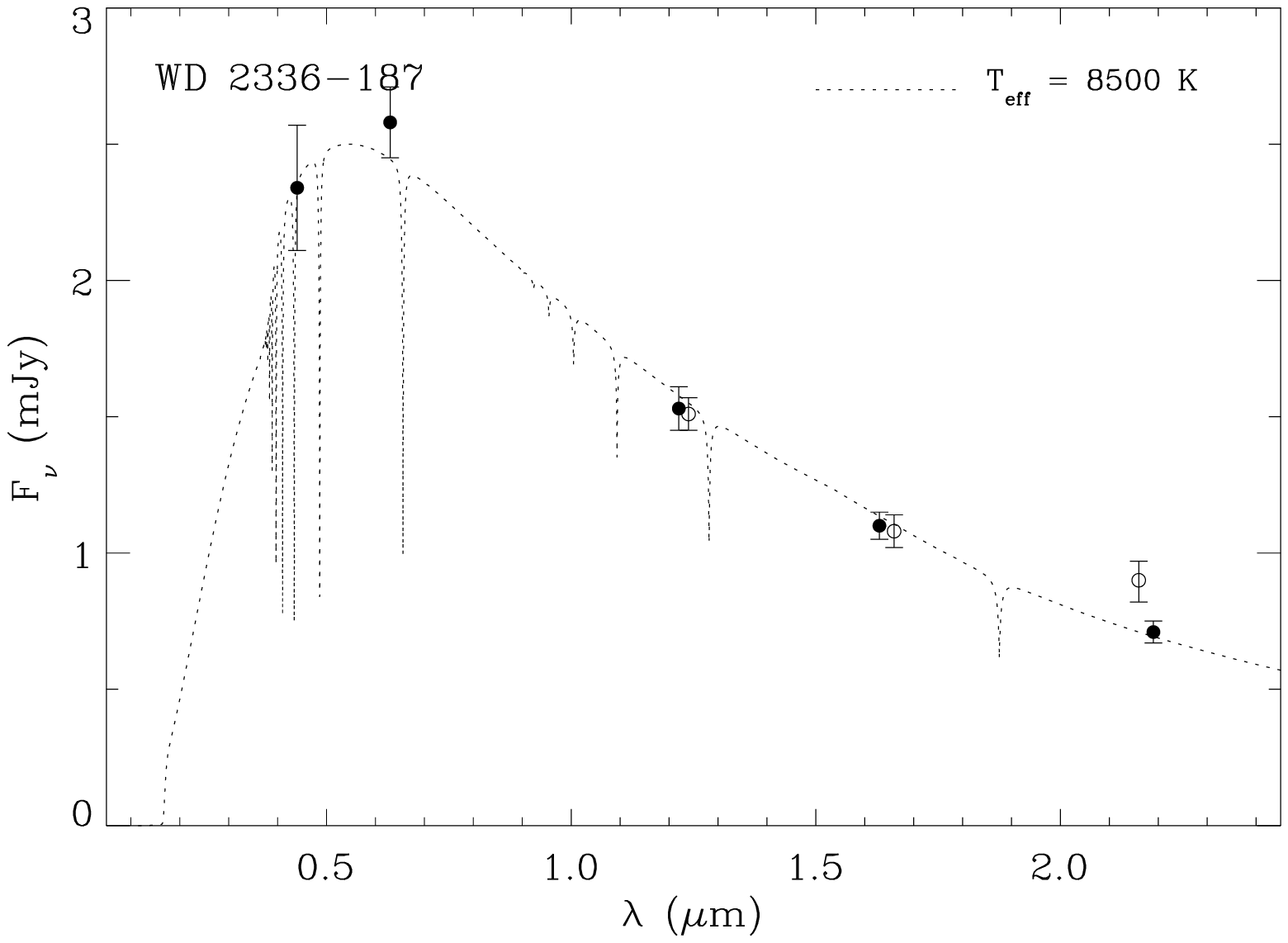}
\caption{Spectral energy distribution of G273-97.  The solid circles are {\em GALEX} far- and
near-ultraviolet, optical $Br$, and IRTF $JHK$ photometry, while the open circles are 2MASS 
$JHK_s$ photometry.
\label{fig51}}
\end{figure*}

\begin{figure*}
\includegraphics[width=140mm]{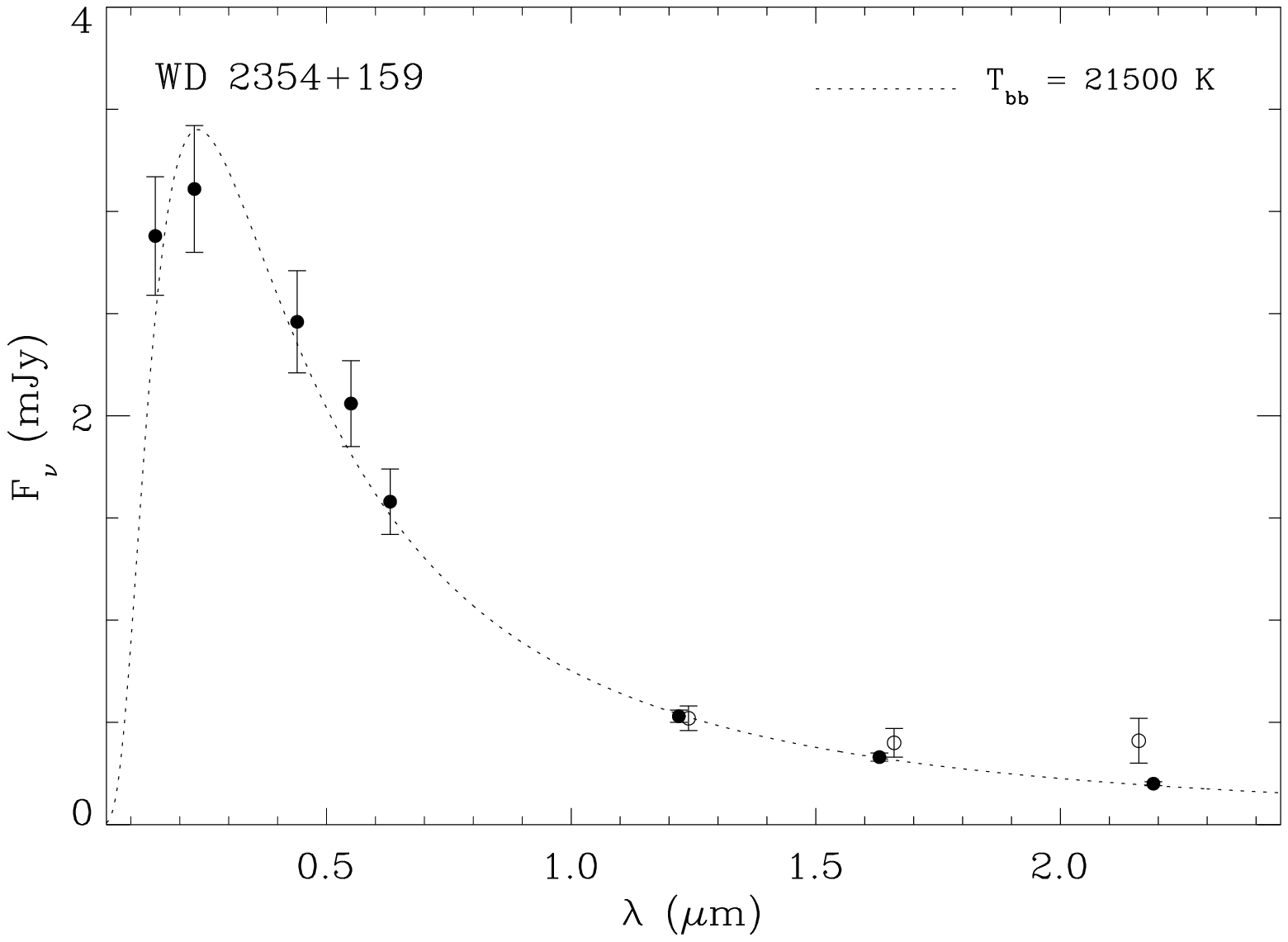}
\caption{Spectral energy distribution of PG 2354$+$159.  The solid circles are {\em GALEX} far- and
near-ultraviolet, optical $BVr$, and IRTF $JHK$ photometry, while the open circles are 2MASS 
$JHK_s$ photometry.
\label{fig52}}
\end{figure*}

However, by bulk, of the 27 suggested 2MASS candidate excesses, there are 17 white dwarfs 
in Table \ref{tbl2} with a confirmed or likely IRTF-measured excess.  Interestingly, 10 of these 
are sufficiently mild that $(J-K)_{\rm IRTF}<0.3$ or $(J-H)_{\rm IRTF}<0.2$, and required shorter
wavelength data to be recognized with confidence.  Five additional sources display relatively 
strong IRTF excesses in all three near-infrared bandpasses; arising from low mass stellar 
companions, these five objects are also fairly unambiguous in 2MASS, where they suffer 
from low S/N $H$- and $K_s$-band fluxes or source confusion.  Figure \ref{fig54} graphically 
represents all the IRTF near-infrared excesses, including the milder and more subtle cases, 
by plotting the 2MASS versus IRTF $J-H$ and $J-K$ measured colors for all white dwarfs with 
photometry in both datasets.  As a rule, the plots show that the 2MASS data overpredicts the 
IRTF colors, especially in $J-K$, and that nine of 22 white dwarfs have $(J-K)_{\rm IRTF}>0.3$ 
as measured by the IRTF and 2MASS respectively.

\subsection{Several Varieties of Near-Infrared Excess at White Dwarfs}

Three metal-contaminated white dwarfs with dust disks and mid-infrared excess emission 
also display excess flux in the IRTF dataset: GD 16, GD 56 and PG 1457$-$086 \citep{far09,
jur07}.  Of the known white dwarfs with orbiting dust and an infrared excess, only about half 
reveal themselves shortward of 3 $\mu$m \citep{far09,kil06}, and, where present, these
near-infrared data constrain the inner dust temperature \citep{jur09}. Nine other metal-polluted 
white dwarfs have IRTF data consistent with photospheric emission, all of which have published 
or pending {\em Spitzer} IRAC observations.

Near-infrared excess due to cool, unevolved dwarf companions is typically evident and strong
already by 1 $\mu$m \citep{far05}.  Yet nearly all five white dwarf - red dwarf binaries confirmed 
by the IRTF observations had uncertain 2MASS photometric excesses which required more 
accurate measurements, primarily to obtain the $I-K$ color of the companion, which in turn 
constrains spectral type (and mass for a given metallicity).  A few cases were followed up in 
order to obtain higher spatial resolution images \citep{far06} than are available via the 2MASS 
image server.

\begin{table*}
\centering
\begin{minipage}{80mm}
\caption{Constraints on Multiple System Components\label{tbl3}} 
\begin{tabular}{@{}lllcc@{}}
\hline

WD		&Companion	&Spectral		&Separation	&$d$\\
		&			&Type		&(arcsec)		&(pc)\\
 
\hline

0023$-$109		&G158-78B			&dM0		&59.4		&92\\
				&G158-78C			&DC8		&...			&...\\
0235$+$064		&PG	0235$+$064B		&dM3.5		&7.8			&62\\
0257$-$005		&KUV 0257$-$005B		&dM4.5		&1.0			&494\\
0518$+$333		&G86-B1AC			&dM3		&7.4			&65\\
0956$+$045		&PG 0956$+$045B		&dM5		&2.0			&114\\
1013$-$010		&G53-38B			&DC8		&...			&...\\
1108$+$325		&TON 60B			&dM3		&0.2			&560\\
1133$+$489		&PG 1133$+$489B		&dM5		&0.1			&370\\
1140$+$004		&WD 1140$+$004B		&dM5		&...			&240\\
1619$+$525		&1619$+$525B		&dM2		&2.6			&300\\
				&1619$+$525C		&dM5		&0.5			&300\\

\hline
\end{tabular}

\end{minipage}
\end{table*}

Few double white dwarf binaries are identified in the near-infrared, but such discoveries are 
not unprecedented \citep{far05,nel05,zuc97}.  The cooler component in a double degenerate 
binary should be outshone by its primary at all wavelengths, in the ultraviolet and optical due to 
effective temperature, and in the infrared owing to radius.  The component with an older cooling 
age should have finished the main sequence first, descending from the higher mass progenitor
star, and becoming the higher mass and smaller radius degenerate star \citep{dob06}.  These
systems can still exhibit an excess from the smaller, cooler component, as can be seen for 
G158-78 and G53-38, where companions have been indicated by independent means.  The 
IRTF photometry tentatively identifies three more double degenerate suspects:  G62-46, GD 
231, and GD 402.

The near-infrared data on the confirmed or likely magnetic DA white dwarfs KPD 0253$+$508
and KUV 2316$+$123, and DQp white dwarfs LHS 2293 and LP 494-12 range from surprising 
to spectacular.  These stars are all presumably single at present, at least until shown otherwise.  
Non-interacting stellar and substellar companions to magnetic white dwarfs are rare (or possibly 
nonexistent; \citealt{lie05b}), but that leaves room for a degenerate companion, as is suspected 
in the case of G62-46.  In any case, non-magnetic models clearly fail to reproduce the spectral 
energy distributions of these stars -- the near-infrared fluxes in particular -- while magnetic models 
are lacking.  The newly acquired IRTF data provides a new impetus to understand magnetic white 
dwarf atmospheres, which are still poorly constrained, especially in the infrared.  Follow up 
spectroscopy is warranted for these cases, ideally with broad wavelength coverage from 
short optical wavelengths through the near-infrared.

\section*{ACKNOWLEDGMENTS}

The author thanks referee T. Von Hippel for comments which improved the quality and
clarity of the manuscript, D. Koester for the use of his models (and B. G\"ansicke for providing 
them), M. Burleigh for many helpful discussions, A. Gianninas and G. Schmidt for sharing their 
optical spectra of PG 0939$+$071 and LHS 2293, respectively.  The data presented here were 
obtained at the Infrared Telescope Facility, which is operated by the University of Hawaii under 
Cooperative Agreement no. NCC 5-538 with NASA, Science Mission Directorate, Planetary 
Astronomy Program.  This publication makes use of data products from the Two Micron All Sky 
Survey, which is a joint project of the University of Massachusetts and the Infrared Processing 
and Analysis Center / California Institute of Technology, funded by NASA and the National 
Science Foundation.  This work includes data taken with the NASA Galaxy Evolution Explorer, 
operated for NASA by the California Institute of Technology under NASA contract NAS5-98034.  
Some data presented herein are part of the Sloan Digital Sky Survey, which is managed by the 
Astrophysical Research Consortium for the Participating Institutions (http://www.sdss.org/).

\appendix

\section{INDIVIDUAL SPECTRAL ENERGY DISTRIBUTIONS}

\clearpage

\clearpage

\clearpage

\clearpage

\clearpage

\clearpage

\clearpage

\clearpage

\clearpage

\clearpage


\clearpage

\clearpage

\clearpage

\clearpage


\clearpage

\clearpage

\clearpage

\clearpage

\clearpage

\clearpage

\clearpage

\clearpage

\clearpage

\clearpage

\clearpage

\label{lastpage}

\end{document}